\documentclass{ws-ijmpa_mod}
\usepackage[super]{cite}
\usepackage{xcolor}
\usepackage[verbose,hypertexnames=false]{hyperref}
\hypersetup{colorlinks=false,allbordercolors=blue,pdfborderstyle={/S/U/W 1}}
\usepackage{graphicx}
\usepackage{subcaption}
\usepackage{amssymb}
\usepackage{comment}

\usepackage{url}

\begin{document}

\markboth{Romy H.S.Budhi et al.}{Preventing Curvature Singularities  in $f(R)$ Dark Energy Models}

%
\catchline{}{}{}{}{}
%

\title{Preventing Curvature Singularities  in $f(R)$ Dark Energy Models}

\author{Romy Hanang Setya Budhi\footnote{corresponding author: romyhanang@ugm.ac.id}}

\address{Physics Department, Universitas Gadjah Mada,\\
Yogyakarta 55281, Indonesia\\
}

\author{Nehla Shobcha}

\address{Physics Department, Universitas Gadjah Mada,\\ 
Yogyakarta 55281, Indonesia\\
}

\author{Norma Sidik Risdianto}

\address{Dept. of Physics Education, Universitas Islam Negeri Sunan Kalijaga,\\
Jl. Marsda Adisucipto 55281, Yogyakarta, Indonesia\\
}

\maketitle

\footnotetext{* Preprint of an article submitted for consideration in \textit{International Journal of Modern Physics A} © 2025 World Scientific Publishing Company. \url{https://www.worldscientific.com/worldscinet/ijmpa}}


\begin{abstract}
The curvature singularity problem in $f(R)$ dark energy models poses a significant challenge to their viability as alternatives to the $\Lambda$CDM paradigm. In this work, we investigate the possibility of resolving this issue by incorporating higher-order corrections that are compatible with the inflationary phase. We analyze the effects of adding $R^2$, $R^{\frac{m+2}{m+1}}$ and $\alpha$-attractor representation in $f(R)$ gravity terms to types of dark energy $f(R)$ models, focusing on their ability to prevent singularities in high-density environments. Our results demonstrate that these corrections can effectively stabilize the models, ensuring their consistency across both inflationary and late-time cosmological scales.
\end{abstract}


\section{Introduction}

Recent cosmological observations, such as those from the Planck satellite and Type Ia supernovae, strongly support the conclusion that the late-time universe is undergoing accelerated expansion \cite{Akrami_2020}. This phenomenon is commonly attributed to the presence of dark energy, a mysterious component that constitutes approximately 70\% of the universe's energy density. The simplest explanation for dark energy is the cosmological constant $\Lambda$, which fits well within the $\Lambda$CDM model. However, the cosmological constant faces theoretical challenges, such as the fine-tuning problem and the coincidence problem \cite{Amendola_2007, Nojiri_2008}. Alternative explanations include dynamical dark energy models, modified gravity theories, and interactions between dark energy and dark matter \cite{Tsujikawa2008, Bamba:2010iy}.
		
		One promising alternative to the cosmological constant is $f(R)$ gravity, a modification of General Relativity (GR) where the gravitational action is generalized to an arbitrary function of the Ricci scalar $R$ \cite{DeFelice_2010, Sotiriou:2008rp}. This theory naturally accommodates dark energy without the need for additional fields, as the modified dynamics of gravity can drive the late-time accelerated expansion. Several successful $f(R)$ models, such as the Hu-Sawicki model \cite{Hu_2007} and the Starobinsky model \cite{Starobinsky:2007hu}, have been proposed and tested against observational data. However, these models often suffer from curvature singularities, where the Ricci scalar $R$ diverges in high-density environments, leading to unphysical predictions \cite{Lee2012, Bamba_2011}. Previous attempts to resolve these singularities include introducing additional terms in the $f(R)$ function or modifying the theory to ensure stability under cosmological perturbations \cite{Appleby:2009uf, Sebastiani2013, Lee2012, Dutta2015, geng2015constraints}.
		
		In this paper, we address the curvature singularity problem in $f(R)$ dark energy models by incorporating higher-order corrections that are compatible with the inflationary phase. Our goal is to demonstrate that such corrections can effectively cure the singularities while maintaining the viability of the models across different cosmological scales \cite{Cognola:2007zu, geng2015constraints, Oikonomou2017, Nojiri_2008, Budhi_Syamputra_2025,  budhi2024constraints}. We explore several modifications, including the addition of $R^2$, $R^{\frac{m+2}{m+1}}$ terms, and the $\alpha$-attractor polynomial extension,  and  analyze their impact on the stability and dynamics of the models \cite{Nehla2025}. By doing so, we aim to identify a viable $f(R)$ gravity model that can consistently describe both the early inflationary epoch and the late-time accelerated expansion, while satisfying all cosmological constraints. The paper is organized as follows: in Section 2, we review the theoretical framework of $f(R)$ gravity and its application to dark energy. In Section 3, we discuss the curvature singularity problem in detail. Sections 4, 5 and 6 present our proposed solutions, and Section 7 concludes with a discussion of the implications of our findings.

\section{Viable $f(R)$ Gravity Theories}\label{sec:2}
\subsection{Overview of $f(R)$ Gravity Models}
		
		In modern cosmology, $f(R)$ gravity theories have gained prominence as a compelling extension of General Relativity (GR), achieved by generalizing the Einstein-Hilbert action. Rather than adopting the Ricci scalar $R$ as in GR, these theories introduce a general function $f(R)$ of the scalar curvature. The action for $f(R)$ gravity is written as
		\begin{equation}
			\label{eq:action}
			S = \frac{1}{2\kappa^2} \int d^{4}x \sqrt{-g} f(R) + S_{M},
		\end{equation}
		where $\kappa^{2} = 8\pi G$, $G$ denotes Newton’s gravitational constant, $g$ is the determinant of the metric tensor $g_{\mu\nu}$, and $S_{M}$ represents the matter action, which includes both relativistic and non-relativistic components. Varying this action with respect to the metric $g_{\mu\nu}$ yields the modified field equations,
		\begin{equation}
			\label{eq:field}
			f_R R_{\mu\nu} - \frac{f}{2}g_{\mu\nu} - \left(\nabla_{\mu}\nabla_{\nu} - g_{\mu\nu}\Box\right) f_R = \kappa^{2}T^{(M)}_{\mu\nu},
		\end{equation}
		where $f_R \equiv df(R)/dR$, $\nabla_{\mu}$ denotes the covariant derivative, $\Box \equiv g^{\mu \nu}\nabla_{\mu} \nabla_{\nu}$ is the d'Alembertian operator, and $T^{(M)}_{\mu\nu}$ is the energy-momentum tensor for matter. Unlike the second-order Einstein field equations in GR, these equations are fourth-order in derivatives of the metric, leading to a richer but more complex dynamical structure \cite{Sotiriou:2008rp}.
		
\subsection{Inflation in $f(R)$ Gravity}
		
		Inflation refers to a brief epoch of accelerated cosmic expansion in the early universe and is now a cornerstone of the standard cosmological paradigm. It resolves several major issues in the original Big Bang framework, including the horizon, flatness, and monopole problems \cite{Guth1981, Starobinsky1980}. While the conventional picture relies on a scalar field (the inflaton) within GR to drive this expansion, $f(R)$ gravity offers a purely geometric alternative for realizing inflationary dynamics \cite{Nojiri2011, DeFelice2010}. In this framework, the inflationary behavior emerges from the additional scalar degree of freedom inherently present in the gravitational sector via the $f(R)$ function \cite{Kuiroukidis2017, KANEDA2010, }.
		
		To facilitate the analysis, it is convenient to work in the Einstein frame, obtained through a conformal transformation of the metric:
		\begin{equation}
			\tilde{g}_{\mu\nu} = \Omega^2 g_{\mu\nu}, \quad \Omega^2 = f_R(R).
		\end{equation}
		Under this transformation, the action becomes
		\begin{equation}
			S_E = \int d^4x \sqrt{-\tilde{g}} \left( \frac{1}{2\kappa^2} \tilde{R} - \frac{1}{2} \tilde{g}^{\mu\nu} \partial_\mu \phi \partial_\nu \phi - V(\phi) \right),
		\end{equation}
		where $\phi$ is the scalar field (often called the scalaron) defined by
		\begin{equation}
			\phi = \sqrt{\frac{3}{2}} \frac{1}{\kappa} \ln f_R(R).
		\end{equation}
		The corresponding scalaron potential is given by
		\begin{equation}
			V(\phi) = \frac{R f_R(R) - f(R)}{2\kappa^2 (f_R(R))^2}.
		\end{equation}
		This formulation establishes the equivalence of $f(R)$ gravity with a scalar-tensor theory in the Einstein frame, where the scalaron $\phi$ effectively plays the role of the inflaton \cite{Magnano1990, Faraoni1999}.
		
		Inflation takes place when the scalaron’s potential energy $V(\phi)$ dominates over its kinetic energy, resulting in a nearly exponential expansion of the scale factor. This regime is achieved under the so-called slow-roll conditions, where the evolution of the scalar field is sufficiently gradual. The slow-roll parameters are defined as
		\begin{equation}
			\epsilon = \frac{1}{2\kappa^2} \left( \frac{V'(\phi)}{V(\phi)} \right)^2, \quad \eta = \frac{1}{\kappa^2} \frac{V''(\phi)}{V(\phi)},
		\end{equation}
		where $V'(\phi) = dV/d\phi$ and $V''(\phi) = d^2V/d\phi^2$. The slow-roll conditions $\epsilon \ll 1$ and $|\eta| \ll 1$ ensure a sustained period of quasi-de Sitter expansion \cite{Linde1982, Lyth2009}.
		
		During inflation, the scale factor evolves approximately as $a \approx \exp(N)$, where the number of e-folds $N$ quantifies the total amount of expansion. It is computed as
		\begin{equation}
			N = \ln{\left(\frac{a_f}{a_i}\right)} \simeq \kappa^2 \int^{\phi^*}_{\phi_{\mathrm{end}}} \frac{V(\phi)}{V'(\phi)} d\phi,
		\end{equation}
		with $\phi^*$ and $\phi_{\mathrm{end}}$ denoting the values of the scalaron field at horizon exit and at the end of inflation, respectively. For a successful inflationary scenario consistent with cosmological observations, the number of e-folds is typically required to lie within the range $55 \leq N \leq 65$.
		
		The inflationary model’s predictions can be characterized by key observables, including the spectral index $n_s$, the tensor-to-scalar ratio $r$, and the amplitude of scalar perturbations $A_s$ at horizon crossing. These quantities are expressed in terms of the slow-roll parameters as
		\begin{equation}
			n_s = 1 - 6\epsilon + 2\eta, \quad r = 16\epsilon, \quad A_s = \frac{\kappa^4 V(\phi)}{24\pi^2 \epsilon}.
		\end{equation}
		To be viable, these predictions must align with current cosmological constraints, such as those reported by the Planck 2018 results: $n_s = 0.9649 \pm 0.0042$, $r < 0.036$, and $A_s \approx 2.1 \times 10^{-9}$ \cite{Akrami_2020}. New measurements from the Atacama Cosmology Telescope (ACT) indicate a higher scalar spectral index ($n_s$) compared to previous Planck results. The combined analysis of ACT and Planck data yields $n_s = 0.9709 \pm 0.0038$. This value further increases to $n_s = 0.9743 \pm 0.0034$ when additional cosmological datasets are included, incorporating CMB observations, baryon acoustic oscillations (BAO), and Dark Energy Spectroscopic Instrument (DESI) measurements \cite{louis2025atacama, calabrese2025atacama}.

\subsection{Viable Dark energy  Models}
				
		For $f(R)$ gravity to serve as a viable framework for explaining late-time cosmic acceleration and the nature of dark energy, it must satisfy a set of theoretical and observational viability conditions. These constraints ensure the physical consistency of the theory and its compatibility with cosmological and astrophysical data. The principal requirements are as follows \cite{Amendola:2015ksp, Bamba:2010iy}. First, the effective gravitational coupling must remain positive, which requires $f_R > 0$, thereby guaranteeing that gravity remains attractive. Second, the stability of scalar perturbations demands that $f_{RR} > 0$, avoiding ghost-like instabilities and ensuring the stability of the scalar degree of freedom. Third, the theory must asymptotically approach the $\Lambda$CDM model at high curvatures, such that $f(R) \rightarrow R - 2\Lambda$, in order to reproduce the correct cosmological behavior at early times. Fourth, the existence of a stable de Sitter solution at low curvature is necessary to account for the present accelerated expansion. Finally, a viable $f(R)$ model must incorporate a chameleon mechanism to suppress deviations from General Relativity in high-density environments, thereby satisfying local gravity constraints such as those tested in the Solar System.
		
		Models that fulfill these criteria are referred to as viable $f(R)$ gravity models. Based on their asymptotic behavior, these models can generally be categorized into two classes: power-law type and exponential type. Representative forms of each class are given by \cite{Starobinsky:2007hu, Tsujikawa2008, Zhang:2005vt, Cognola:2007zu, Linder:2009jz, Chen2019}:
		\begin{eqnarray}
			\label{eq:star}
			f^{p}(R) &=& R - \lambda_s R_{ch} \left[1 - \left(1 + \frac{R^2}{R_{ch}^2}\right)^{-n}\right], \\
			\label{eq:exp}
			f^{e}(R) &=& R - \lambda_e R_{ch}\left(1 - e^{-R/R_{ch}}\right),
		\end{eqnarray}
		where $\lambda_s$, $\lambda_e$, and $n$ are dimensionless parameters, and $R_{ch}$ denotes a characteristic curvature scale. In the high-curvature limit, both models approach $f(R) \rightarrow R - \lambda R_{ch}$, effectively mimicking the $\Lambda$CDM model. In this limit, $\lambda R_{ch}$ plays the role of an effective cosmological constant, and $R_{ch}$ is typically of the order $R_{ch} \sim \Lambda_{obs}/\lambda$, where $\Lambda_{obs} = \kappa^2 \rho_{DE}^{(0)}$ is defined by the present dark energy density. Consequently, a smaller $R_{ch}$ is required for larger values of $\lambda$ to match observational constraints, leading to a larger ratio $R/R_{ch}$. In the limit where $\lambda \gg 1$ and $R_{ch} \rightarrow 0$, $f(R)$ gravity becomes indistinguishable from the $\Lambda$CDM model.
		
		In this context, the Hu-Sawicki model \cite{Hu:2007nk} and the Starobinsky model \cite{Starobinsky:2007hu} are examples of the power-law class, while the Tsujikawa model \cite{Tsujikawa2008}, the exponential gravity models \cite{Zhang:2005vt, Cognola:2007zu, Linder:2009jz, Bamba:2010iy, Bamba:2011yr}, and the Appleby-Battye (AB) model \cite{Appleby:2007vb, Appleby:2009uf} fall into the exponential-type category.
		
		The viability of these $f(R)$ gravity models has far-reaching implications for cosmology. They provide an alternative to the cosmological constant for explaining the accelerated expansion of the universe, relying instead on a dynamical mechanism that naturally emerges from the modified gravitational action. Furthermore, the presence of the chameleon mechanism allows $f(R)$ theories to satisfy stringent local gravitational constraints, enhancing their applicability across both cosmological and astrophysical scales. Future observations, particularly those involving gravitational waves and large-scale structure surveys, are expected to offer critical tests for distinguishing $f(R)$ gravity from other modified gravity frameworks \cite{Sotiriou:2008rp}.
		
		In the high-curvature regime, where the scalar curvature $R$ is much larger than the characteristic scale $R_{ch}$, the functional forms of the power-law and exponential models simplify as follows:
		\begin{align}
			f^p(R) &= R - \lambda R_{\text{ch}} \left[1 - \left(\frac{R_{\text{ch}}}{R} \right)^{2n} \right], \label{eq:fp} \\
			f^e(R) &= R - \lambda R_{\text{ch}} \left(1 - e^{-R/R_{\text{ch}}} \right). \label{eq:fe}
		\end{align}
		These simplified forms will be utilized in the subsequent analysis to study the emergence of curvature singularities in dense astrophysical environments within the context of $f(R)$ dark energy models.

\section{Curvature singularity in $f(R)$ gravity dark energy model}
		
		Starting from the field equations given in Eq.~\ref{eq:field}, one can derive the trace equation by contracting with the metric tensor. This yields:
		\begin{equation}\label{eq:trace}
			Rf_R - 2f - R + 3\square f_R = \kappa^2 T,
		\end{equation}
		where $T$ is the trace of the matter energy-momentum tensor $T_{\mu\nu}^{(M)}$. This equation reduces to the standard trace equation in General Relativity in the limit $f(R) = 0$.
		
		In a localized region of spacetime that can be approximated as a perfect, homogeneous, and isotropic fluid, and where the non-relativistic matter density $\rho_m$ is significantly larger than the background density $\rho_{ch}$, the trace of the energy-momentum tensor can be assumed to evolve homogeneously with time. This behavior is modeled as \cite{Lee2012, Dutta2015}:
		\begin{equation}
			T(t) = -T_0 \left( 1 + \frac{t}{t_{ch}} \right),
		\end{equation}
		where $T_0 = \rho_m^{(0)} - 3P_m^{(0)} \simeq \rho_m^{(0)}$ is the initial local energy density and $t_{ch}$ represents the characteristic timescale over which this density evolves.
		
		
		Assuming that the gravitational field of the object is sufficiently weak, the background spacetime can be approximated by a flat Minkowski metric. Under this approximation, the d'Alembert operator in Eq. \ref{eq:trace} reduces to a second-order time derivative, thereby simplifying the trace equation to a more tractable form:
		\begin{equation}\label{eq:diff}
			\Box f_R = -\partial_t^2(f_R) = \frac{1}{3}[\kappa^2 T - Rf_R + 2f + R].
		\end{equation}
		To further facilitate the analysis, we introduce a change of variables from $R$ and $t$ to dimensionless variables $y = \beta (R_{\text{ch}} / R)$ and $\tau = \gamma^{-1} t$, where $\beta$ and $\gamma$ are scaling parameters. This transformation allows the time evolution of the system to be expressed in terms of $\tau$, with derivatives denoted by a prime ($'$). It is important to emphasize that the functional forms of $y$ and $\gamma$ depend on the specific $f(R)$ model considered. To reflect this model dependence, subscripts are used accordingly.
		
		\begin{figure}[htbp] 
			\centering 
			\begin{subfigure}[b]{0.45\textwidth} 
				\includegraphics[width=\textwidth]{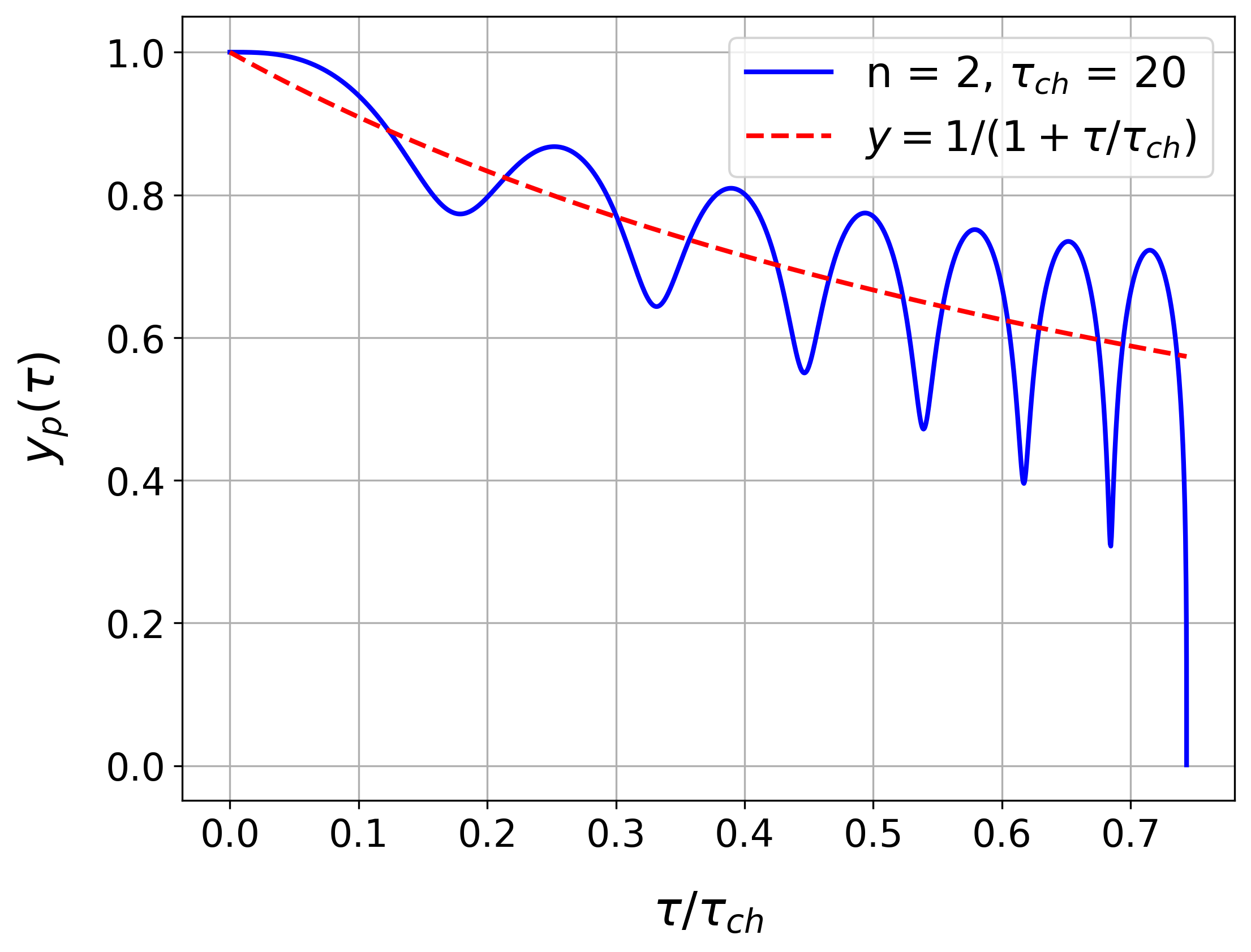} 
				\caption{} 
				\label{fig:subfig1_ori}
			\end{subfigure}
			\hfill 
			\begin{subfigure}[b]{0.45\textwidth} 
				\includegraphics[width=\textwidth]{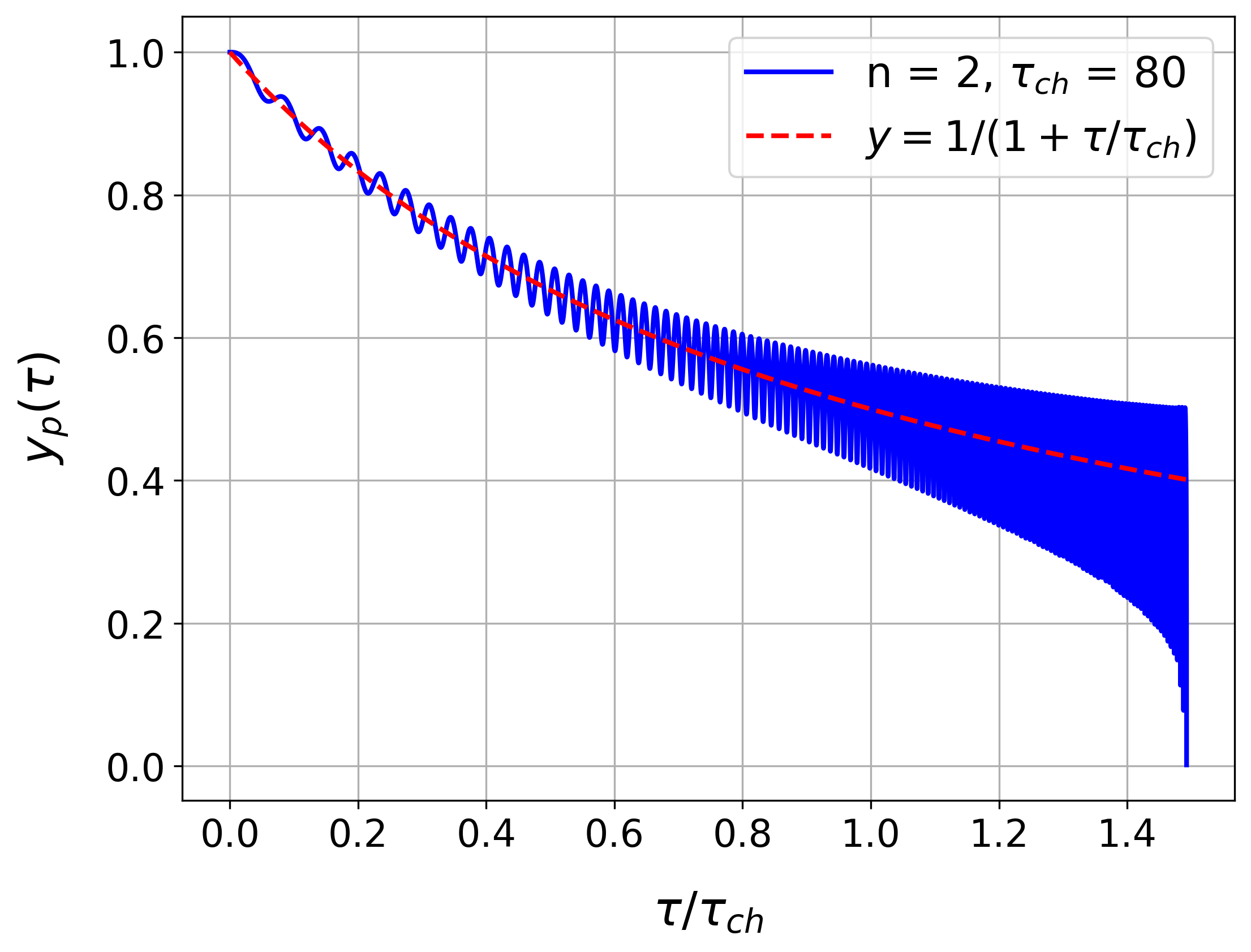} 
				\caption{} 
				\label{fig:subfig2_ori}
			\end{subfigure}
			
			\vspace{1em} 
			
			\begin{subfigure}[b]{0.45\textwidth} 
				\includegraphics[width=\textwidth]{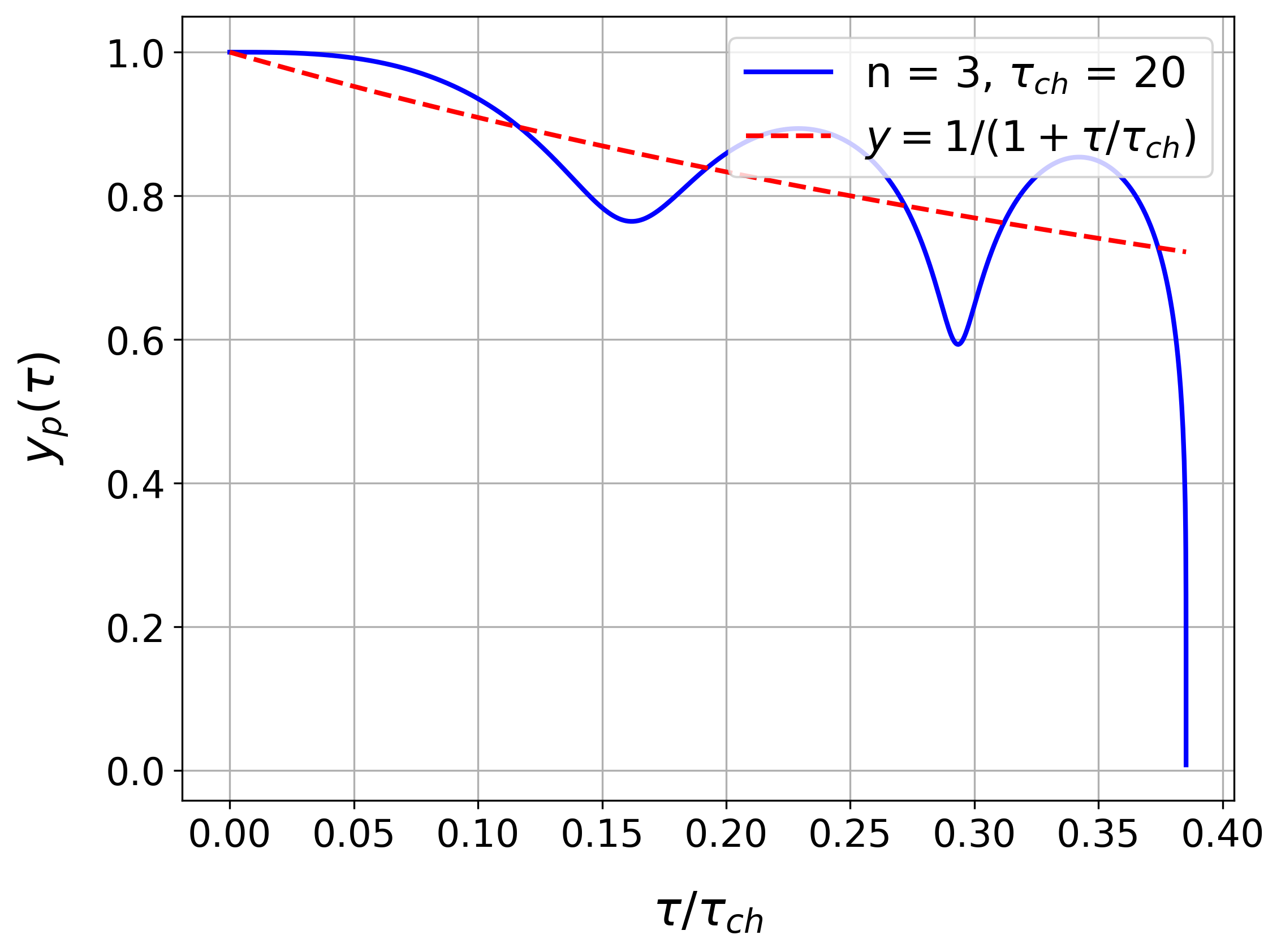} 
				\caption{} 
				\label{fig:subfig3_ori}
			\end{subfigure}
			\hfill 
			\begin{subfigure}[b]{0.45\textwidth} 
				\includegraphics[width=\textwidth]{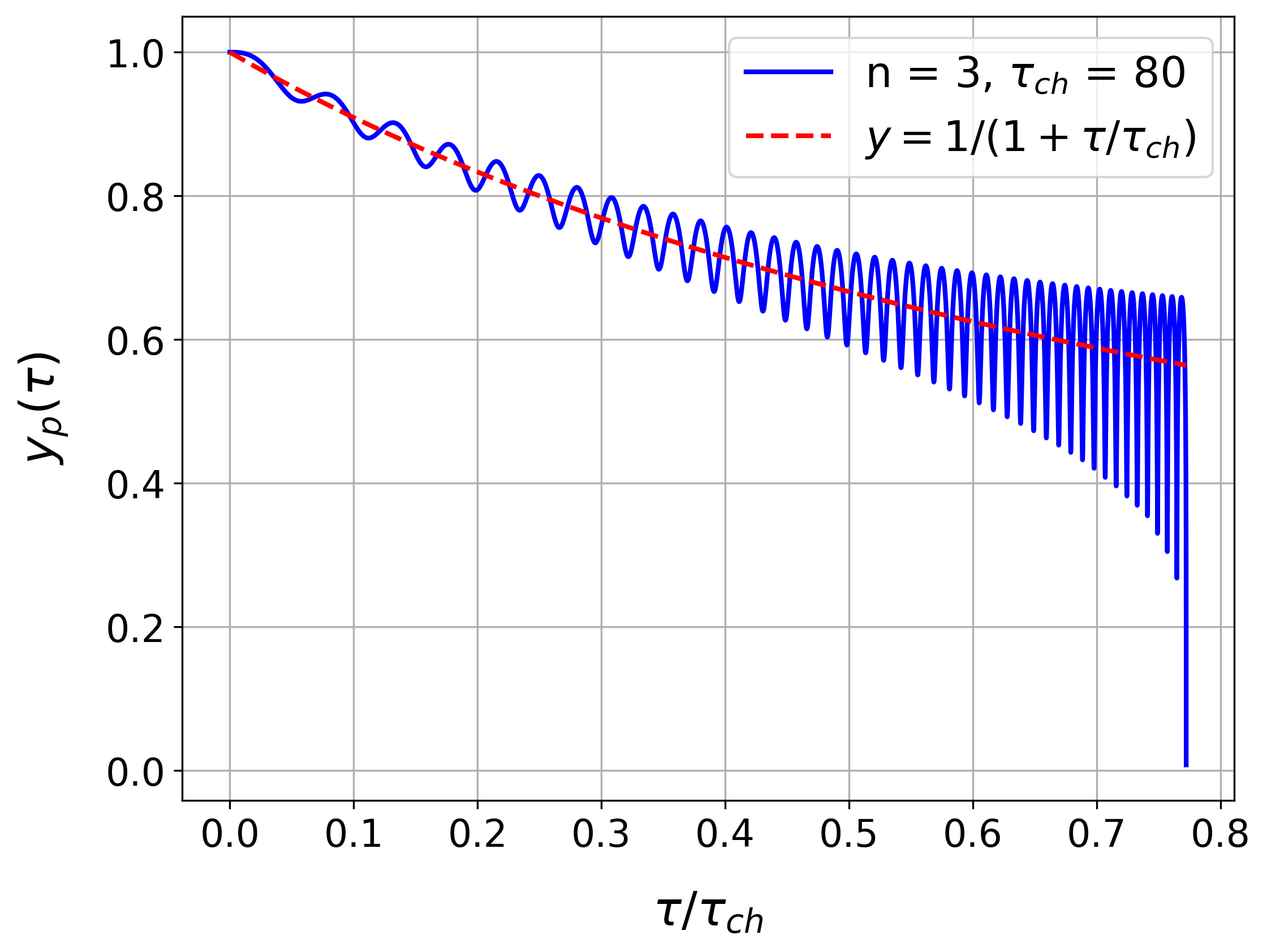} 
				\caption{} 
				\label{fig:subfig4_ori}
			\end{subfigure}
			
			\caption{Oscillations of $y_p(\tau)$ in the power-law dark energy model for various values of $n$ and $\tau_{\text{ch}}$ in dense area ($\beta \gg 1$).} 
			\label{fig:fp}
		\end{figure}
		
		For the power-law type $f(R)$ dark energy model, the dimensionless variable is defined as $y_p = \beta (R_{\text{ch}} / R)$. The corresponding equation of motion takes the following form:
		\begin{equation}\label{eq:diff1}
			y_p'' + 2n \frac{{y_p'}^2}{y_p} + \frac{(\kappa^2 T_0) \gamma_p^2 \beta^{2n+1}}{6n \lambda (2n + 1)} y_p^{-2n} \left[ \left(1 + \frac{\tau}{\tau_{\text{ch}}}\right) - \frac{1}{y_p} \right] = 0,
		\end{equation}
		where the scaling parameters are given by
		\begin{equation}
			\beta = \frac{\kappa^2 T_0}{R_{\text{ch}}}, \quad \gamma_p^2 = \frac{6n(2n+1) \lambda}{R_{\text{ch}}} \beta^{-(2n+2)}.
		\end{equation}
		Substituting these into Eq. \eqref{eq:diff1}, the equation simplifies to:
		\begin{equation}\label{eq:diff2}
			y_p'' + 2n \frac{{y_p'}^2}{y_p} + y_p^{-2n} \left[\left(1 + \frac{\tau}{\tau_{\text{ch}}}\right) - y_p^{-1} \right] = 0,
		\end{equation}
		where $\tau_{\text{ch}} = \gamma_p^{-1} t_{\text{ch}}$ represents the rescaled characteristic timescale.
		
		The parameter $\beta$ is chosen to satisfy $\beta = \kappa^2 T_0 / R_{\text{ch}} \simeq \rho_m / \rho_c \gg 1$, ensuring that $y$ remains finite and physically meaningful in high-density environments. For instance, for a matter density of $\rho_m \sim 10^{-24}\, \text{g/cm}^3$ and a characteristic curvature scale $R_{\text{ch}} \propto R_0 \sim 10^{-35}\, \text{s}^{-2}$, we find $\beta \sim 1.67 \times 10^5$. Assuming $\lambda \simeq 1$, the scaling parameter $\gamma_p$ evaluates to approximately $\gamma_p \sim 519\, \text{s}$ for $n = 2$ and $\gamma_p \sim 0.004\, \text{s}$ for $n = 3$.
		
		Equation \eqref{eq:diff2} can be solved numerically using the initial conditions $y_p(0) = \kappa^2 T_0 / R \simeq 1$ and $y_p'(0) = 0$. These conditions model the initial configuration of the high-density environment, such as during gravitational collapse. The differential equation is nonlinear due to the terms $2n y_p'^2 / y_p$ and $y_p^{-2n}$, precluding closed-form solutions. Nevertheless, numerical solutions reveal that the system exhibits oscillatory behavior.
		The equilibrium trajectory of the system can be determined by setting $y_p'' = 0$ and $y_p' = 0$. Under these conditions, the equilibrium line is found to be $y(\tau) = 1 / (1 + \tau/\tau_{\text{ch}})$, which is independent of $n$ and asymptotically approaches zero as $\tau \to \infty$. This equilibrium curve is included in the graphical analysis.
		The oscillatory dynamics can be interpreted by examining the roles of the nonlinear terms. The term $2n y_p'^2 / y_p$ is always positive for $y_p \geq 0$ and $n > 0$, thus acting as a damping component that drives $y_p''$ negative. Meanwhile, the term
		$
		y_p^{-2n} \left[ \left(1 + \frac{\tau}{\tau_{\text{ch}}} \right) - y_p^{-1} \right]
		$
		acts as a time-dependent restoring force, driving the system toward the equilibrium configuration whenever deviations occur.
		
		Numerical results for Eq. \ref{eq:diff2} are illustrated in Fig. \ref{fig:fp}, showing that the oscillations of $y_p$ are centered around the equilibrium curve $y = 1/(1 + \tau/\tau_{\text{ch}})$. As the value of $n$ increases, the frequency of oscillation decreases. Since $y_p \propto 1/R$, a curvature singularity emerges as $y_p \to 0$. The figure indicates that the power-law dark energy model leads to curvature singularities at finite values of $\tau/\tau_{\text{ch}}$, depending on the value of $\tau_{\text{ch}}$. These singularities occur at later times for larger $\tau_{\text{ch}}$.
		For example, when $\tau_{\text{ch}} = 20$, the singularity occurs at $\tau/\tau_{\text{ch}} \approx 0.7434$ for $n = 2$ and $\tau/\tau_{\text{ch}} \approx 0.3853$ for $n = 3$. In contrast, for $\tau_{\text{ch}} = 80$, the singularities appear at $\tau/\tau_{\text{ch}} \approx 1.4927$ for $n = 2$ and $\tau/\tau_{\text{ch}} \approx 0.7722$ for $n = 3$. These results suggest that in high-density environments, curvature singularities can develop within a relatively short timescale, given by $t_{\text{sing}} = \gamma_p \tau = \alpha \gamma_p \tau_{\text{ch}}$, where $\alpha$ is a constant of order unity.

		\begin{figure}[htbp!] 
			\centering 
			\begin{subfigure}[b]{0.45\textwidth} 
				\includegraphics[width=\textwidth]{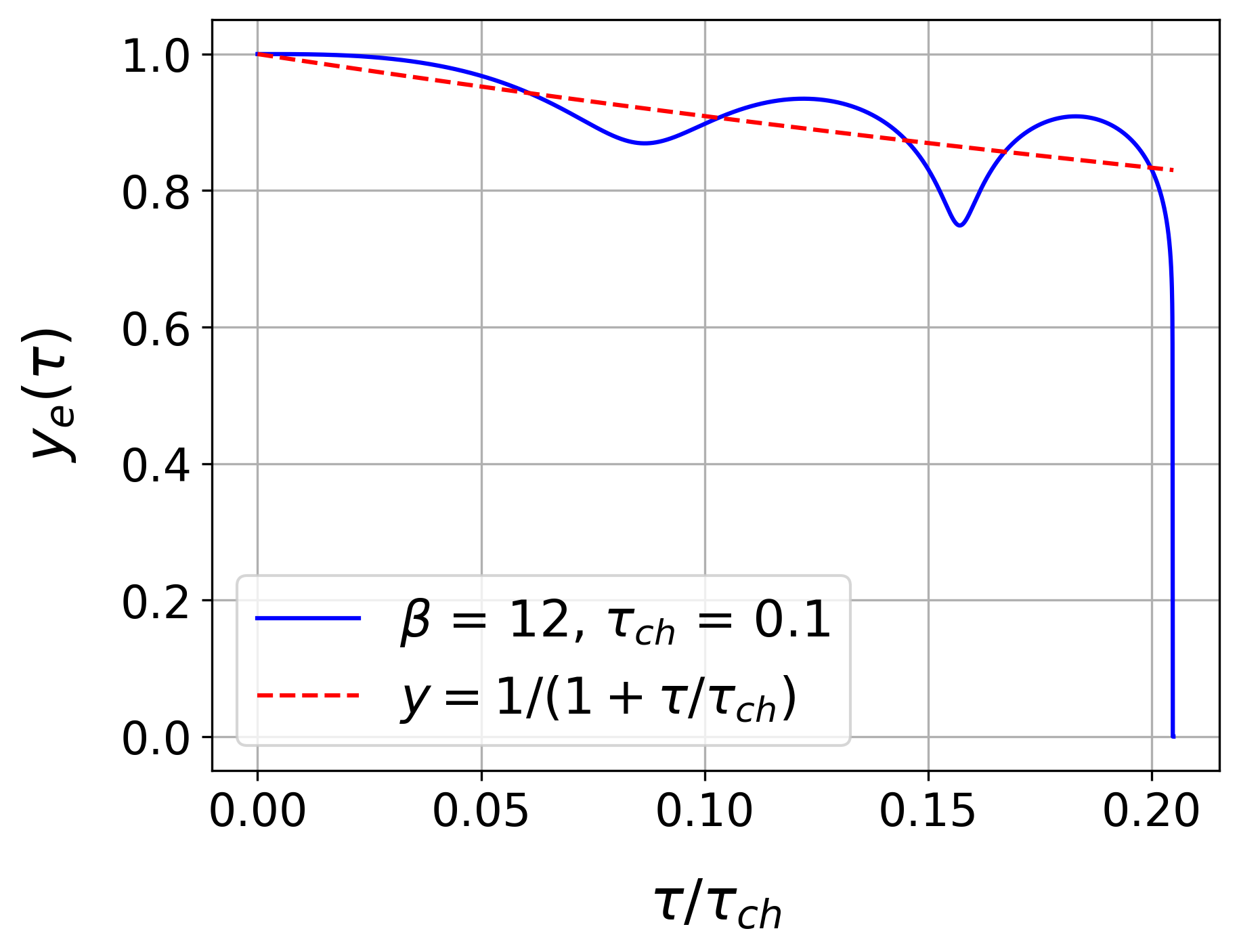} 
				\caption{} 
				\label{fig:subfig1_ori2}
			\end{subfigure}
			\hfill 
			\begin{subfigure}[b]{0.45\textwidth} 
				\includegraphics[width=\textwidth]{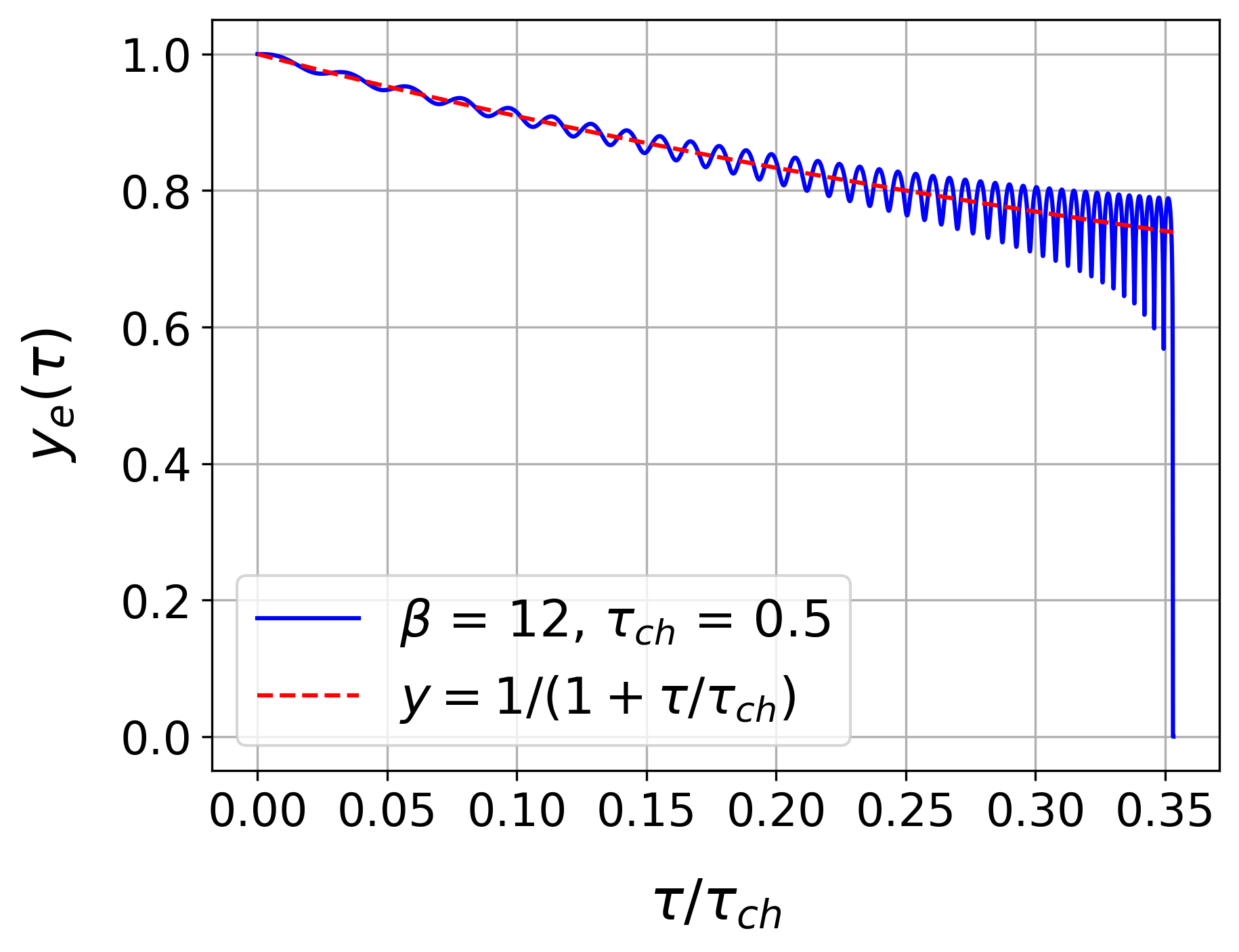} 
				\caption{} 
				\label{fig:subfig21_ori2}
			\end{subfigure}
			
			\vspace{1em} 
			
			\begin{subfigure}[b]{0.45\textwidth} 
				\includegraphics[width=\textwidth]{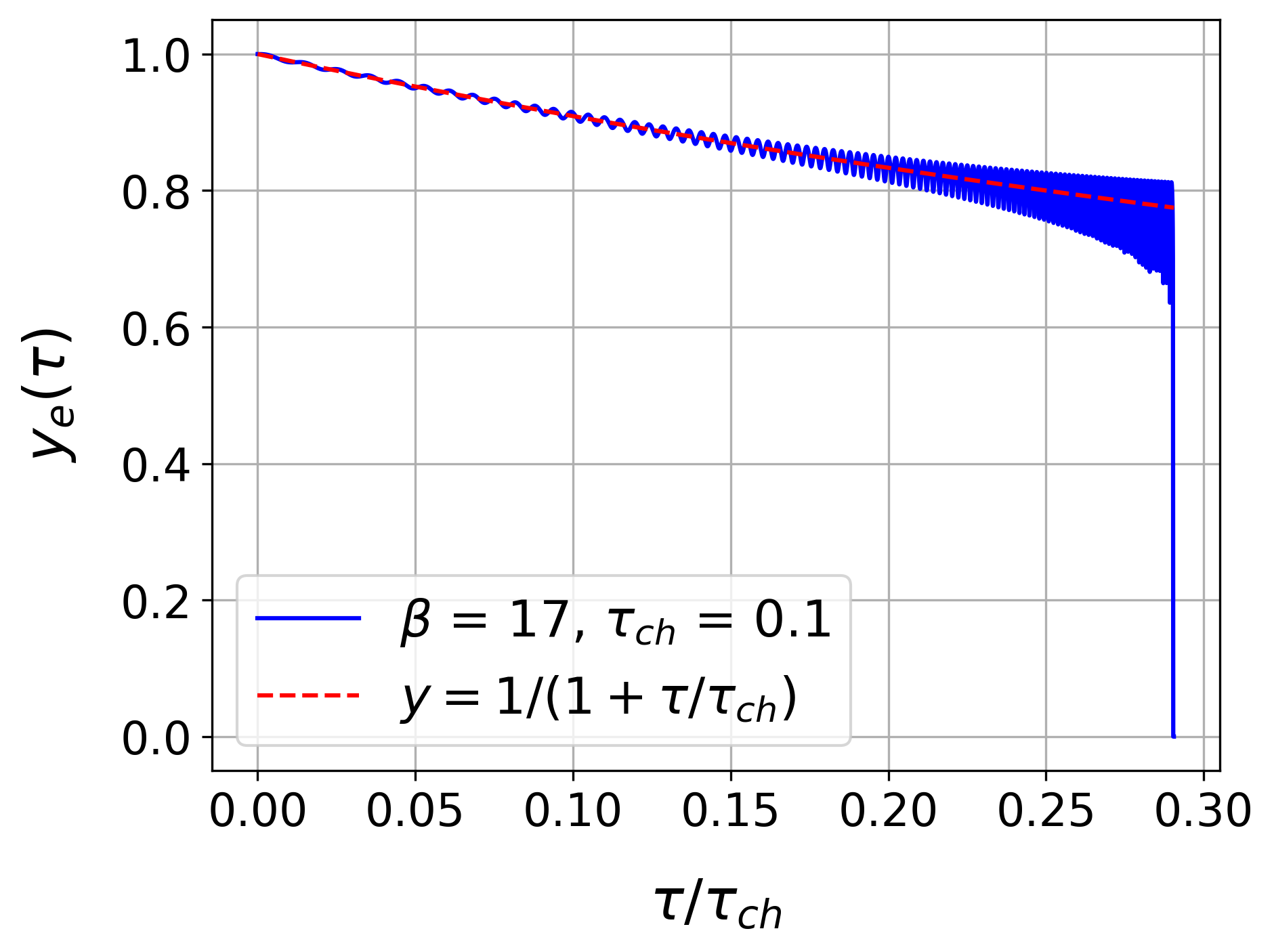} 
				\caption{} 
				\label{fig:subfig31_ori2}
			\end{subfigure}
			\hfill 
			\begin{subfigure}[b]{0.45\textwidth} 
				\includegraphics[width=\textwidth]{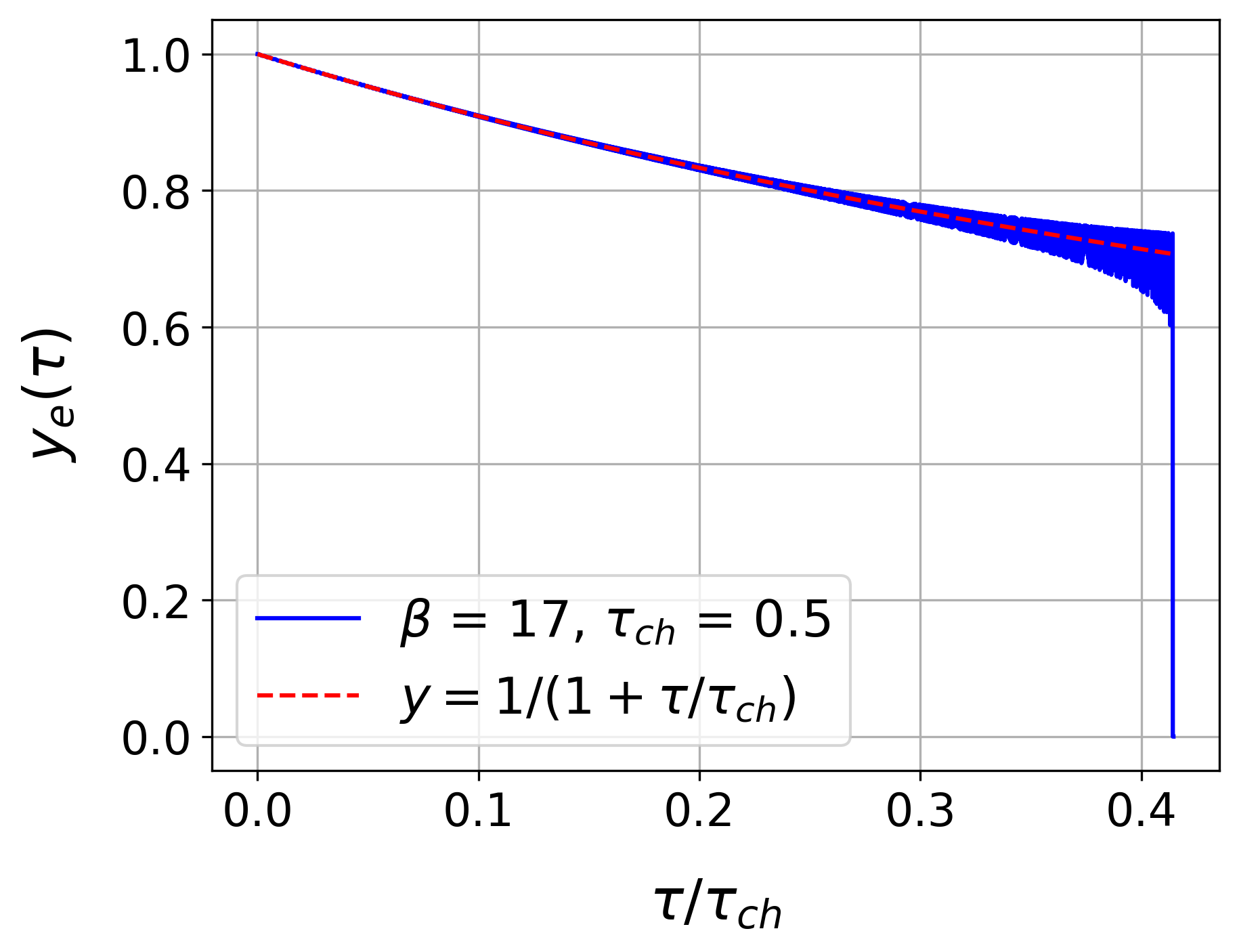} 
				\caption{} 
				\label{fig:subfig41_ori2}
			\end{subfigure}
			
			\caption{Oscillations of $y_e(\tau)$ in the exponential-law dark energy model for various values of $\beta$ and $\tau_{\text{ch}}$.} 
			\label{fig:fe}
		\end{figure}
		
		For the exponential-type $f(R)$ dark energy model, Eq. \ref{eq:fe} is substituted into Eq. \ref{eq:diff}. To distinguish this model from the power-law case, we define the dimensionless variable $y_e = \beta (R_{\text{ch}}/R)$ and rescale the time variable using $\gamma_e$, yielding $\tau = \gamma_e^{-1} t$. With these definitions, the governing differential equation becomes
		\begin{equation} \label{eq:ye}
			y_e'' +  (\beta y_e^{-1} - 2) \frac{{y_e'}^2}{y_e} + y_e^2 e^{\beta y_e^{-1}} \left[\left(1 + \frac{\tau}{\tau_{\text{ch}}} \right) - y_e^{-1} \right] = 0,
		\end{equation}
		where
		\begin{equation}
			\gamma_e^2 = \frac{3 \lambda}{R_{\text{ch}}}.
		\end{equation}
		
		To analyze this equation, we impose initial conditions that represent the physical state at the onset of the high-density regime. Specifically, we set $y_e(0) = \kappa^2 T_0 / R$, normalized to unity, and $y_e'(0) = 0$. These conditions ensure that the system begins in equilibrium prior to the emergence of curvature oscillations.
		Numerical solutions of Eq. \ref{eq:ye} are shown in Fig. \ref{fig:fe}. The dynamics are governed by the interplay between a damping term, $(\beta y_e^{-1} - 2) y_e'^{2} / y_e$, and a restoring term, $y_e^2 e^{\beta y_e^{-1}} \left[\left(1 + \tau/\tau_{\text{ch}}\right) - y_e^{-1}\right]$. The damping term, which scales with $\beta$, gradually suppresses the oscillations. For values of $\beta$ in the range of 10–20, the damping is moderate, allowing oscillations to persist for some time before decay sets in.
		
		The equilibrium configuration of the system is defined by the trajectory $y_e(\tau) = 1 / (1 + \tau / \tau_{\text{ch}})$. As $\tau \to \infty$, this trajectory asymptotically approaches zero, implying that the Ricci scalar $R$ diverges and a curvature singularity emerges. However, the numerical results show that the system exhibits damped oscillations around this equilibrium path before reaching the singular state. The amplitude and frequency of these oscillations depend on both $\beta$ and the characteristic timescale $\tau_{\text{ch}}$.
		
		The behavior of the exponential-type model stands in contrast to the power-law $f(R)$ model, where oscillations are generally more pronounced and the singularity tends to occur later for larger values of $\tau_{\text{ch}}$. This distinction arises due to the presence of the exponential term $e^{\beta y_e^{-1}}$ in the restoring force, which significantly enhances the system’s sensitivity to variations in $R$. For moderate values of $\beta$ (in the range 10–20), representing a curvature regime that is not excessively high, the exponential model exhibits stronger damping and earlier singularity onset compared to the power-law model.
		
		The parameter $\gamma_e$ plays a central role in setting the temporal scale of the system’s evolution. Given its definition, $\gamma_e^2 = 3\lambda / R_{\text{ch}}$, and using representative values $\lambda \sim 1$ and $R_{\text{ch}} \sim 10^{-35} \, \text{s}^{-2}$, we estimate $\gamma_e \sim \sqrt{3 \times 10^{35}} \, \text{s} \approx 1.73 \times 10^{17} \, \text{s}$.
		The singularity in the exponential-type model arises as $y_e \to 0$, corresponding to $R \to \infty$. As shown in Fig. \ref{fig:fe}, this occurs at finite values of $\tau / \tau_{\text{ch}}$, with the precise timing dependent on the parameters $\beta$ and $\tau_{\text{ch}}$. For example, when $\tau_{\text{ch}} = 0.1$, the singularity occurs at $\tau / \tau_{\text{ch}} \approx 0.2048$ for $\beta = 12$. For $\tau_{\text{ch}} = 0.5$, it occurs at $\tau / \tau_{\text{ch}} \approx 0.4146$ for $\beta = 17$.
		These findings imply that the exponential-type $f(R)$ dark energy model predicts curvature singularities within a finite, though relatively long, timescale given by $t_{\text{sing}} = \gamma_e \tau = \alpha \gamma_e \tau_{\text{ch}}$, where $\alpha \in \mathcal{O}(10^{-1})$. Based on the numerical simulations, we estimate the singularity time to lie within the range $t_{\text{sing}} = [10^{15} - 10^{16}] \, \text{s}$, which remains below the current age of the universe, $t_U \simeq 4 \times 10^{17} \, \text{s}$.

\section{Solving singularity by adding $R^{2}$ term}
		
		Inflationary dynamics within the framework of $ F(R) $ gravity have been the subject of extensive investigation over several decades \cite{Starobinsky1980, BarrowCotsakis1988, BerkinMaeda1990, SaidovZhuk2010, Huang2014, Sebastiani2013, Motohashi2015, Broy2015}. Among the various models proposed, the Starobinsky model, characterized by the functional form \cite{Starobinsky1980}:
		\begin{equation}
			f(R) = R + \frac{R^2}{6M^2},
		\end{equation}
		stands out as one of the most successful and widely studied. This model, despite its simplicity, exhibits exceptional predictive accuracy in describing inflationary phenomena.
		The Starobinsky model provides well-defined predictions for key inflationary observables, such as the scalar spectral index $ n_s $ and the tensor-to-scalar ratio $ r $, which are expressed as:
		\begin{equation}
			n_s \approx 1 - \frac{2}{N}, \quad r \approx \frac{12}{N^2}.
		\end{equation}
		For typical values of the e-folding number $ N \approx 50-60 $, these predictions yield:
		\begin{equation}
			n_s \approx 0.960 - 0.967, \quad r \approx 0.003 - 0.004,
		\end{equation}
		which are in excellent agreement with observational data from the Planck satellite. The inflaton mass scale $M \approx 1.13 \times 10^{-5} M_{\text{Pl}}$ is determined by the amplitude of scalar perturbations. Remarkably, the predictions for $n_s$ and $r$ remain robust against variations in $M$, demonstrating the model's stability across different parameter regimes. 
		
		Recent joint analyses incorporating data from Planck, ACT, and DESI  yield a scalar spectral index of $n_s = 0.974 \pm 0.003$ \cite{louis2025atacama, calabrese2025atacama}. Notably, this value positions the Starobinsky inflationary model at the boundary of the $2\sigma$ confidence interval derived from the combined observational constraints, under the assumption of approximately 60 e-folds of inflation\cite{calabrese2025atacama}. This proximity has sparked renewed interest and debate within the cosmology community, prompting critical reassessments of canonical inflationary models and motivating the exploration of alternative theoretical frameworks in light of the high-precision measurements from ACT \cite{berera2504early, dioguardi2025fractional, salvio2025independent, gao2025non}. Nevertheless, a more refined treatment of the Starobinsky scenario, as presented in \cite{DREES2025139612}, demonstrates that the model remains compatible with current observational data at the $2\sigma$ level for e-folding numbers $N \geq 58$.
		
		\begin{figure}[htbp] 
			\centering 
			\begin{subfigure}[b]{0.45\textwidth} 
				\includegraphics[width=\textwidth]{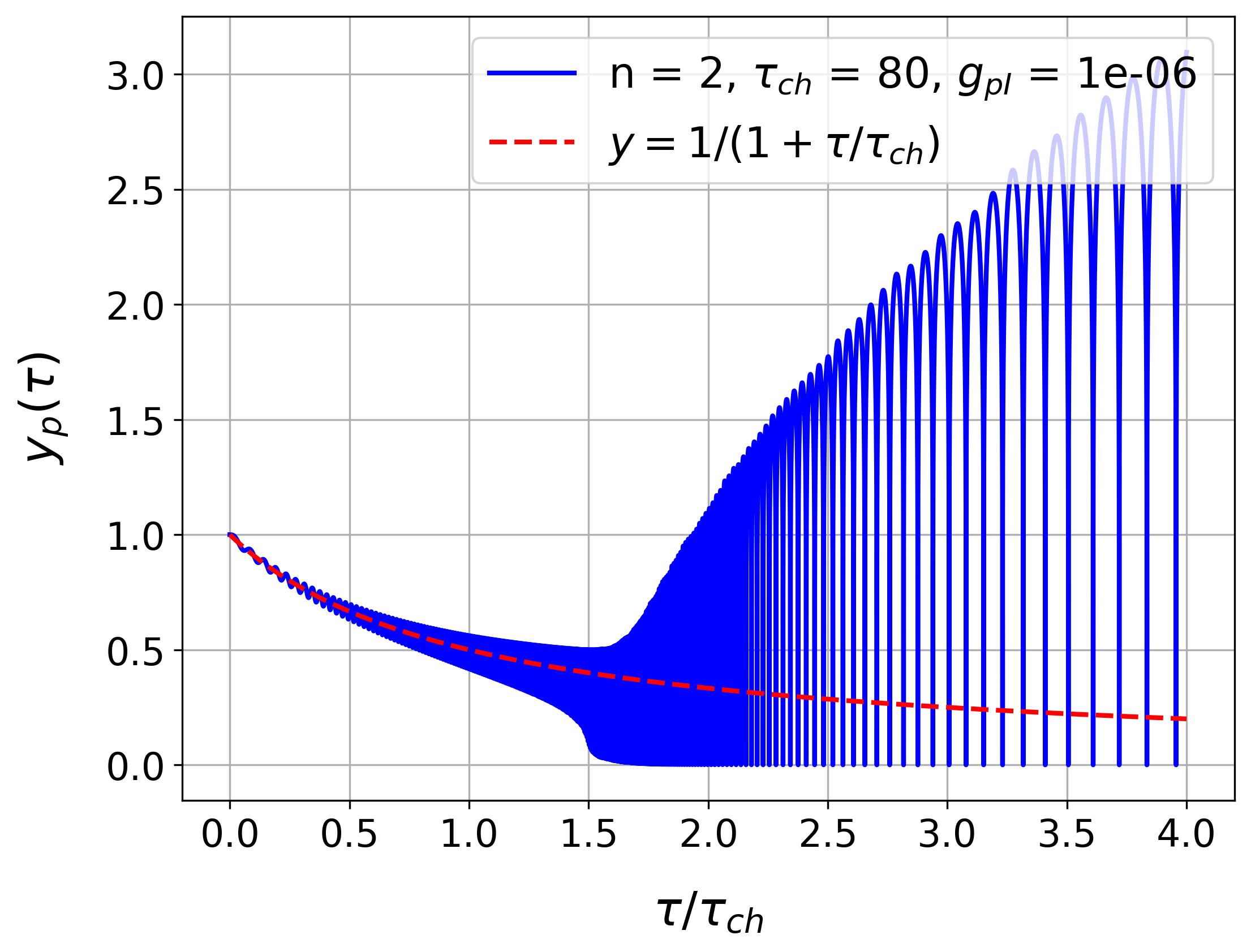} 
				\caption{} 
				\label{fig:subfig1_R2_2}
			\end{subfigure}
			\hfill 
			\begin{subfigure}[b]{0.45\textwidth} 
				\includegraphics[width=\textwidth]{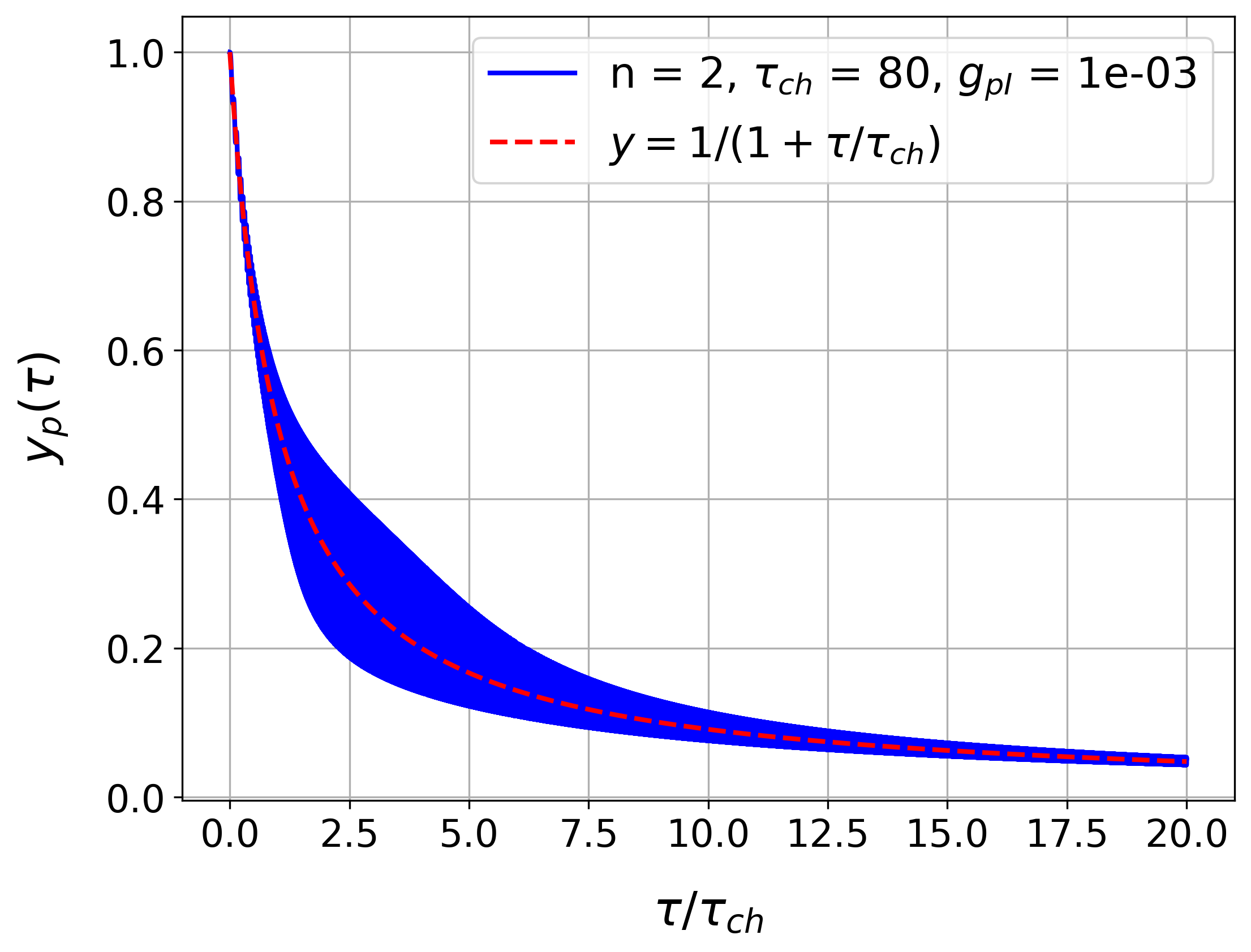} 
				\caption{} 
				\label{fig:subfig2_R2_2}
			\end{subfigure}
			
			\vspace{1em} 

			\begin{subfigure}[b]{0.45\textwidth} 
				\includegraphics[width=\textwidth]{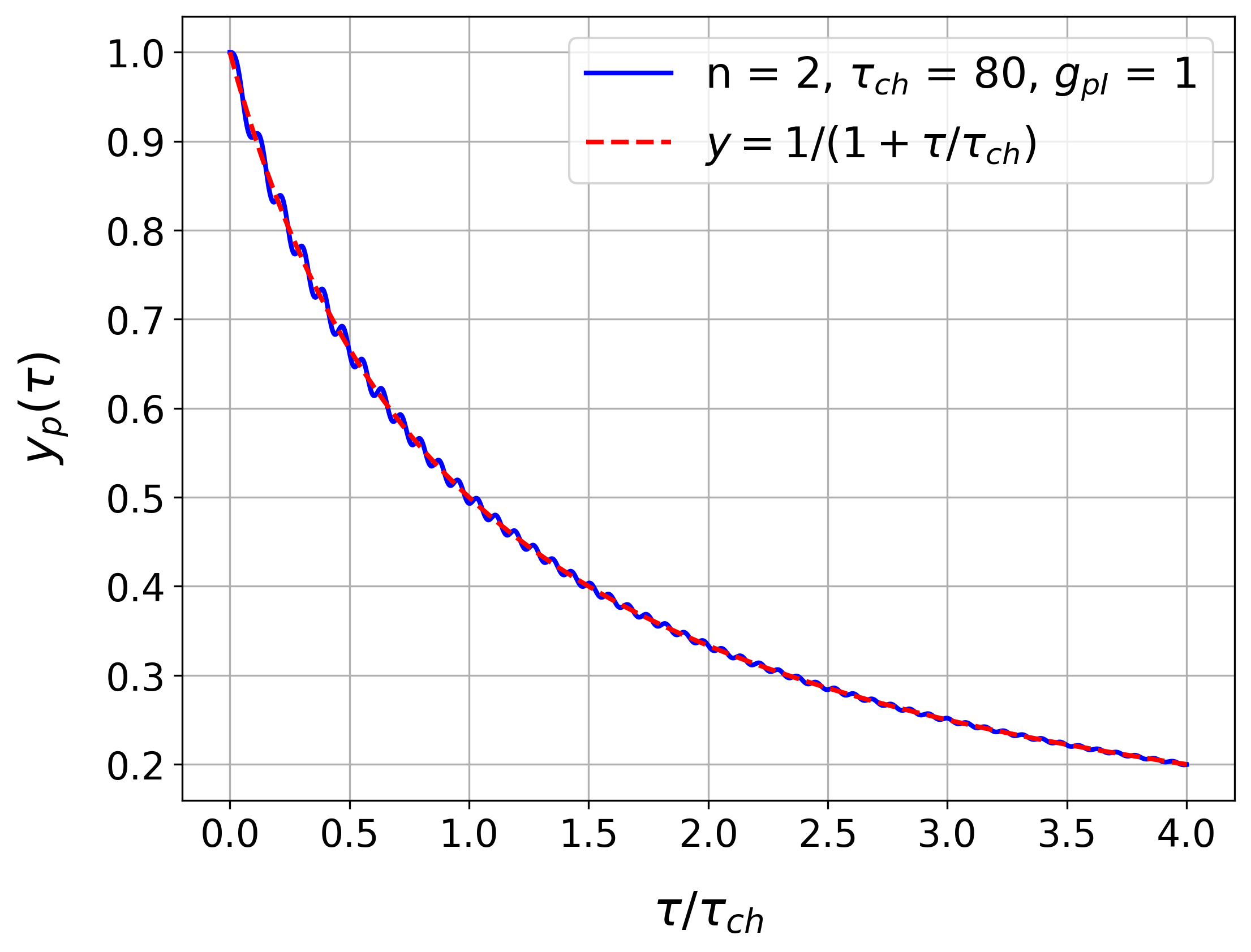} 
				\caption{} 
				\label{fig:subfig3_R2_2}
			\end{subfigure}
			\hfill 
			\begin{subfigure}[b]{0.45\textwidth} 
				\includegraphics[width=\textwidth]{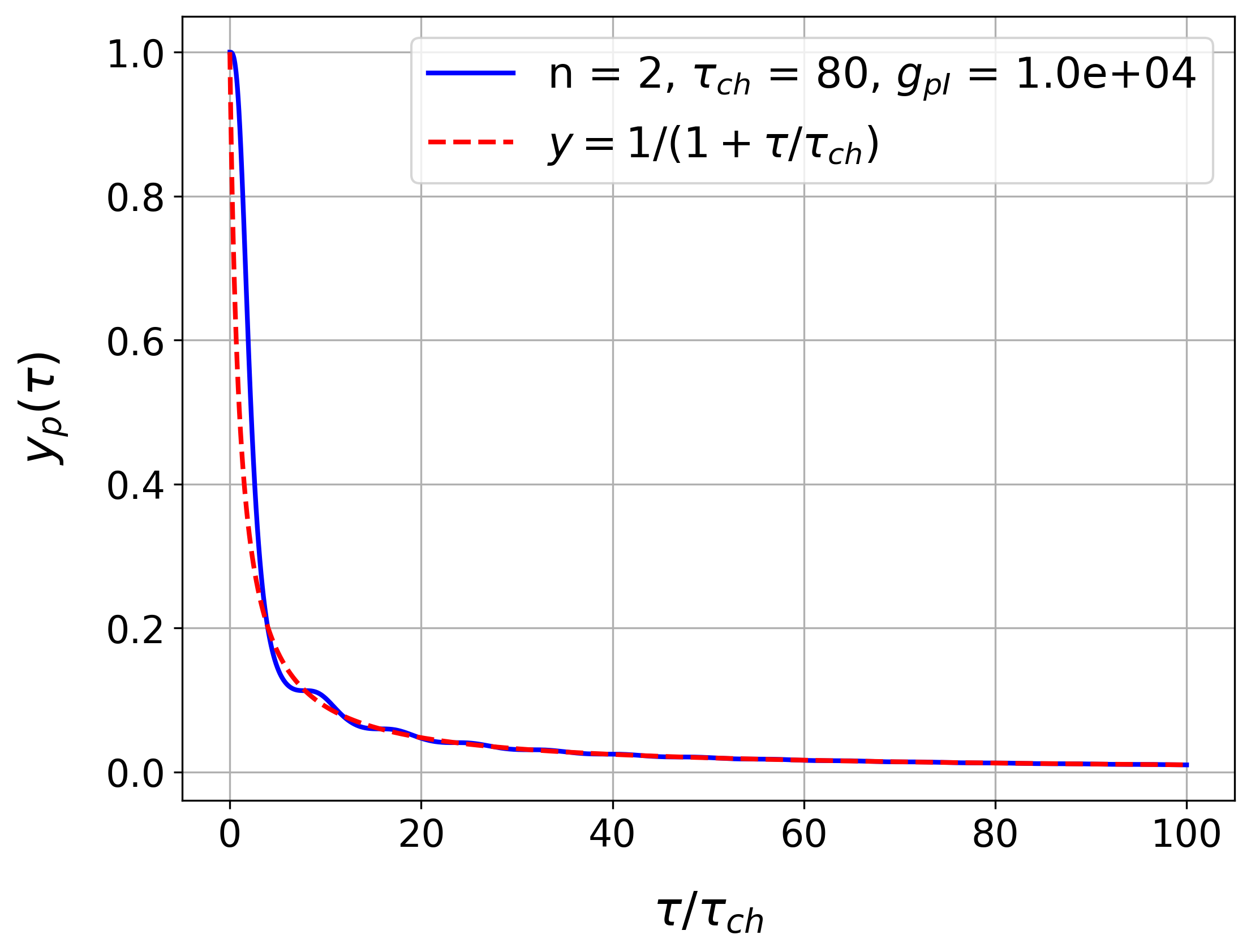} 
				\caption{} 
				\label{fig:subfig4_R2_2}
			\end{subfigure}

			\caption{Oscillations of $y_p(\tau)$ in the power-law dark energy model with $R^2$ correction in dense area ($\beta \gg 1$) for various values of  $g_{pI}$  .} 
			\label{fig:fp_R2}
		\end{figure}

		Regarding curvature singularities, studies \cite{Lee2012, Bamba_2011} have shown they can be avoided by introducing additional terms $R^m/M^{2(m-1)}$ with $1 < m \leq 2$ in $f(R)$ dark energy models. The Starobinsky model itself has been proven to prevent such singularities \cite{Lee2012, Appleby:2009uf}.
		
		In the high-curvature regime $R \gg R_{\rm ch}$, the power-law $f(R)$ dark energy model with Starobinsky correction becomes:
		\begin{equation} \label{eq:fp_star}
			f(R) = R - \lambda R_{\text{ch}} \left[1 - \left(\frac{R_{\text{ch}}}{R} \right)^{2n} \right] + \frac{R^2}{6 m_R^2},
		\end{equation}
		where $m_R$ represents an adjustable mass scale. The corresponding differential equation derived from Eq.~\ref{eq:trace} is:
		\begin{equation} \label{diff1_R2}
			y_p'' + 2n \frac{{y_p'}^2}{y_p} +g_{pI} y_p^{-(2n+2)}\left[y_p'' - \frac{2 {y_p'}^2}{y_p} \right] + y_p^{-2n} \left[\left(1 + \frac{\tau}{\tau_{\rm ch}} \right) - y_p^{-1} \right] = 0,
		\end{equation}
		with the coefficient:
		\begin{equation}\label{eq:gp1}
			g_{pI} = \frac{R_{\rm ch}}{m_R^2 n \lambda (2n+1)} \beta^{2n+2}.
		\end{equation}
		Here we maintain the rescaling parameters $\beta$ and $\gamma_p$ used in deriving Eq.~\ref{eq:diff1}.
		
		The inclusion of the $R^2$ correction term modifies the evolution of the curvature scalar $R$ in dense environments. Equation \ref{diff1_R2} governs the dynamics of the rescaled curvature variable $y_p = \beta (R_{\mathrm{ch}} / R)$ and incorporates the higher-order contribution from the Starobinsky term. Relative to Equation \ref{eq:diff1}, the primary distinction lies in the additional term $g_{pI} y_p^{-(2n+2)} \left[y_p^{\prime\prime} - \frac{2 y_p^{\prime 2}}{y_p} \right]$, which directly originates from the $R^2$ component.
		
		To understand its physical impact, recall that Equation \ref{eq:diff1} describes an oscillatory system driven by a damping term $2n \frac{y_p^{\prime 2}}{y_p}$ and a restoring force $y_p^{-2n} \left[ \left(1 + \frac{\tau}{\tau_{\mathrm{ch}}}\right) - y_p^{-1} \right]$. This system oscillates around the attractor solution $y_p(\tau) = 1 / (1 + \tau / \tau_{\mathrm{ch}})$, with singularities appearing when $y_p \to 0$ (i.e., $R \to \infty$). The $R^2$ correction modifies this evolution by introducing a feedback mechanism that couples the second derivative $y_p^{\prime\prime}$ and the nonlinear term $y_p^{\prime 2} / y_p$ via a factor $g_{pI} y_p^{-(2n+2)}$. Since the exponent is negative, this term becomes increasingly dominant as $y_p$ decreases—precisely in the regime where singularities are most threatening.
		Physically, the correction acts as a stabilizer: when $y_p^{\prime\prime}$ is positive, the opposing term $-2 y_p^{\prime 2}/y_p$ can reduce the rate of divergence, dampening the trajectory toward $y_p = 0$. Conversely, when $y_p^{\prime\prime}$ is negative, the term reinforces the restoring effect, countering further deviation. As a result, the $R^2$ term serves as a regulator that may prevent the onset of curvature singularities.
		
		The magnitude of this regulation depends critically on the parameter $g_{pI}$. When $g_{pI}$ is small, the dynamics remain governed by the original oscillatory terms, allowing singularities to develop. As $g_{pI}$ increases, the influence of the stabilizing term grows, reducing the oscillation amplitude and potentially preventing $y_p$ from reaching zero. This reveals a clear threshold effect: for sufficiently large $g_{pI}$, the curvature evolution is regularized.
		
		Numerical solutions to Equation \ref{diff1_R2} are displayed in Figure \ref{fig:fp_R2}, where $n = 2$ and $\tau_{\mathrm{ch}} = 80$, enabling direct comparison with Figure \ref{fig:fp}. The figure highlights the impact of varying $g_{pI}$. For $g_{pI} \ll 1$, the curvature scalar undergoes pronounced oscillations, and the singularity time $\tau_{\mathrm{sing}}$ is highly sensitive to $g_{pI}$. Specifically, $\tau_{\mathrm{sing}} \simeq 505.76 \tau_{\mathrm{ch}}$ for $g_{pI} = 10^{-4}$, $\tau_{\mathrm{sing}} \simeq 55.76 \tau_{\mathrm{ch}}$ for $g_{pI} = 10^{-8}$, and $\tau_{\mathrm{sing}} \simeq 4.33 \tau_{\mathrm{ch}}$ for $g_{pI} = 10^{-12}$. As $g_{pI}$ increases, the oscillations are progressively damped, and $y_p(\tau)$ approaches the equilibrium trajectory $y = 1 / (1 + \tau / \tau_{\mathrm{ch}})$. When $g_{pI} > 1$, singularities are fully suppressed.
		
		These results confirm that the $R^2$ correction term can effectively resolve curvature singularities in power-law $f(R)$ dark energy models, provided that $g_{pI} > 1$. For $\lambda = 1$, $R_{\mathrm{ch}} \simeq 4\Lambda \simeq 2.024 \times 10^{-83} M_{\mathrm{Pl}}^2$, and $m_R = M$, one finds $g_{pI} \simeq 10^{-44}$ for $n = 2$, $g_{pI} \simeq 10^{-34}$ for $n = 3$, and $g_{pI} \simeq 10^{-24}$ for $n = 4$. These values indicate that the Starobinsky inflationary scale, $M \approx 1.13 \times 10^{-5} M_{\mathrm{Pl}}$, is too large to suppress singularities in this context. Achieving $g_{pI} > 1$ requires lower mass scales: $m_R \simeq 1.28 \times 10^{-27} M_{\mathrm{Pl}}$ for $n = 2$, $m_R \simeq 1.26 \times 10^{-22} M_{\mathrm{Pl}}$ for $n = 3$, and $m_R \simeq 1.37 \times 10^{-17} M_{\mathrm{Pl}}$ for $n = 4$. These findings underscore the crucial role of the $R^2$ term in stabilizing curvature dynamics and preventing singularity formation in modified gravity theories.

		
		\begin{figure}[htbp] 
			\centering 
			\begin{subfigure}[b]{0.45\textwidth} 
				\includegraphics[width=\textwidth]{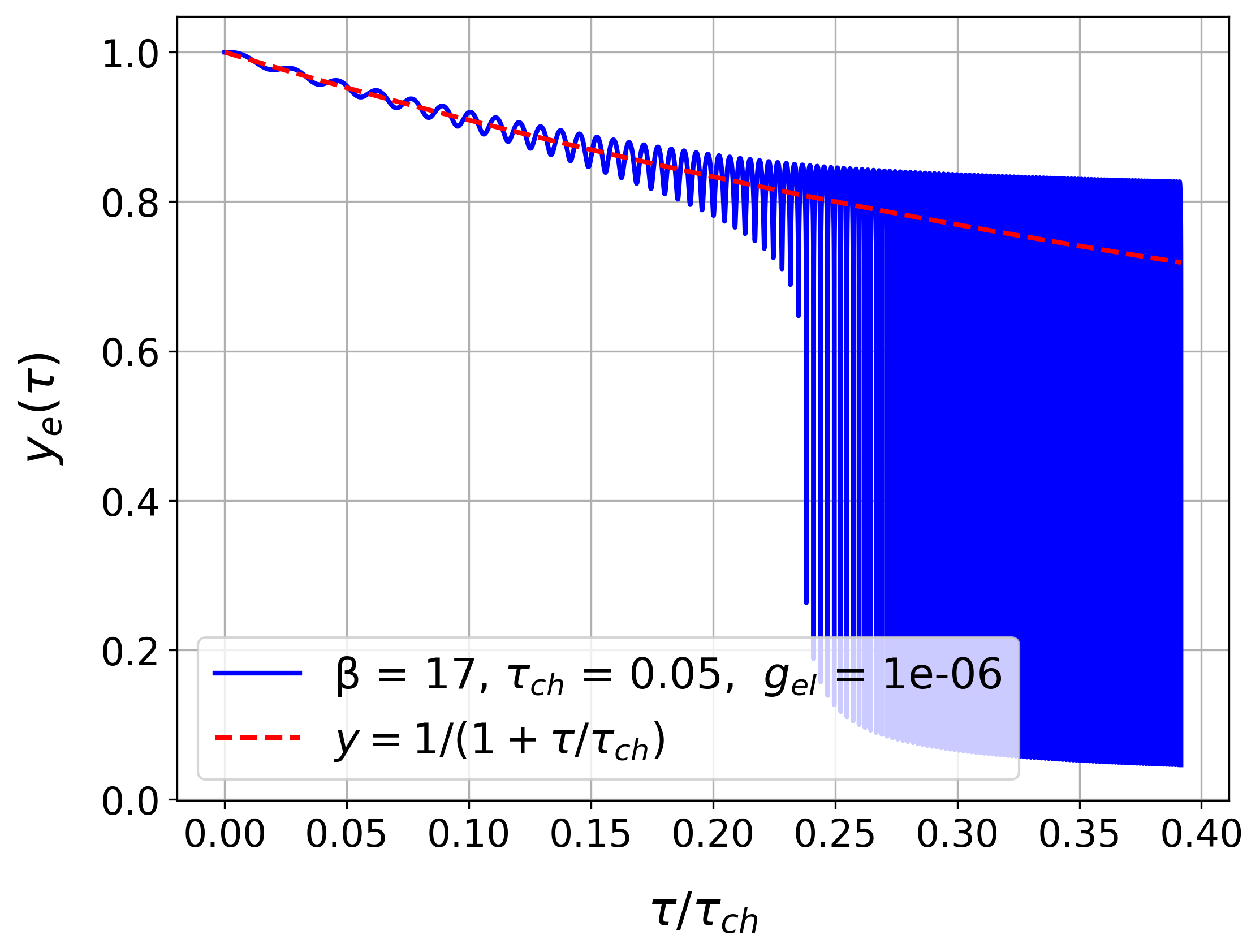} 
				\caption{} 
				\label{fig:subfig1_R2_1}
			\end{subfigure}
			\hfill 
			\begin{subfigure}[b]{0.45\textwidth} 
				\includegraphics[width=\textwidth]{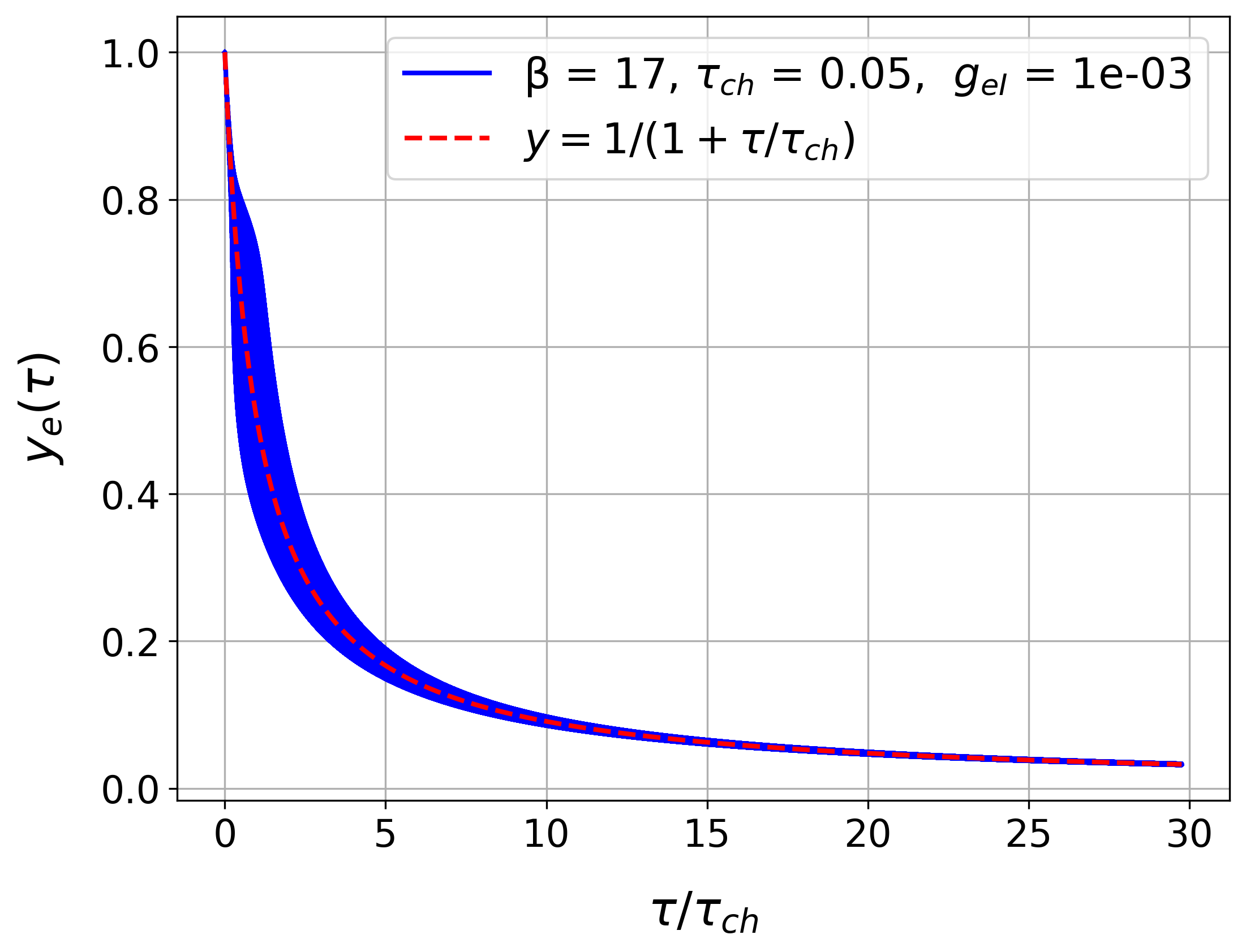} 
				\caption{} 
				\label{fig:subfig2_R2_1}
			\end{subfigure}
			
			\vspace{1em} 
			
			\begin{subfigure}[b]{0.45\textwidth} 
				\includegraphics[width=\textwidth]{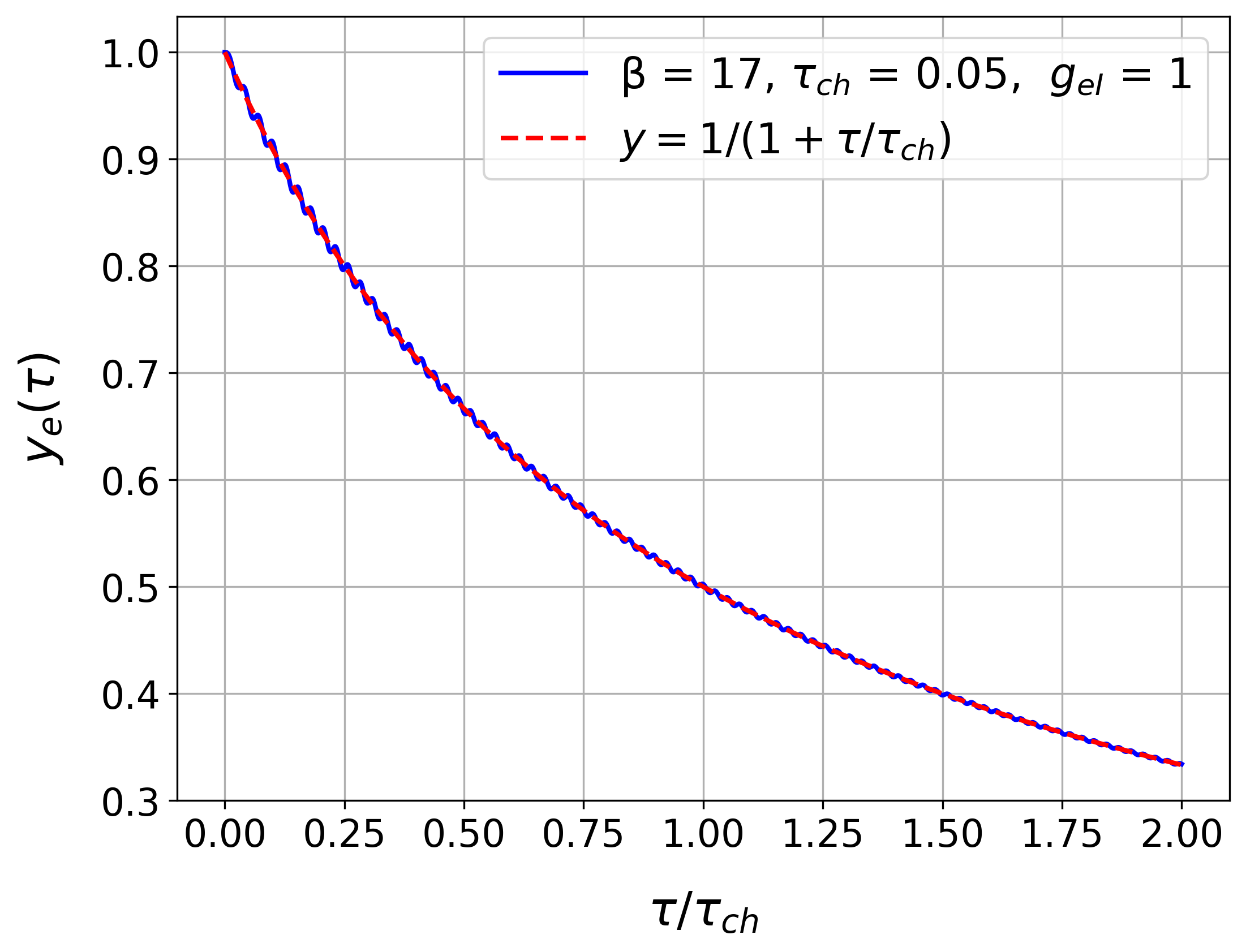} 
				\caption{} 
				\label{fig:subfig3_R2_1}
			\end{subfigure}
			\hfill 
			\begin{subfigure}[b]{0.45\textwidth} 
				\includegraphics[width=\textwidth]{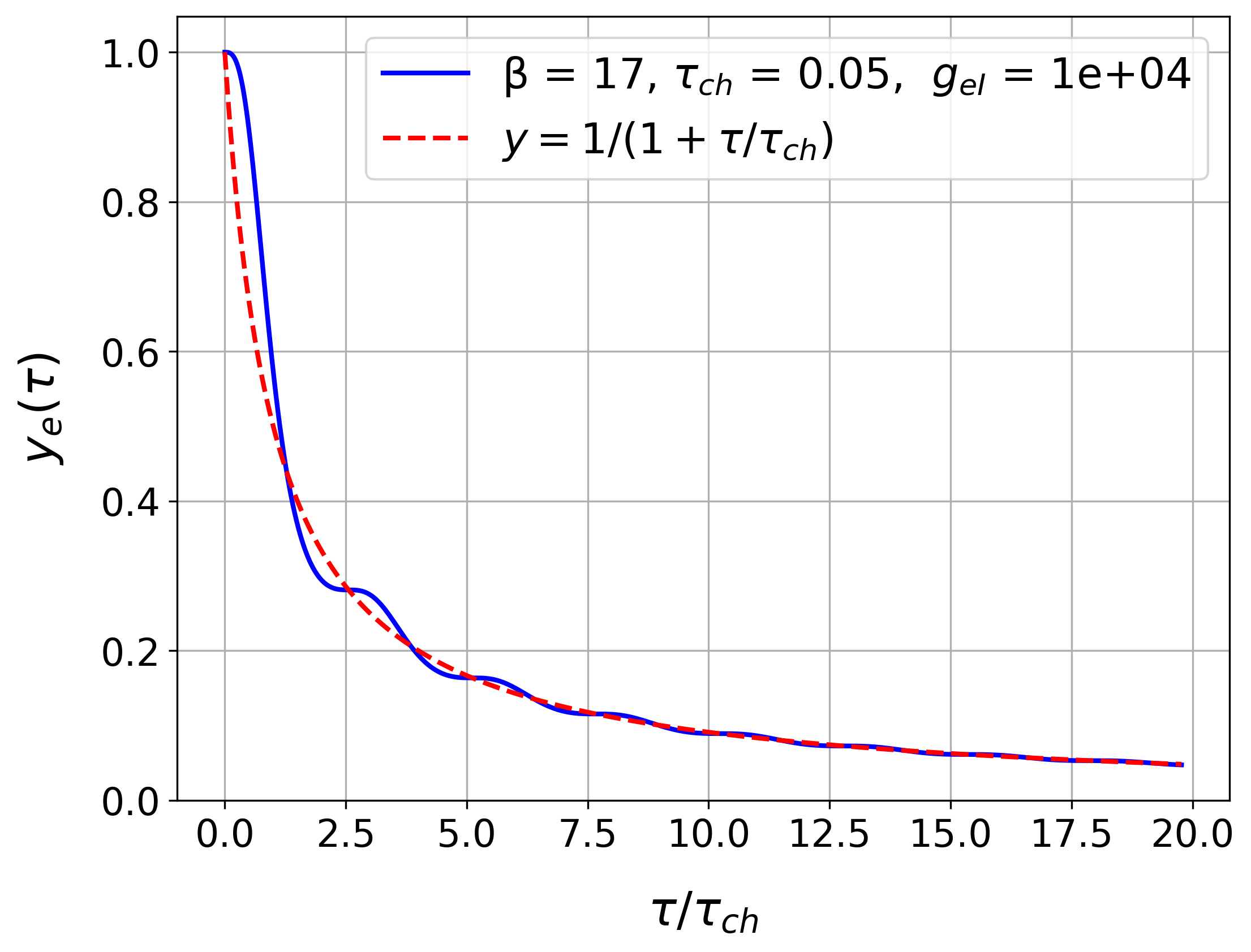} 
				\caption{} 
				\label{fig:subfig4_R2_1}
			\end{subfigure}

			\caption{Oscillations of $y_e(\tau)$ in the exponential-law dark energy model with $R^2$ correction for various values  $g_{eI}$.} 
			\label{fig:fe_R2}
		\end{figure}

		In the regime where $R \geq R_{\mathrm{ch}}$, the exponential-law model takes the form
		\begin{equation}
			f(R) = R - \lambda R_{\mathrm{ch}} \left(1 - e^{-R/R_{\mathrm{ch}}} \right) + \frac{R^2}{6 m_R^2},
		\end{equation}
		where the final term represents the $R^2$ correction. The corresponding differential equation derived from Equation \ref{eq:trace} is given by
		\begin{equation}\label{diff:ye_R2}
			y_e'' + (\beta y_e^{-1} - 2) \frac{{y_e'}^2}{y_e} + g_{eI} e^{\beta \left(\frac{1}{y_e} - 1 \right)} \left(y_e'' - \frac{2 {y_e'}^2}{y_e} \right) + y_e^2 e^{\frac{\beta}{y_e}} \left[\left(1 + \frac{\tau}{\tau_{\mathrm{ch}}} \right) - y_e^{-1} \right] = 0,
		\end{equation}
		where the coefficient $g_{eI}$ is defined as
		\begin{equation}\label{eq:4.9}
			g_{eI}:= \frac{2 R_{\mathrm{ch}}}{3 m_R^2 \lambda}  e^\beta.
		\end{equation}
		Here, $\beta$ and $\gamma_e$ retain their definitions as introduced in the uncorrected model.
		
		The third term in Equation \ref{diff:ye_R2}, given by $g_{eI} e^{\beta(1/y_e - 1)} \left(y_e^{\prime\prime} - \frac{2y_e^{\prime 2}}{y_e}\right)$, originates from the $R^2$ correction applied to the exponential-type $f(R)$ dark energy model. This term plays a crucial role in regulating the curvature dynamics and mitigating singular behavior through several interrelated mechanisms. The exponential prefactor $e^{\beta(1/y_e - 1)}$ renders this term highly sensitive to the curvature regime: as $y_e \to 0$ (corresponding to $R \to \infty$), the term increases rapidly, thereby dominating the evolution near potential singularities.
		The differential structure $y_e^{\prime\prime} - 2y_e^{\prime 2}/y_e$ acts as a feedback mechanism. When $y_e^{\prime\prime} > 0$, indicating an accelerating collapse toward a singularity, the negative contribution $-2y_e^{\prime 2}/y_e$ can counterbalance this acceleration. Conversely, when $y_e^{\prime\prime} < 0$, the term can reinforce the restoring force, inhibiting divergent behavior. This response contrasts with the polynomial prefactor in the power-law model of Equation \ref{diff1_R2}, where the exponential term here enables stronger and more localized regulation of curvature growth, even for relatively small values of the coupling constant $g_{eI}$.
		
		The numerical solutions illustrated in Figure \ref{fig:fe_R2} elucidate the impact of the $R^2$ correction on the system's behavior. For small values of $g_{eI}$ ($g_{eI} \ll 1$), the evolution of $y_e(\tau)$ exhibits damped oscillations around the equilibrium trajectory $y_e = 1/(1 + \tau/\tau_{\rm ch})$. Compared to the uncorrected scenario, the oscillation amplitudes are significantly suppressed by the exponential correction. As $g_{eI}$ exceeds unity, the singular behavior is entirely suppressed: the function $y_e(\tau)$ asymptotically approaches the equilibrium solution without ever reaching zero, indicating complete stabilization of the curvature scalar.
		
		This parameter dependence yields important physical insights. In particular, achieving $g_{eI} \geq 1$ necessitates the mass scale $m_R$ of the $R^2$ correction to be on the order of $10^{-38} M_{\rm Pl}$, assuming typical values of $\lambda \sim 1$ and $R_{\rm ch} \sim 4\Lambda$. This required scale is dramatically lower than the inflationary mass scale associated with the Starobinsky model, $M \sim 10^{-5} M_{\rm Pl}$. Thus, these findings suggest that effective singularity resolution via $R^2$ corrections in the exponential $f(R)$ model operates at energy scales far below those characteristic of early-universe inflation.

\section{Solving singularity by adding $R^{\frac{m+2}{m+1}}$ term}
		
		We examine whether $f(R)$ models capable of driving inflation can simultaneously avoid curvature singularities in the context of $f(R)$ dark energy. Our analysis focuses on a generalized $f(R)$ gravity framework that extends the Starobinsky scalar potential \cite{Sebastiani2013}, specifically considering the case where the potential takes the form $V(\phi) \simeq \gamma \exp(-m\phi)$ with $m > 0$. In the high-curvature regime ($R \gg \gamma$), this yields:
		\begin{equation}
			f(R \gg \gamma) \simeq \gamma \left(\frac{m+1}{m+2}\right) \left( \frac{1}{4(m+2)} \right)^{\frac{1}{1+m}}  \left( \frac{R}{\gamma} \right)^{\frac{m+2}{m+1}}.
		\end{equation}
		The corresponding primordial power spectrum amplitude, spectral indices, and tensor-to-scalar ratio are given by:
		\begin{equation}
			\Delta_{\mathcal{R}}^2 \simeq \frac{\kappa^2 \gamma \, e^{\frac{2}{3} N m^2}}{8\pi^2 m^2}, \quad
			n_s \simeq 1 - \frac{2m^2}{3}, \quad r \simeq \frac{16m^2}{3}.
		\end{equation}
		These predictions remain consistent with Planck data \cite{Akrami_2020} for $m = 0.1,0.2$, with $\gamma\simeq 10^{-11}  M_{\rm Pl}^2$, indicating that Starobinsky-like $R^2$ gravity models can successfully drive this inflationary scenario.
		
		\begin{figure}[htbp] 
			\centering 
			\begin{subfigure}[b]{0.32\textwidth} 
				\includegraphics[width=\textwidth]{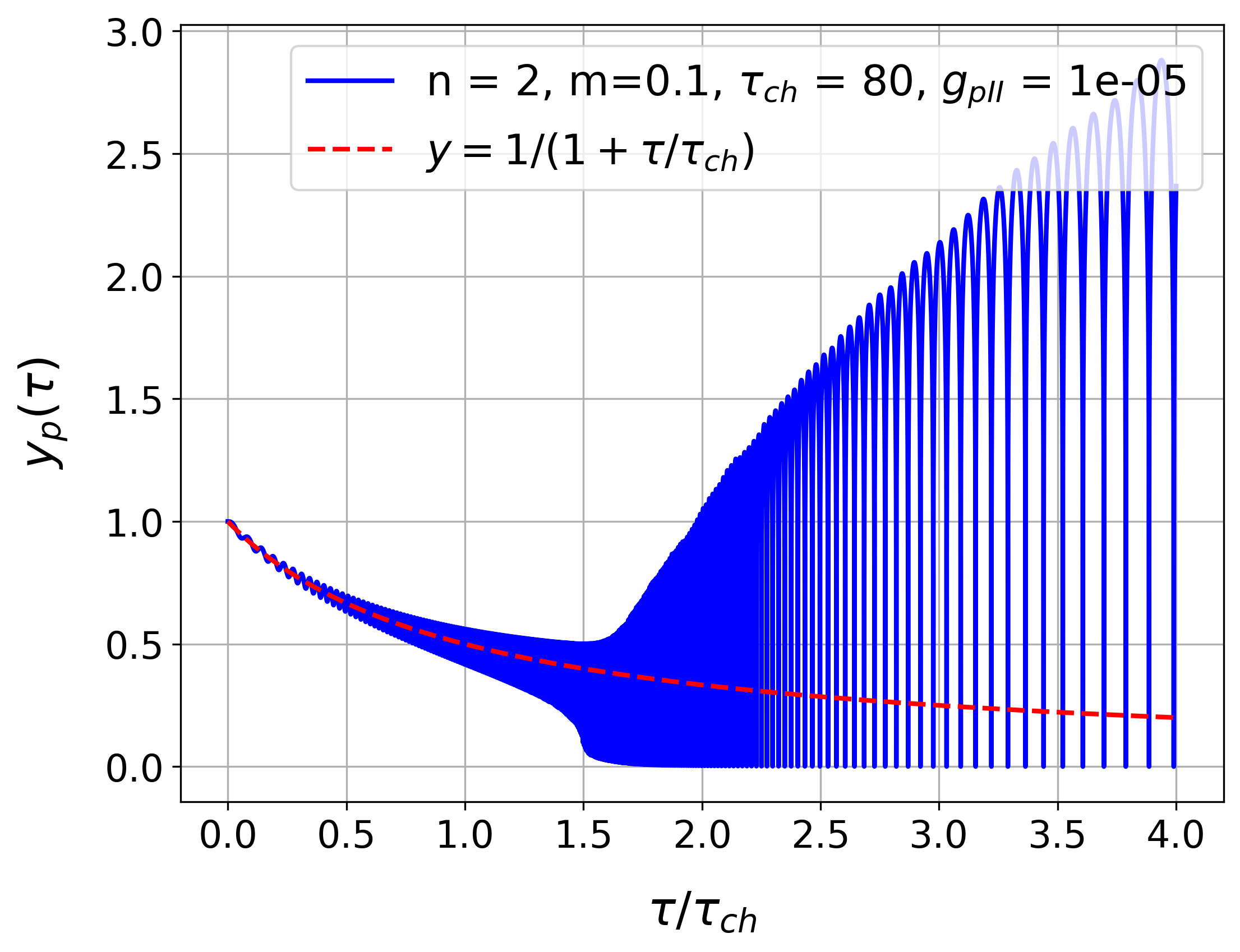} 
				\caption{} 
				\label{fig:subfig1_m1}
			\end{subfigure}
			\hfill 
			\begin{subfigure}[b]{0.32\textwidth} 
				\includegraphics[width=\textwidth]{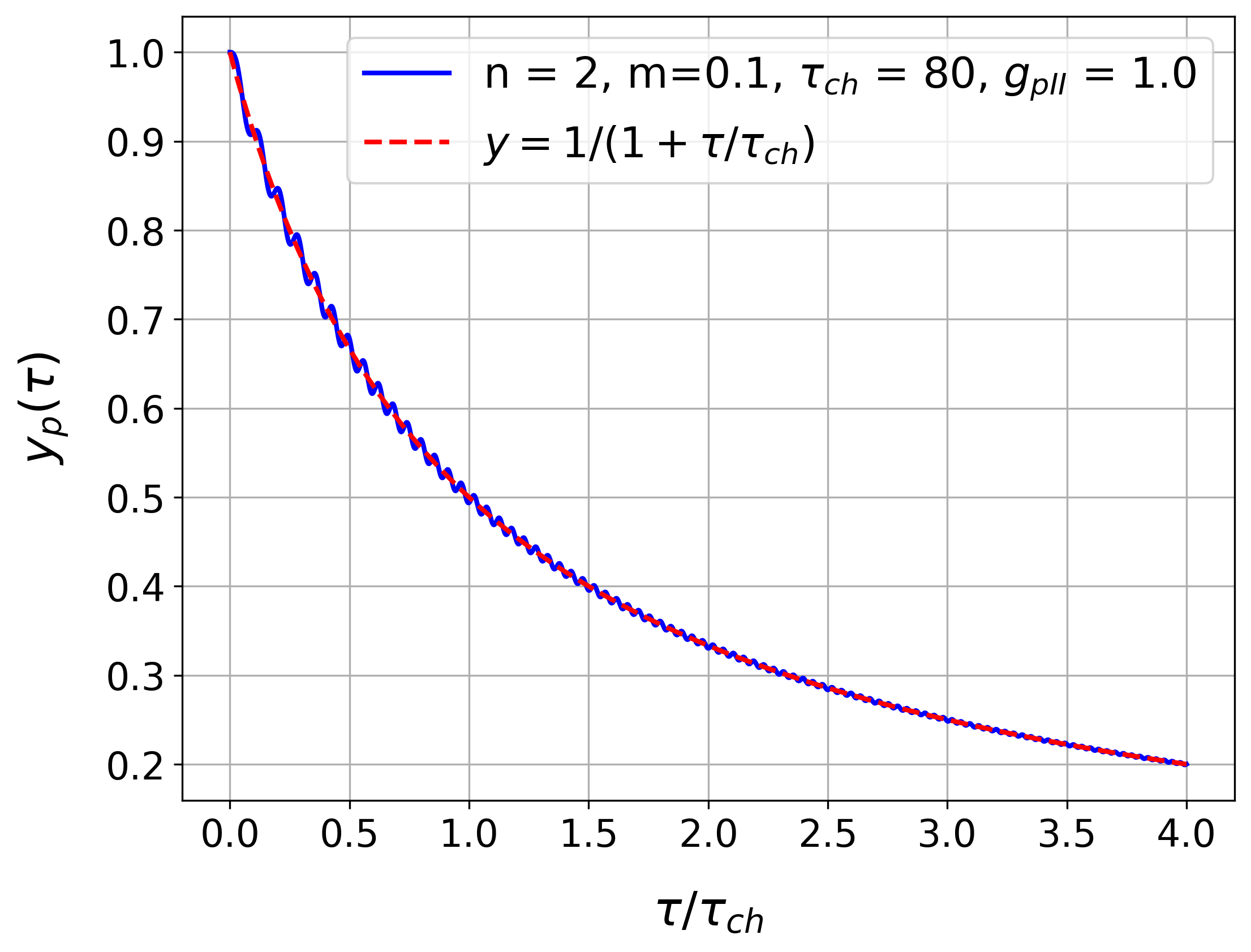} 
				\caption{} 
				\label{fig:subfig2_m1}
			\end{subfigure}
			\hfill 
			\begin{subfigure}[b]{0.32\textwidth} 
				\includegraphics[width=\textwidth]{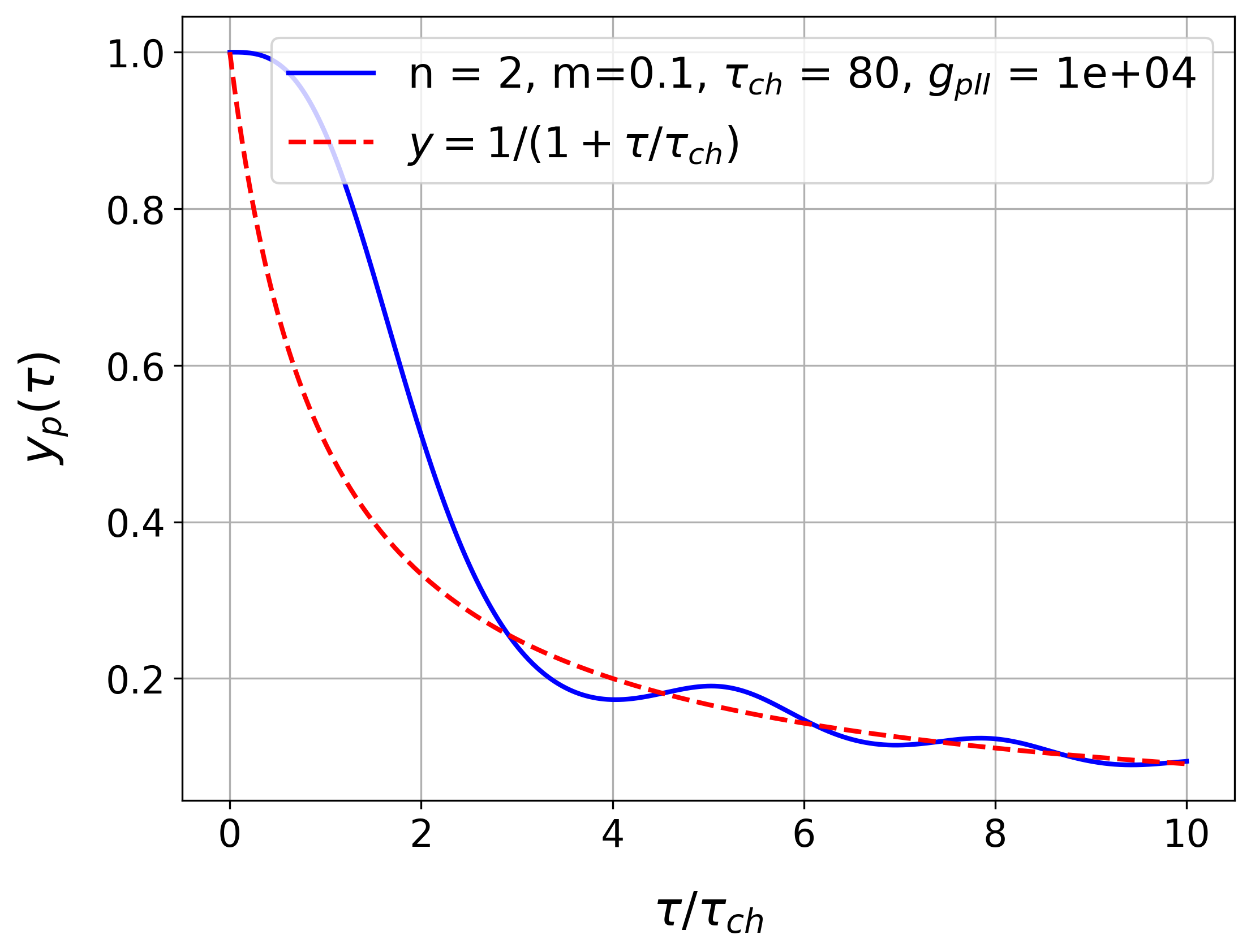} 
				\caption{} 
				\label{fig:subfig3_m1}
			\end{subfigure}
			
			\vspace{1em} 

			\begin{subfigure}[b]{0.32\textwidth} 
				\includegraphics[width=\textwidth]{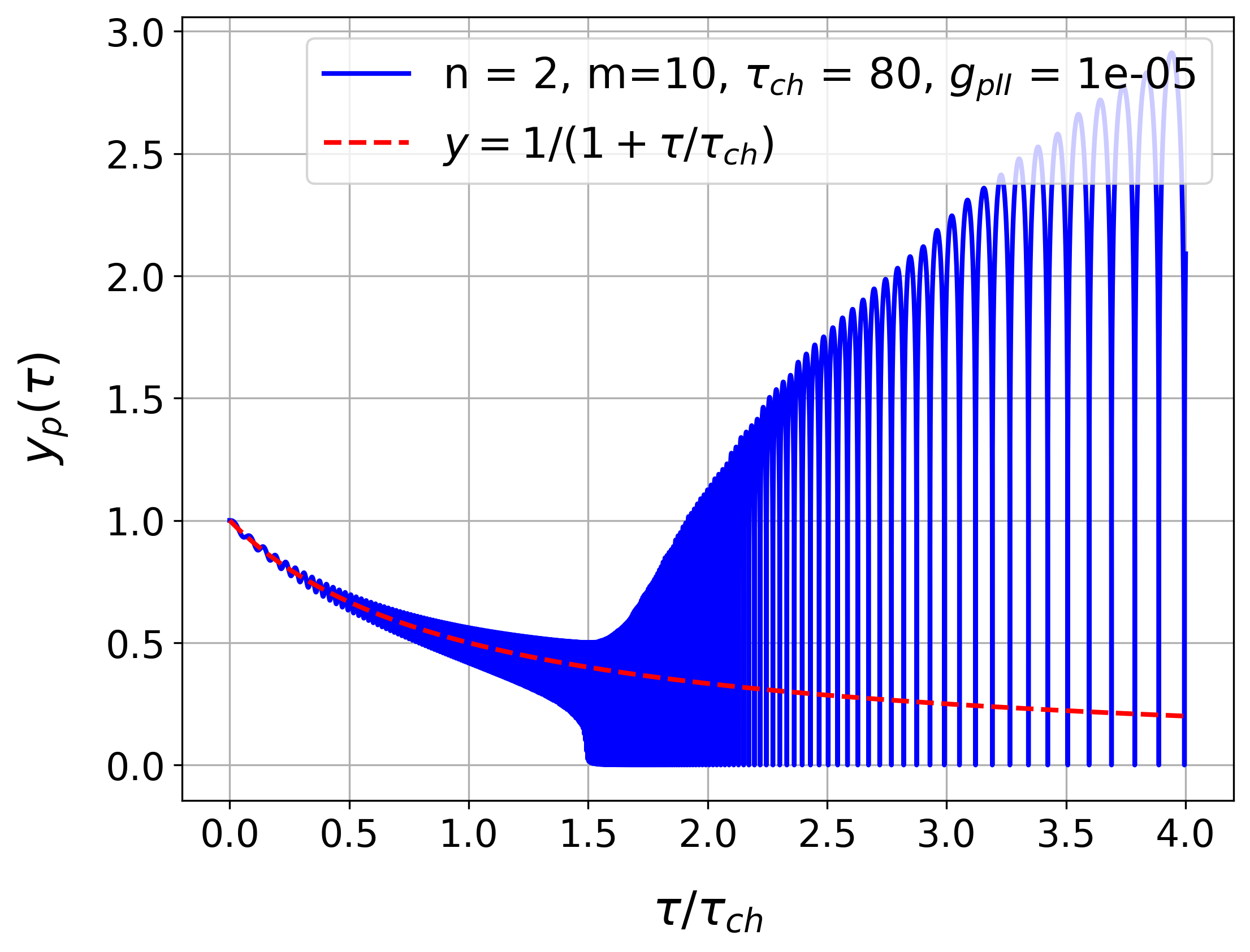} 
				\caption{} 
				\label{fig:subfig4_m1}
			\end{subfigure}
			\hfill 
			\begin{subfigure}[b]{0.32\textwidth} 
				\includegraphics[width=\textwidth]{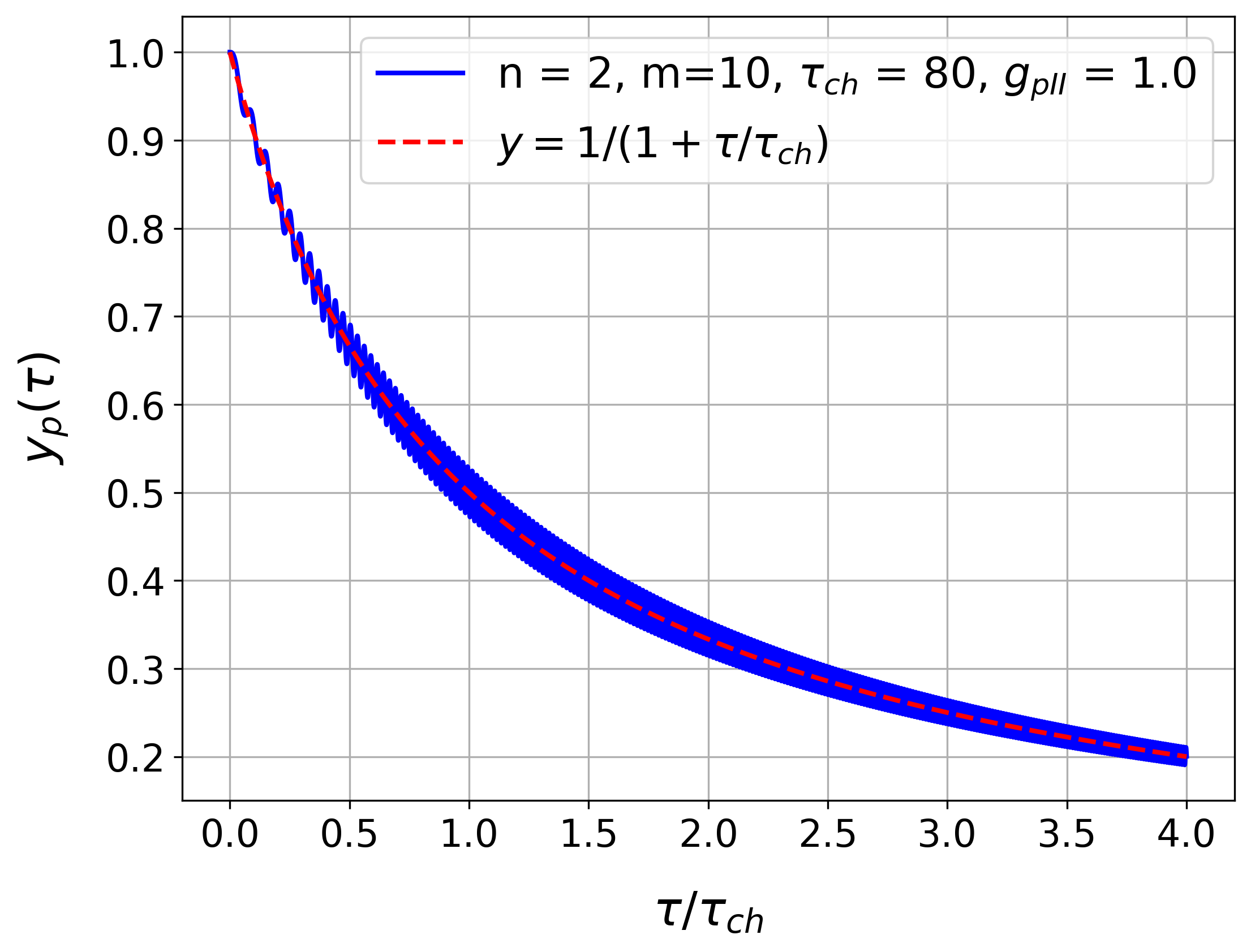} 
				\caption{} 
				\label{fig:subfig5_m1}
			\end{subfigure}
			\hfill 
			\begin{subfigure}[b]{0.32\textwidth} 
				\includegraphics[width=\textwidth]{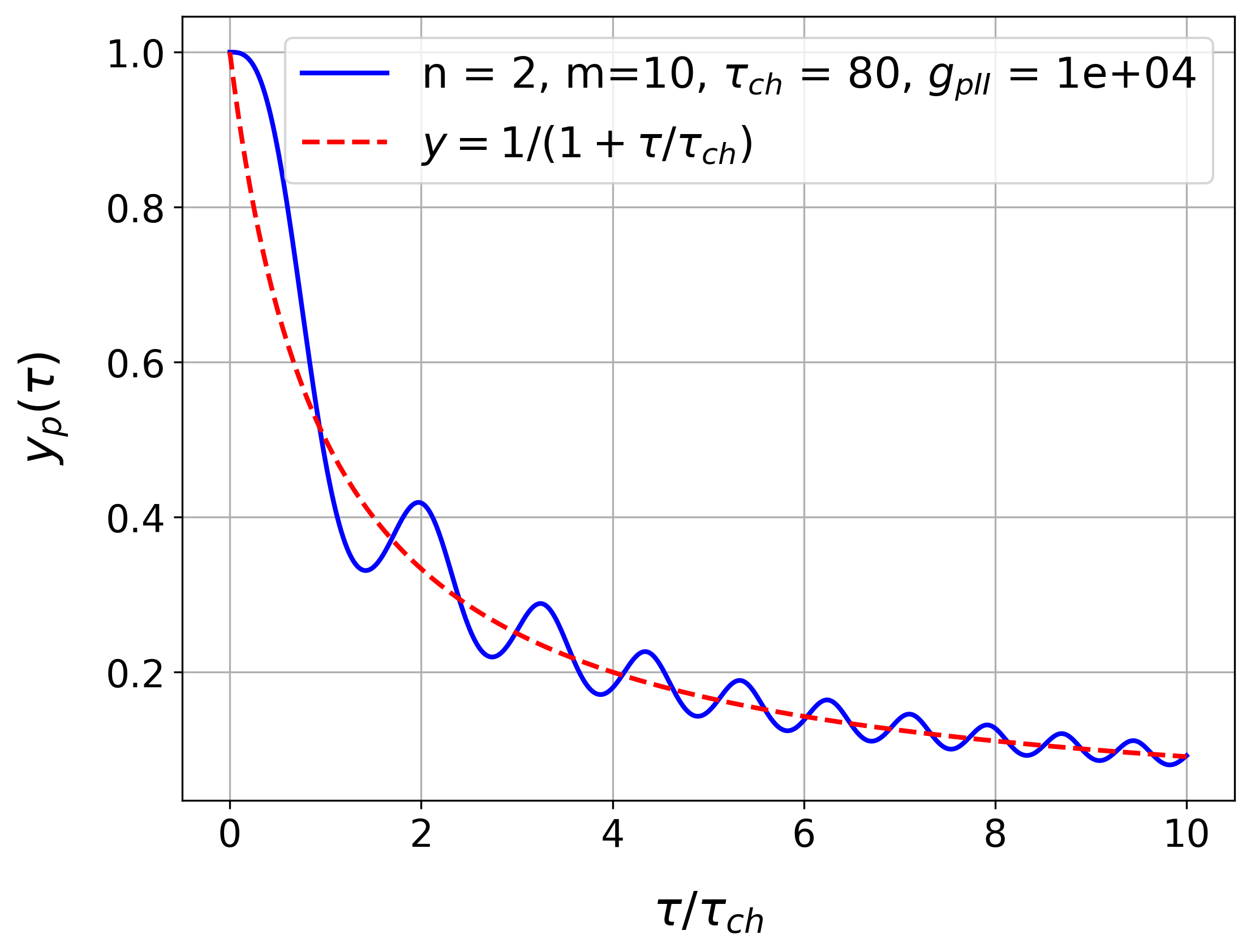} 
				\caption{} 
				\label{fig:subfig6_m1}
			\end{subfigure}

			\caption{Oscillations of $y_p(\tau)$ in the power-law dark energy model with $R^{\frac{m+2}{m+1}}$ correction for various values  $g_{pII}$.} 
			\label{fig:fp_Rm}
		\end{figure}
		
		For the power-law $f(R)$ dark energy model with $R^{\frac{m+2}{m+1}}$ correction in the limit $R \gg R_{\rm ch}$, we obtain:
		\begin{equation} \label{eq:5.4}
			f(R) = R - \lambda R_\text{ch} \left[1 - \left(\frac{R_{ch}}{R} \right)^{2n} \right] + \Gamma R^{\frac{m+2}{m+1}},
		\end{equation}
		where 
		\begin{equation}
			\Gamma := \gamma \left(\frac{m+1}{m+2}\right) \left(\frac{1}{4 (m+2)}\right)^{\frac{1}{m+1}} \left(\frac{1}{\gamma}\right)^{\frac{m+2}{m+1}}.
		\end{equation}
		Here, $m$ represents an arbitrary positive real number, not necessarily integer-valued. The correction term $R^{\frac{m+2}{m+1}}$ approaches $R^2$ for small $m$ and converges to $R$ for large $m$. From this, we derive the differential equation:
		\begin{equation}\label{eq:yp2}
			y_p'' + 2n \frac{{y_p'}^2}{y_p} + \frac{1}{m+1}  g_{pII} y_p^{-(2n+2)}\left[y_p'' - \frac{(m+2)}{(m+1)} \frac{{y_p'}^2}{y_p} \right] +  y_p^{-2n} \left[\left(1 + \frac{\tau}{\tau_{ch}} \right) - y_p^{-1} \right]  = 0
		\end{equation}
		with
		\begin{equation}\label{eq:gp2}
			g_{pII} = \frac{(m+2)}{(m+1)} \frac{\beta^{2n+1}}{2 \lambda n (2 n + 1)}  \Gamma (R_{ch} \beta)^{\frac{1}{m+1}}.
		\end{equation}
		Notably, Eq.~\eqref{eq:yp2} reduces to Eq.~\eqref{eq:fp_star}, while Eq.~\eqref{eq:gp2} converges to Eq.~\eqref{eq:gp1} as $m \rightarrow 0$.
		
		Following our established methodology, we numerically solve the differential equation for $y_p(\tau)$ using initial conditions $y_p(0)=1$ and $y_p'(0)=0$. Figure~\ref{fig:fp_Rm} presents these solutions for various $m$ and $g_{pII}$ values in the high-density regime ($\beta \simeq 10^5$), with fixed parameters $n=2$ and $\tau_{ch}=80$ to enable direct comparison with the $R^2$ correction case shown in Fig.~\ref{fig:fp_R2}.
		
		The results reveal several key features. First, the power-law parameter $m$ exhibits relatively weak influence on the solutions, with minimal differences between small ($m=0.1$) and large ($m=10$) values. The $m=0.1$ case closely resembles the $R^2$ correction results, confirming that the $R^2$ scenario represents a special case of our more general $R^{\frac{m+2}{m+1}}$ framework.
		
		While $m$ appears primarily relevant for inflationary-scale corrections, we find that curvature singularities can be avoided even for large $m$ values. However, larger $m$ values do enhance both the amplitude and frequency of $y_p(\tau)$ oscillations. The parameter $g_{pII}$ emerges as the dominant factor controlling singularity formation - when $g_{pII} \geq 1$, both oscillation amplitude and frequency decrease significantly, with $y_p(\tau)$ asymptotically approaching the equilibrium solution $y = 1/(1+\tau/\tau_{ch})$. This behavior robustly prevents curvature singularities at finite times.


		\begin{figure}[htbp] 
			\centering 
			\begin{subfigure}[b]{0.32\textwidth} 
				\includegraphics[width=\textwidth]{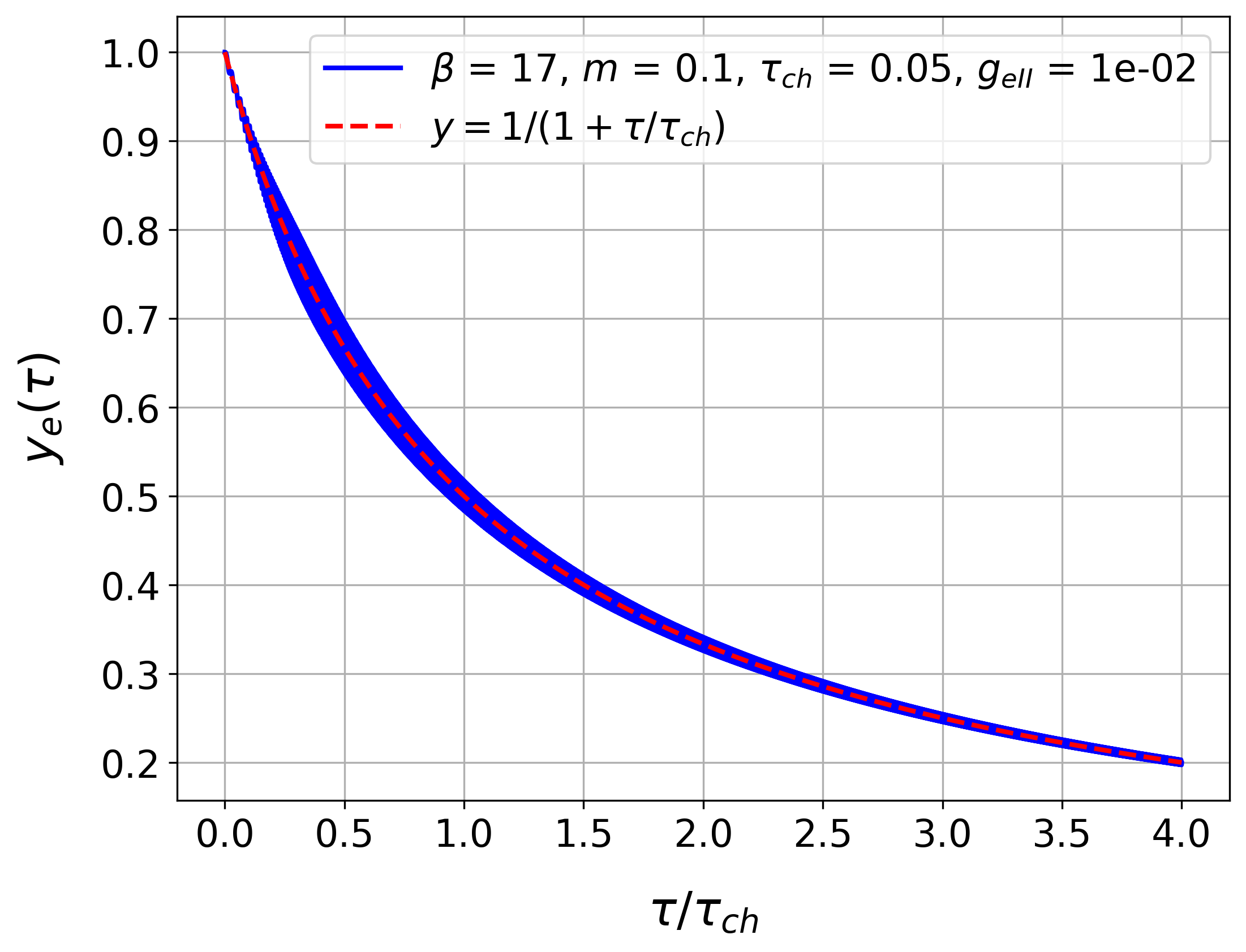} 
				\caption{} 
				\label{fig:subfig1_m2}
			\end{subfigure}
			\hfill 
			\begin{subfigure}[b]{0.32\textwidth} 
				\includegraphics[width=\textwidth]{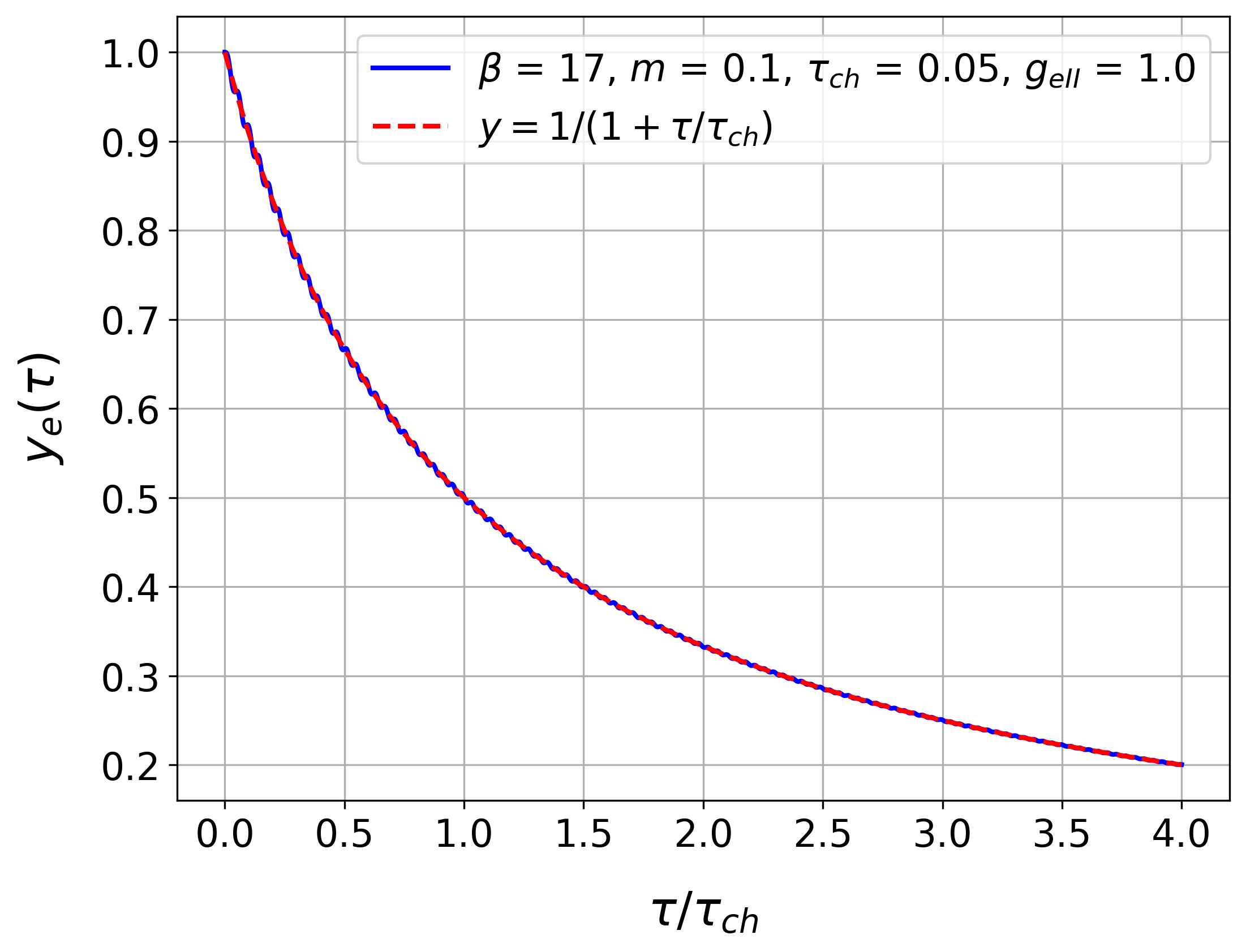} 
				\caption{} 
				\label{fig:subfig2_m2}
			\end{subfigure}
			\hfill
			\begin{subfigure}[b]{0.32\textwidth} 
				\includegraphics[width=\textwidth]{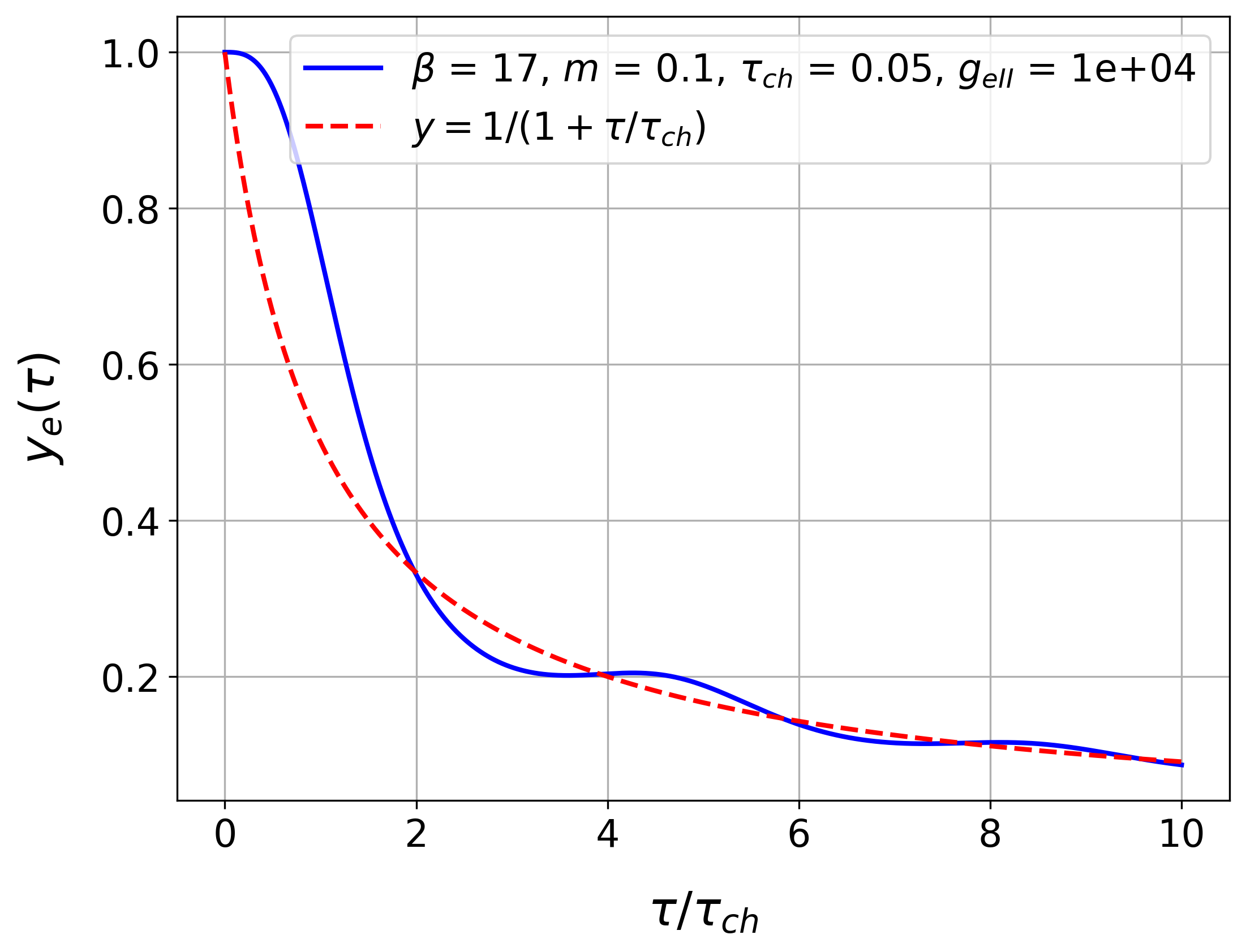} 
				\caption{} 
				\label{fig:subfig3_m2}
			\end{subfigure}
			
			\vspace{1em} 
			
			\begin{subfigure}[b]{0.32\textwidth} 
				\includegraphics[width=\textwidth]{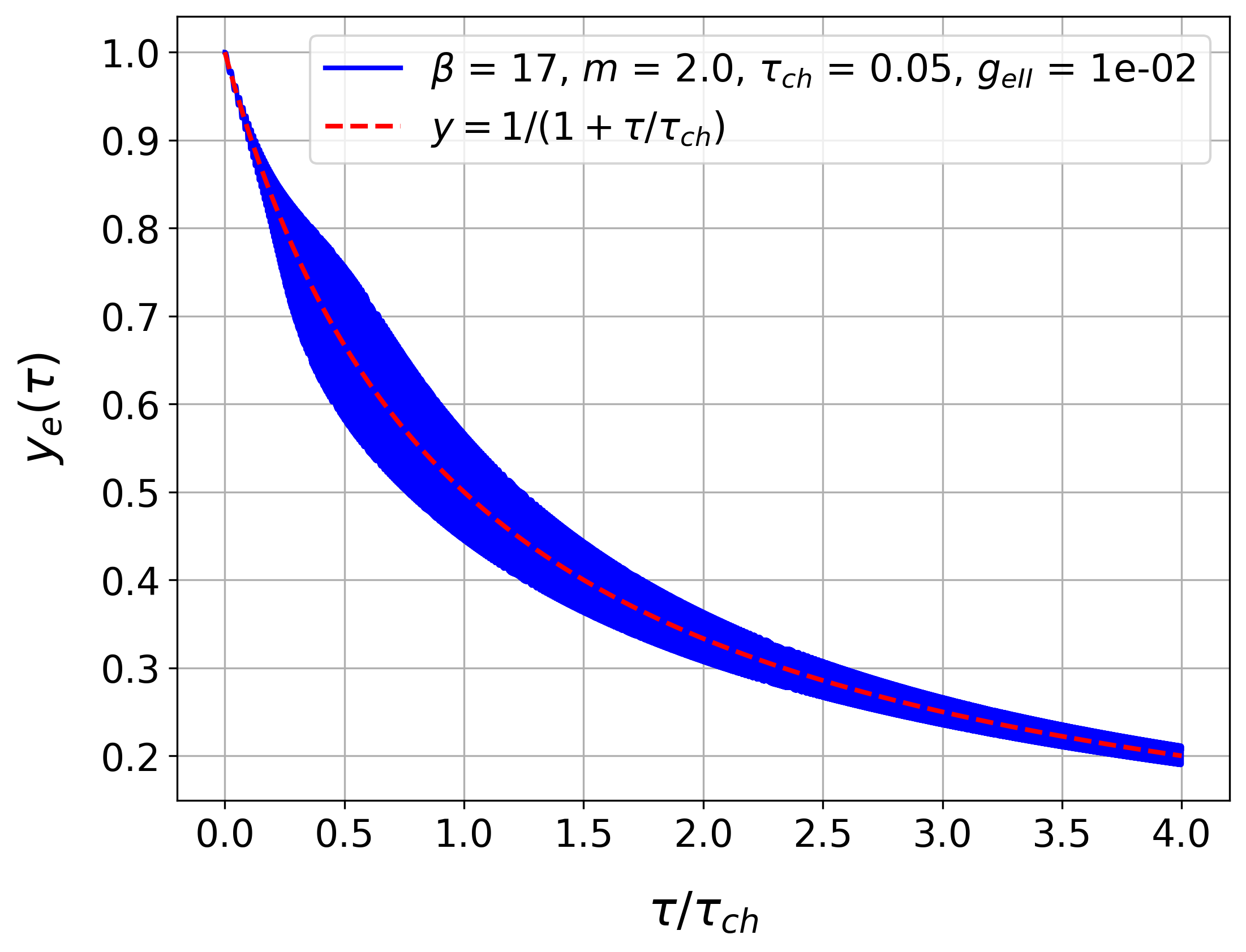} 
				\caption{} 
				\label{fig:subfig4_m2}
			\end{subfigure}
			\hfill 
			\begin{subfigure}[b]{0.32\textwidth} 
				\includegraphics[width=\textwidth]{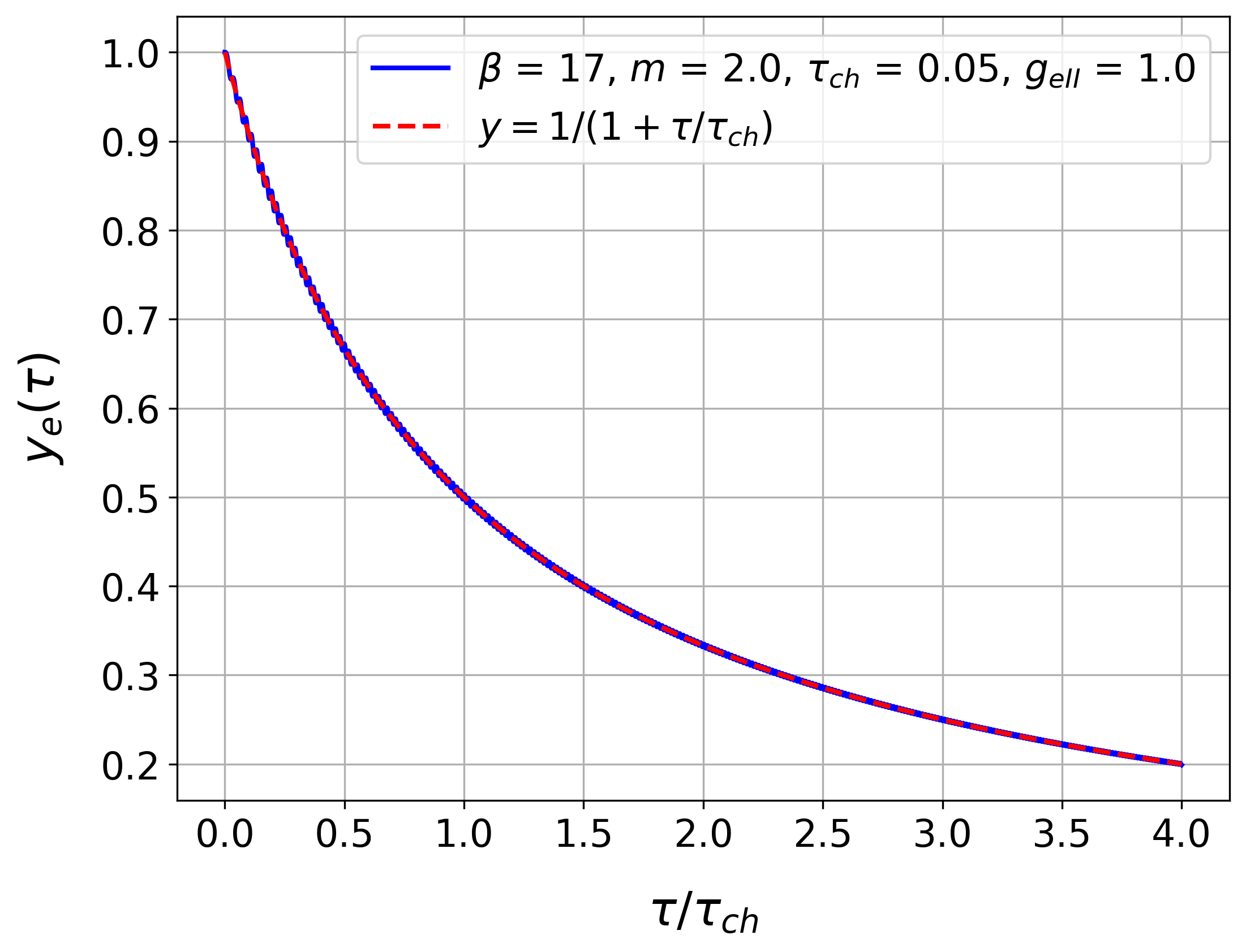} 
				\caption{} 
				\label{fig:subfig5_m2}
			\end{subfigure}
			\hfill 
			\begin{subfigure}[b]{0.32\textwidth} 
				\includegraphics[width=\textwidth]{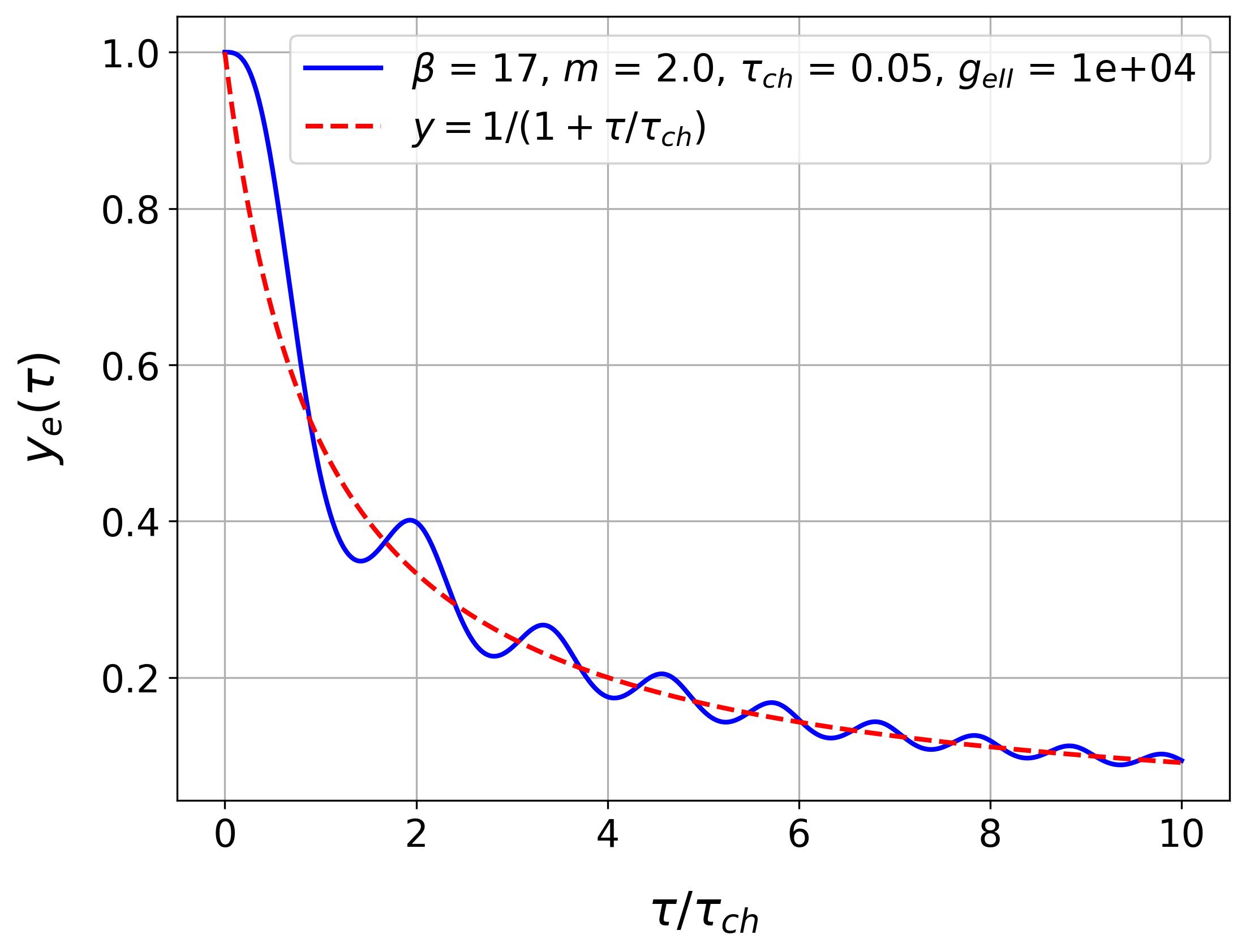} 
				\caption{} 
				\label{fig:subfig6_m2}
			\end{subfigure}

			\caption{Oscillations of $y_e(\tau)$ in the exponential-law dark energy model with $R^{\frac{m+2}{m+1}}$ correction for various values  $g_{eII}$.} 
			\label{fig:fe_Rm}
		\end{figure}
		
		The exponential-law $f(R)$ dark energy model incorporating an $R^{\frac{m+2}{m+1}}$ correction term takes the following form:
		\begin{equation} \label{eq:5.8}
			f(R) = R - \lambda R_{\mathrm{ch}} \left(1 - e^{-R/R_{\mathrm{ch}}} \right) +  \Gamma R^{\frac{m+2}{m+1}},
		\end{equation}
		which yields the corresponding differential equation derived from Eq.~\ref{eq:trace}:
		\begin{align}\label{eq:5.9}
			y_e'' &+  (\beta y_e^{-1} - 2) \frac{{y_e'}^2}{y_e}  + \frac{(m+2)}{(m+1)^2} \frac{3}{2} g_{eII} y_e^{\frac{m}{m+1}} e^{\beta \left(\frac{1}{y_e} - 1 \right)} \left[ y_e'' - \frac{(m+2)}{(m+1)} \frac{{y_e'}^2}{y_e} \right] \nonumber \\
			& + y_e^2 e^{\frac{\beta}{y_e}} \left[\left(1 + \frac{\tau}{\tau_{ch}} \right) - y_e^{-1} \right] = 0,
		\end{align}
		where the parameter $g_{eII}$ is defined as:
		\begin{equation}\label{eq:5.10}
			g_{eII} = \frac{2}{3} \frac{\Gamma (R_{ch} \beta)^{\frac{m+2}{m+1}}}{3 R_{ch} \beta^2 \lambda} e^\beta. 
		\end{equation}
		In the limit $m \rightarrow 0$, Eqs.~\eqref{eq:5.9} and \eqref{eq:5.10} reduce to Eqs.~\eqref{diff:ye_R2} and \eqref{eq:4.9}, respectively, recovering the $R^2$ correction case.
		
		Figure~\ref{fig:fe_Rm} presents numerical solutions to the differential equation \eqref{eq:5.9} for model parameters $\beta=17$ and $\tau_{ch}=0.05$, chosen to enable direct comparison with the $R^2$ correction results shown in Figure~\ref{fig:fe_R2}. Following the same convention as in the power-law dark energy analysis, we display results for small $m$ values ($m=0.1$) in the left panels (a, c, e) and large $m$ values ($m=10$) in the right panels (b, d, f).
		
		For the small $m$ case ($m=0.1$), the numerical solutions show excellent agreement with the $R^2$ correction results from Figure~\ref{fig:fe_R2}. The differential term containing the $g_{eII}$ factor proves particularly effective at regularizing curvature singularities that would otherwise occur when $y_e(\tau) \rightarrow 0$. Furthermore, when $g_{eII}\geq 1$, we observe significant suppression of both the oscillation frequency and amplitude of $y_e(\tau)$.
		
		In the large-$m$ regime ($m = 10$), where the term $R^{\frac{m+2}{m+1}}$ asymptotically approaches $R^n$ with $n \sim 1$, we observe qualitatively similar behavior. The $g_{eII}$ term continues to effectively regularize curvature singularities while suppressing both the amplitude and frequency of oscillations when $g_{eII} \geq 1$. In contrast, for $g_{eII} < 1$, the system exhibits amplified oscillatory behavior, with increased amplitude and frequency, potentially resulting in finite-$\tau$ singularities that arise more rapidly than in the small-$m$ case. These findings demonstrate that $R^n$ corrections within the range $1 < n \leq 2$ can successfully mitigate curvature singularities in exponential-law dark energy models, provided the condition $g_{eII} \geq 1$ is satisfied.

\section{Solving singularity by adding $\alpha$-attractor term}
		
		Among inflationary models proposed to address the initial singularity problem, the $\alpha$-attractor framework stands out for its capacity to unify diverse inflationary scenarios, including the well-established Starobinsky model \cite{Kallosh2013, Kallosh2013a, Kallosh2015, Galante2014}. This class of models comprises two principal variants: (1) the T-model with potential $V_T \propto \tanh^2(\varphi / \sqrt{6\alpha})$, and (2) the E-model characterized by $V_E \propto (1 - e^{-\sqrt{2/3\alpha}\varphi})^2$. Remarkably, despite their differing functional forms, both variants yield nearly identical predictions for inflationary observables.
	
		The $\alpha$-attractor models generate universal predictions for the scalar spectral index $n_s$ and tensor-to-scalar ratio $r$:
		\begin{equation}
			n_s = 1 - \frac{2}{N}, \quad r = \frac{12\alpha}{N^2},
		\end{equation}
		where $\alpha > 0$ represents a dimensionless parameter controlling model flexibility, and $N$ denotes the required e-folding number. When $\alpha = 1$, these predictions align precisely with those of the Starobinsky and Higgs inflation models, while the $\alpha \to \infty$ limit recovers chaotic quadratic inflation. 
		
		\begin{figure}[htbp!] 
			\centering 
			\begin{subfigure}[b]{0.32\textwidth} 
				\includegraphics[width=\textwidth]{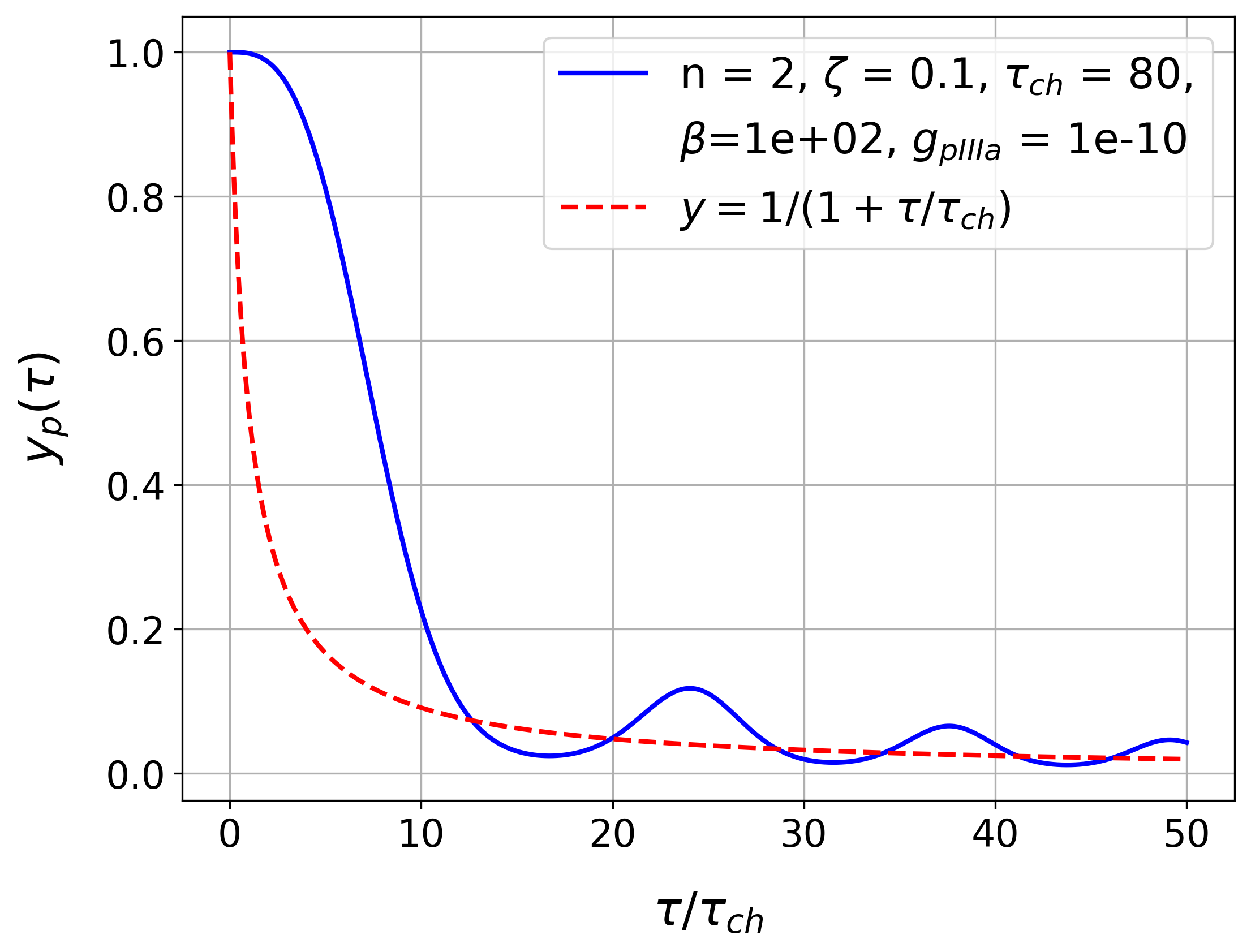} 
				\caption{} 
				\label{fig:subfig1_alpha}
			\end{subfigure}
			\hfill 
			\begin{subfigure}[b]{0.32\textwidth} 
				\includegraphics[width=\textwidth]{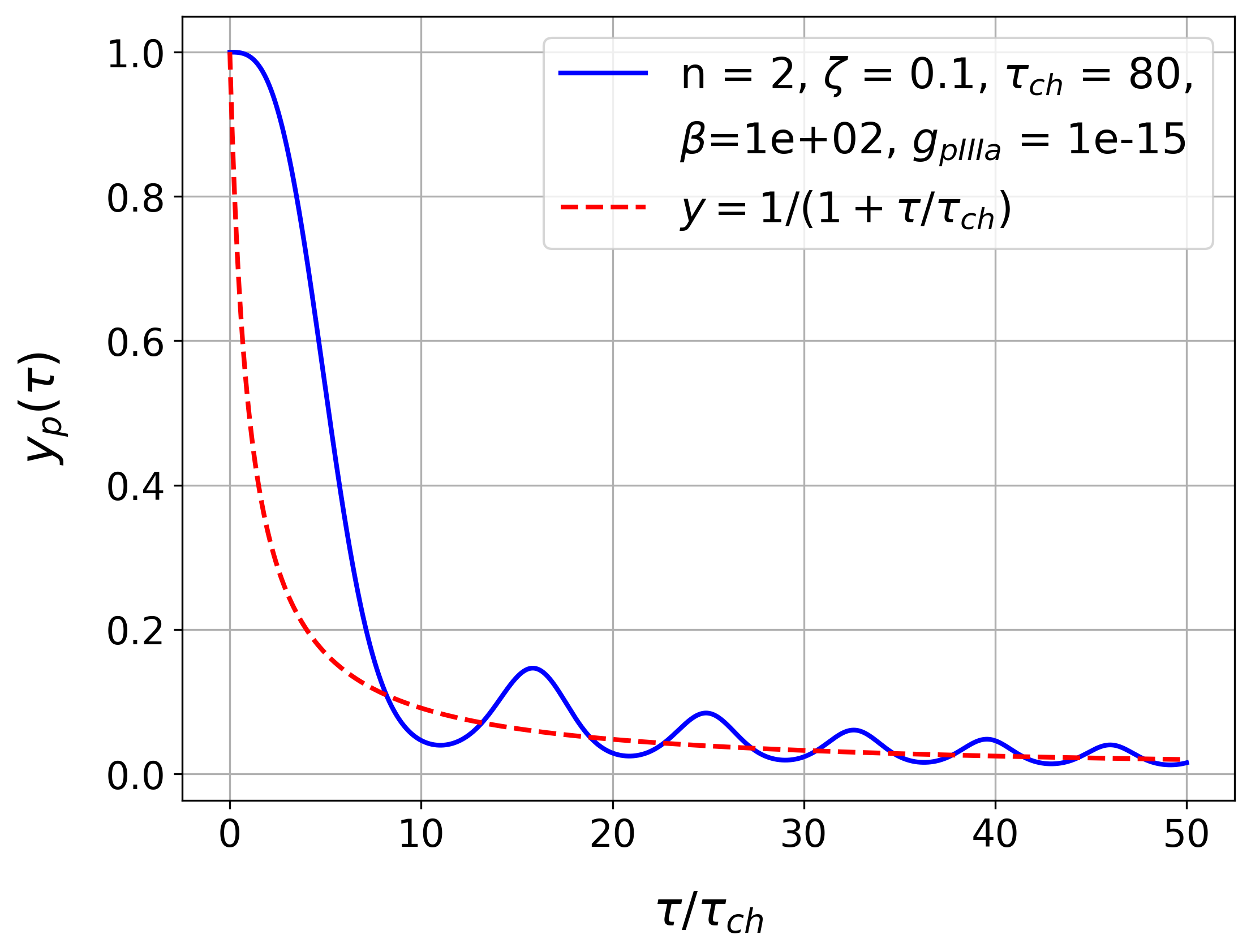} 
				\caption{} 
				\label{fig:subfig2_alpha}
			\end{subfigure}
			\hfill
			\begin{subfigure}[b]{0.32\textwidth} 
				\includegraphics[width=\textwidth]{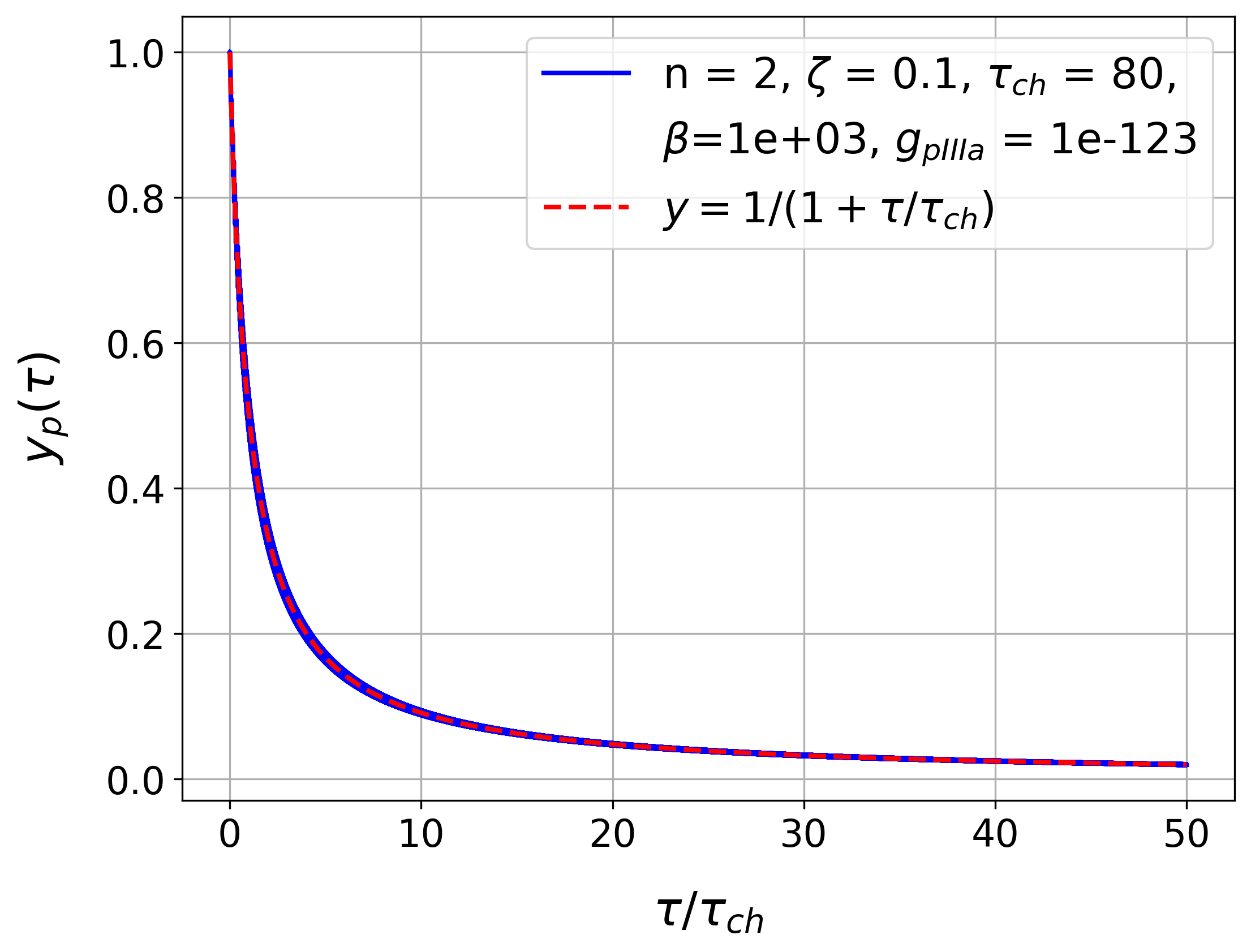} 
				\caption{} 
				\label{fig:subfig3_alpha}
			\end{subfigure}
			
			\vspace{1em} 
			
			\begin{subfigure}[b]{0.32\textwidth} 
				\includegraphics[width=\textwidth]{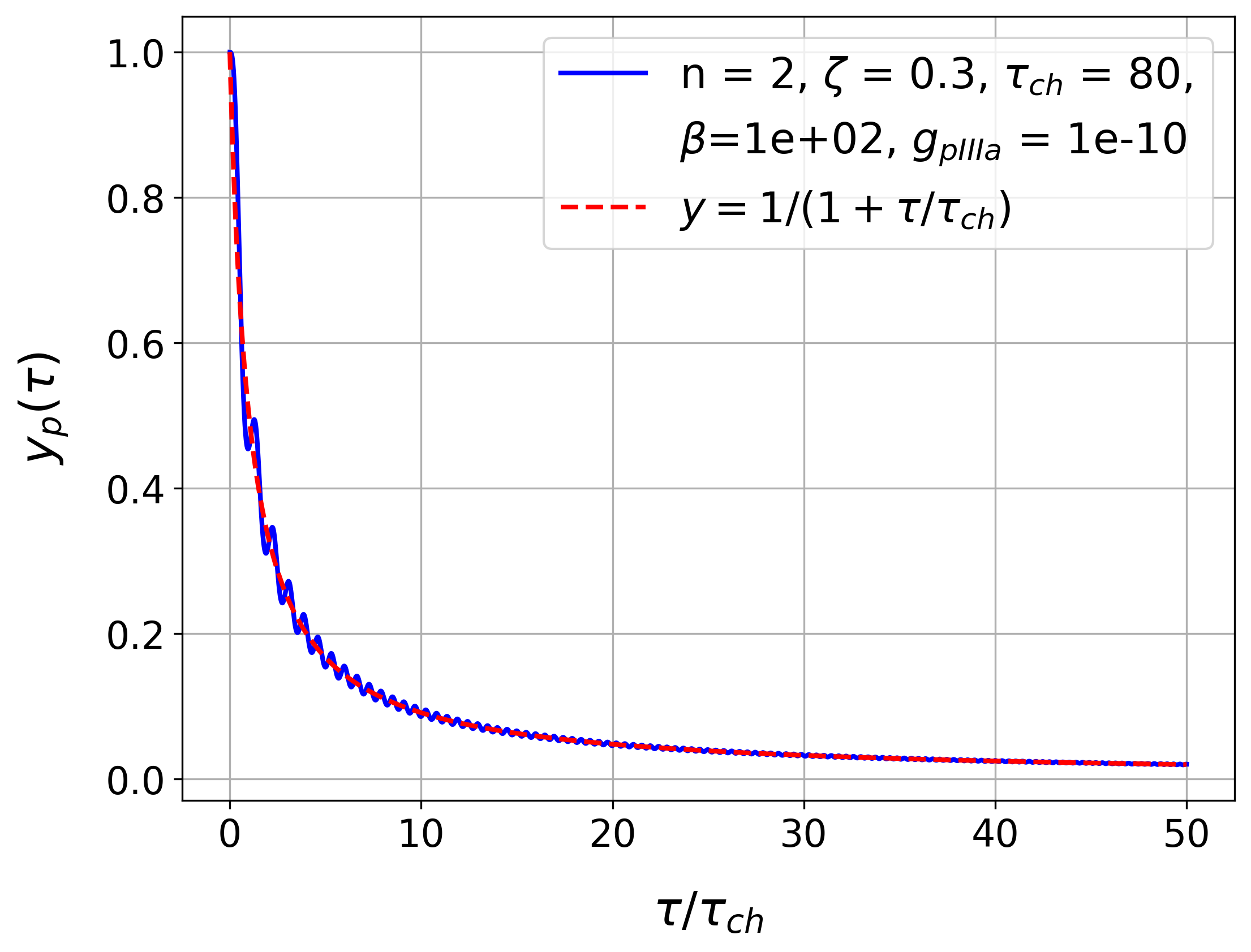} 
				\caption{} 
				\label{fig:subfig4_alpha}
			\end{subfigure}
			\hfill 
			\begin{subfigure}[b]{0.32\textwidth} 
				\includegraphics[width=\textwidth]{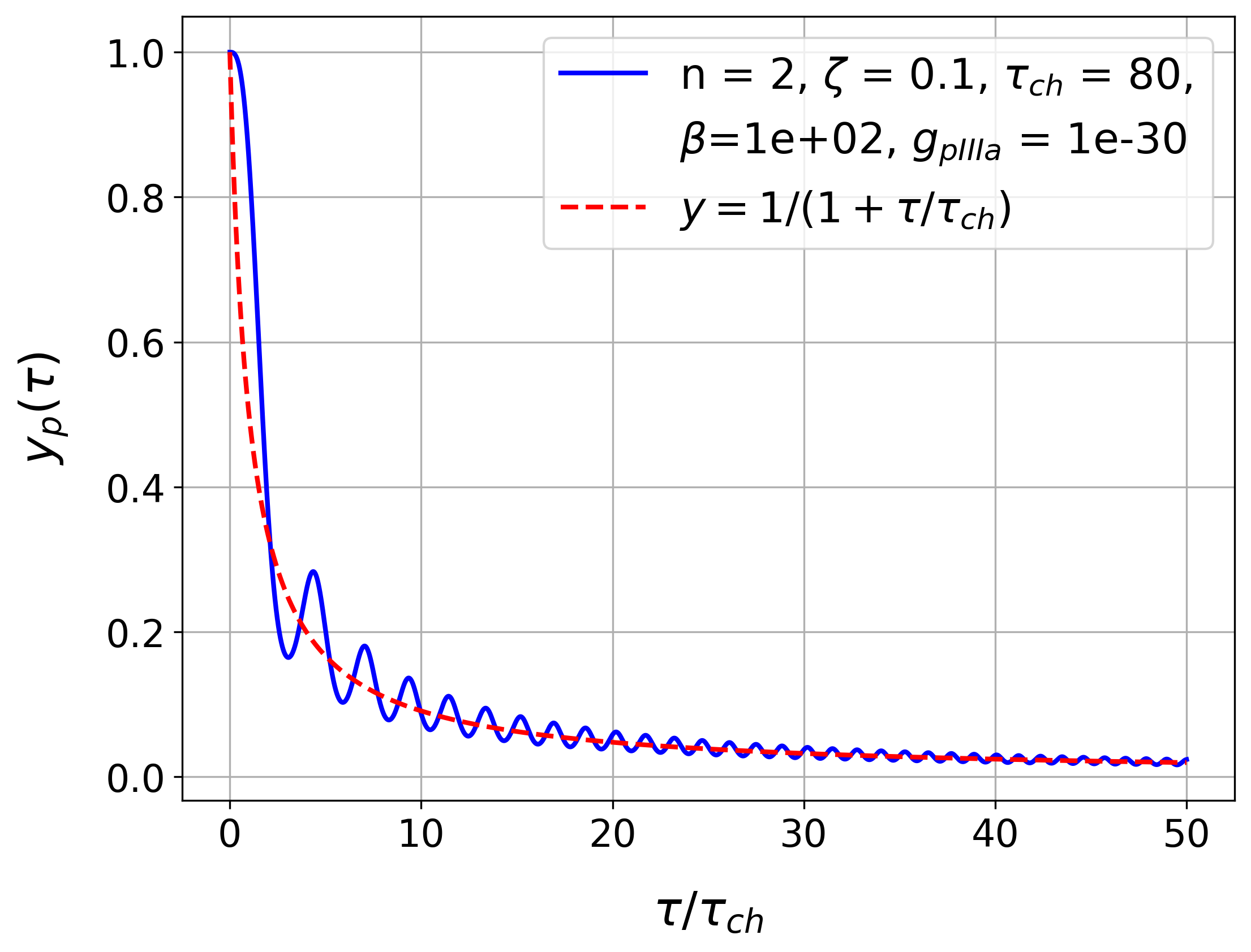} 
				\caption{} 
				\label{fig:subfig5_alpha}
			\end{subfigure}
			\hfill 
			\begin{subfigure}[b]{0.32\textwidth} 
				\includegraphics[width=\textwidth]{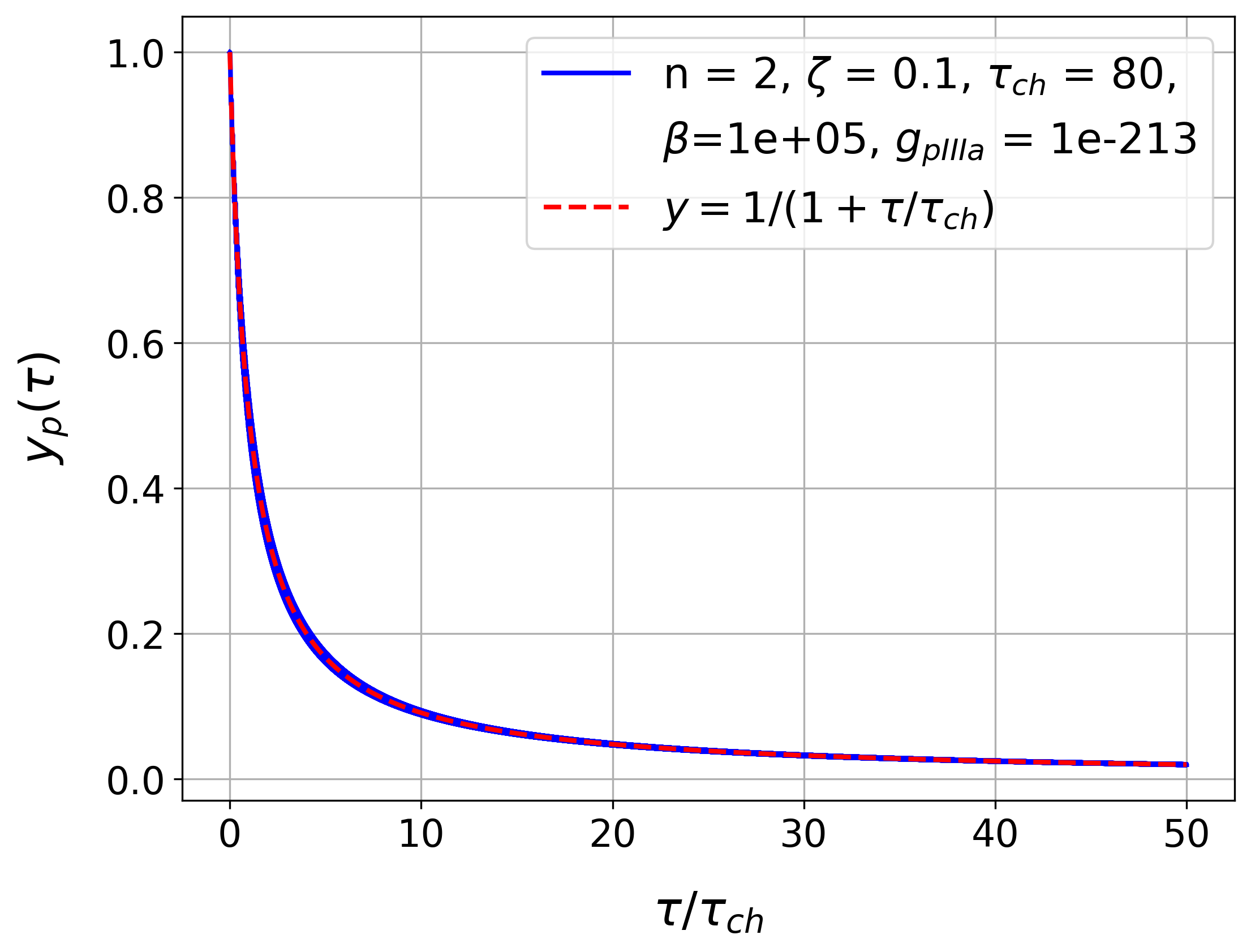} 
				\caption{} 
				\label{fig:subfig6_alpha}
			\end{subfigure}
			
			\caption{Oscillations of $y_p(\tau)$ in the power-law dark energy model with $\alpha$-attractor correction .} 
			\label{fig:fp_alpha}
		\end{figure}
		
		Within $f(R)$ gravity, the $\alpha$-attractor model corresponds to a polynomial modification of the Starobinsky model in the high-curvature regime \cite{miranda2017reconstructing}:
		\begin{equation}\label{polcorrStaro}
			f(R) = R + \left[\frac{(1 - \zeta)^2}{3M_\alpha^2}\right]^\zeta R^{\zeta + 1} + \frac{(1 - \zeta)^2}{6M_\alpha^2}R^2,
		\end{equation}
		where
		\begin{equation}
			\zeta \equiv 1 - \frac{1}{\sqrt{\alpha}}.
		\end{equation}
		Analysis of $n_s$ and $r$ as functions of $\zeta$ reveals excellent consistency with Planck observational constraints for $-0.02 < \zeta \lesssim 0.8$ (corresponding to $0.96 < \alpha \lesssim 25$). The upper limit $\zeta \approx 0.8$ corresponds to an $R^{1.8}$ correction to the Starobinsky potential, demonstrating the model's capacity to encompass diverse inflationary dynamics. A recent analysis in \cite{addazi2025curvature} investigated the impact of adding an $R^{3/2}$ correction to the Starobinsky model. The study found that this extension improves agreement with ACT data, without affecting the model’s consistency with other observational results. Interestingly, this type of modification is closely related to the $\alpha$-attractor class, suggesting that such models are worth further theoretical and phenomenological study as possible refinements of standard inflationary theory.

		The power-law dark energy model incorporating $\alpha$-attractor corrections takes the form:
		\begin{equation}\label{alpha_p}
			f(R) \simeq - \lambda R_{ch} \left[1 - \left(\frac{R_{ch}}{R} \right)^{2n} \right] + \left[\frac{(1 - \zeta)^2}{3 M_\alpha^2} \right]^\zeta R^{\zeta  + 1} + \frac{(1 - \zeta )^2}{6 M_\alpha^2} R^2
		\end{equation}
		The associated differential equation derived from Eq.~\eqref{eq:trace} becomes:
		\begin{align} \label{eq:6.5}
			y_p'' &+ 2n \frac{{y_p'}^2}{y_p}   + \frac{g_{pIIIa}}{6} y_p^{-(2n+2)} \left(y_p'' - 2 \frac{{y_p'}^2}{y_p} \right)  +\zeta g_{pIIIb} y_p^{-(2n + \zeta + 1)} \left[y_p'' - (\zeta + 1) \frac{{y_p'}^2}{y_p} \right] \nonumber \\
			&  + y_p^{-2n} \left[\left(1 + \frac{\tau}{\tau_{ch}} \right) - y_p^{-1} \right]  = 0
		\end{align}
		with coefficients defined as:
		\begin{align}
			g_{pIIIa} & = (1 - \zeta)^2 \frac{R_{ch}}{\lambda n (2n +1) M_\alpha^2} \beta^{2n + 2} \\
			g_{pIIIb} &= \left(\frac{(1 - \zeta)^2 (R_{ch} \beta)}{3 M_\alpha^2} \right)^\zeta \frac{ (\zeta + 1) \beta^{2n+1}}{2 \lambda n (2n +1)}  = \frac{(1+\zeta)}{2 \cdot 3^\zeta} \left( \frac{\lambda n (2n +1)}{\beta^{2n + 1}} \right)^{\zeta-1} (g_{pIIIa})^\zeta.
		\end{align}
		This differential equation reveals how $\alpha$-attractor corrections modify curvature oscillations through terms containing $g_{pIIIa}$ when $\zeta \neq 1$.
		
		Numerical solutions of Eq.~\eqref{eq:6.5} with initial conditions $y_p(0)=1$ and $y'_p(0)=0$ are presented in Figure~\ref{fig:fp_alpha} for various parameter combinations ($n, \zeta, \tau_{ch}, \beta, g_{pIIIa}$). Throughout our analysis, we maintain $0 < g_{pIIIa} \leq 1$ and $\beta \gg 1$ to emphasize the $\alpha$-attractor correction's influence. Parameter dependencies are organized as: $\zeta$ in panels (a) and (d), $g_{pIIIa}$ in (b) and (e), and $g_{pIIIb} \simeq 1$ in (c) and (f).

		Panels (a) and (d) reveal that significant curvature fluctuations occur primarily for $0 < \zeta \ll 1$, with amplitude decreasing sharply as $\zeta$ increases. This suggests that $\alpha$-attractor terms with $\zeta \gtrsim 0.3$ (equivalent to Starobinsky inflation with $R^{1.3}$ corrections) effectively suppress both curvature fluctuations and singularities in the power-law dark energy model. Panels (b) and (e) demonstrate that smaller $g_{pIIIa}$ values reduce fluctuation amplitudes while increasing their frequency. When $0 < g_{pIIIa} \ll 1$ (yielding $g_{pIIIb} \simeq 1$), the solution $y_p(\tau)$ closely follows the equilibrium curve $y = 1/(1 + \tau/\tau_{ch})$ for all $\beta \gg 1$ cases.
		The condition $g_{pIIIb} \simeq 1$ emerges as crucial for avoiding curvature singularities at arbitrary $\beta \gg 1$, allowing us to express the mass parameter as $M_\alpha^2/R_c = f(\beta, \zeta, n, \lambda)$. This dependence makes $M_\alpha^2$ particularly sensitive to the local density parameter $\beta$, resulting in values distinct from those constrained purely by inflationary considerations.

		The exponential-law dark energy model with $\alpha$-attractor corrections is given by:
		\begin{equation}
			F(R) = - \lambda R_{ch} (1 - \mathrm{e}^{-\frac{R}{R_{ch}}}) +  \left[\frac{(1 - \zeta)^2}{3 M_\alpha^2} \right]^\zeta R^{\zeta + 1} + \frac{(1 - \zeta)^2}{6 M_\alpha^2} R^2
		\end{equation}
		with corresponding curvature fluctuation equation:
		\begin{align}\label{eq:6.9}
			y_e'' &+ (\beta y_e^{-1} - 2) \frac{{y_e'}^2}{y_e} +   \frac{g_{eIIIa}}{2} \mathrm{e}^{\beta \left(\frac{1}{y_e} - 1\right)} \bigg(y_e'' - \frac{2 {y_e'}^2}{y_e} \bigg)  + g_{eIIIb} y_e^{-(\zeta - 1)} \mathrm{e}^{\beta \left(\frac{1}{y_e} - 1\right)} \bigg[y_e'' - (\zeta + 1) \frac{{y_e'}^2}{y_e} \bigg] \nonumber \\
			&  + y_e^2 e^{\frac{\beta}{y_e}} \left[\left(1 + \frac{\tau}{\tau_{ch}} \right) - y_e^{-1} \right] = 0,
		\end{align}
		where
		\begin{align}
			g_{eIIIa} &= (1 - \zeta)^2 \frac{2 R_{ch}}{3 \lambda M_\alpha^2} \mathrm{e}^\beta \\
			g_{eIIIb} &= \bigg(\frac{(1 - \zeta)^2 (R_{ch} \beta)}{3 M_\alpha^2} \bigg)^\zeta  \frac{(\zeta + 1)\mathrm{e}^\beta}{\beta \lambda}  = \frac{(1+\zeta)}{ 2^{\zeta}}  \left( \frac{ \lambda \beta}{  \mathrm{e}^\beta} \right)^{\zeta - 1} (g_{eIIIa})^\zeta.
		\end{align}
		\begin{figure}[htbp!] 
			\centering 
			\begin{subfigure}[b]{0.32\textwidth} 
				\includegraphics[width=\textwidth]{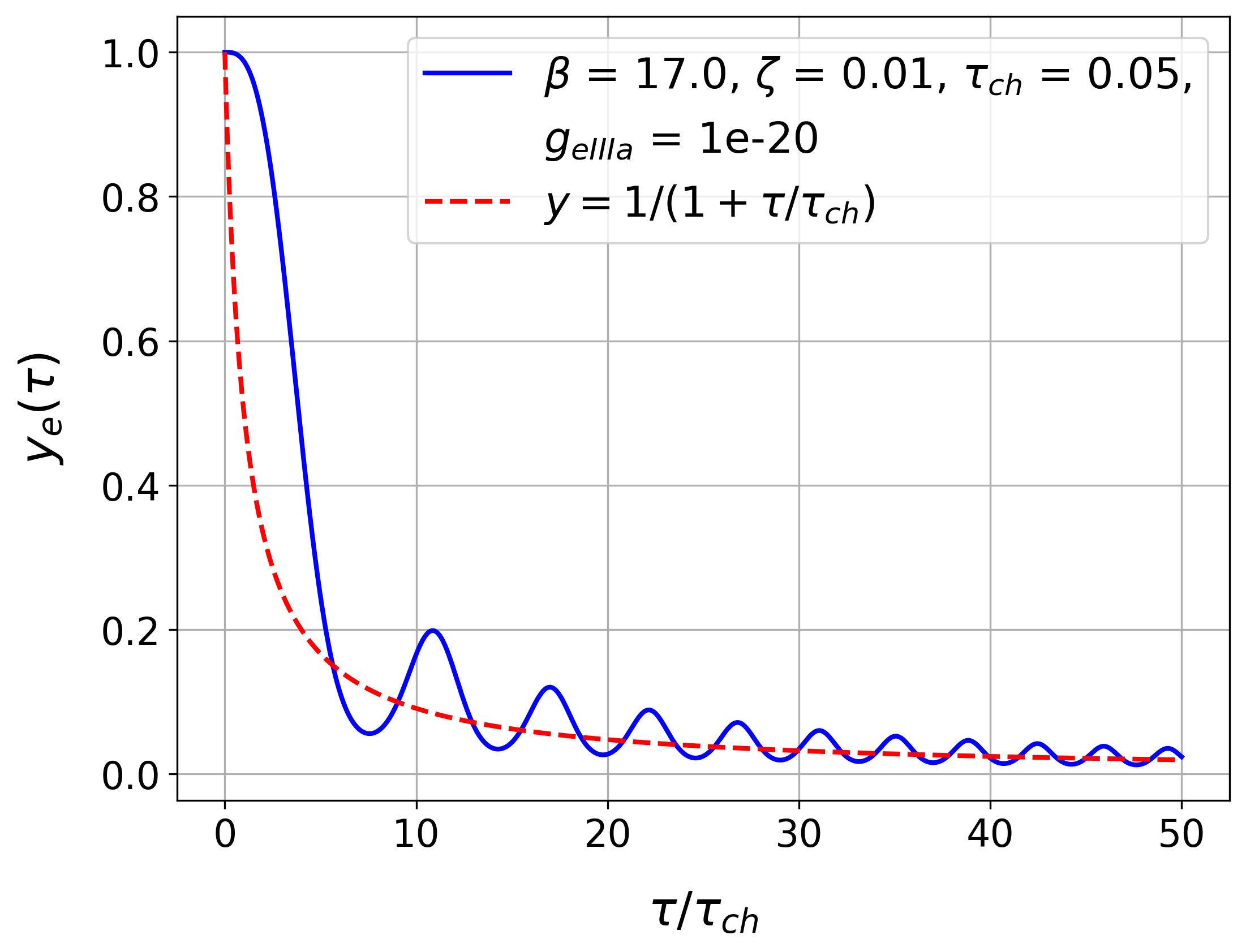} 
				\caption{} 
				\label{fig:subfig1_alpha2}
			\end{subfigure}
			\hfill 
			\begin{subfigure}[b]{0.32\textwidth} 
				\includegraphics[width=\textwidth]{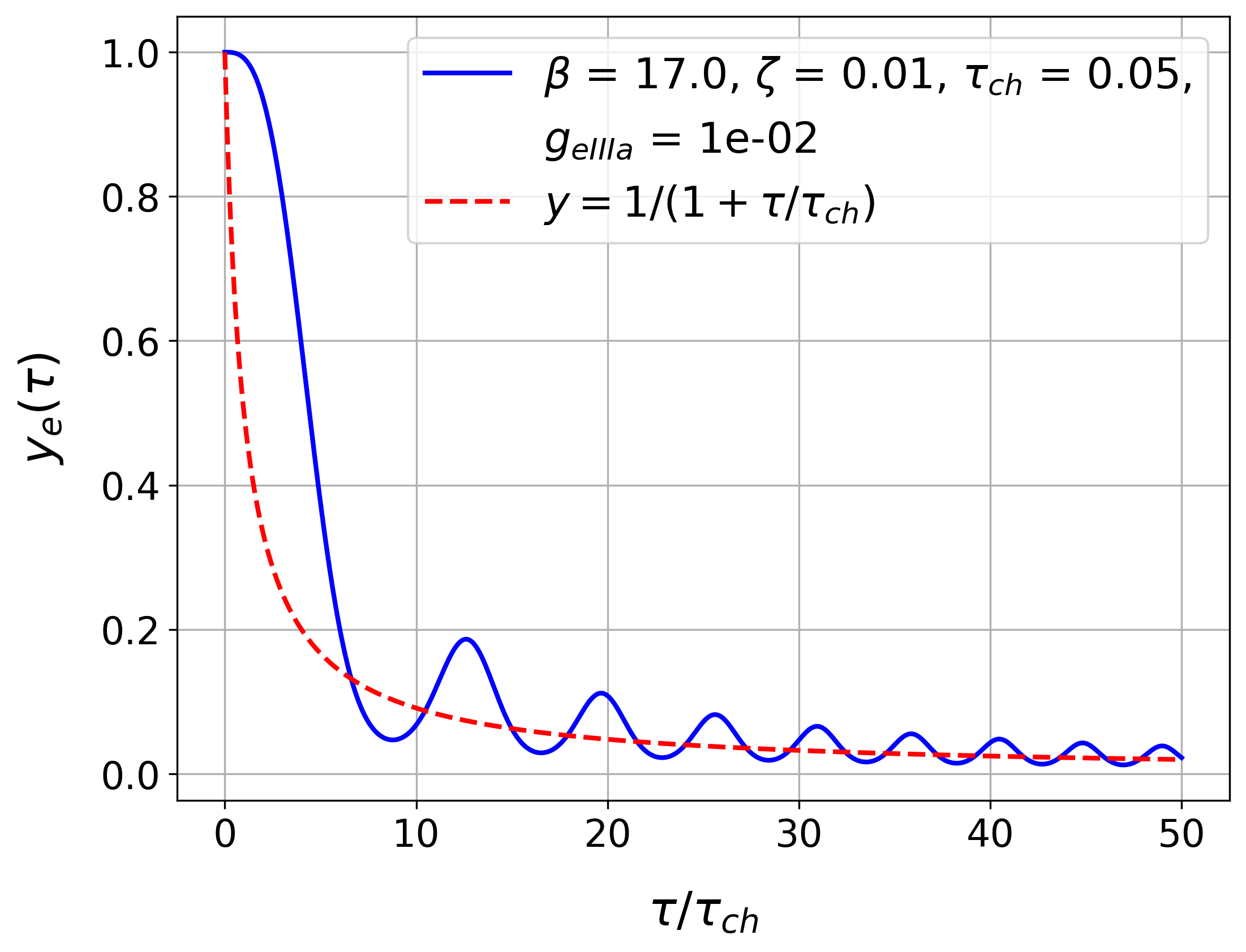} 
				\caption{} 
				\label{fig:subfig2_alpha2}
			\end{subfigure}
			\hfill 
			\begin{subfigure}[b]{0.32\textwidth} 
				\includegraphics[width=\textwidth]{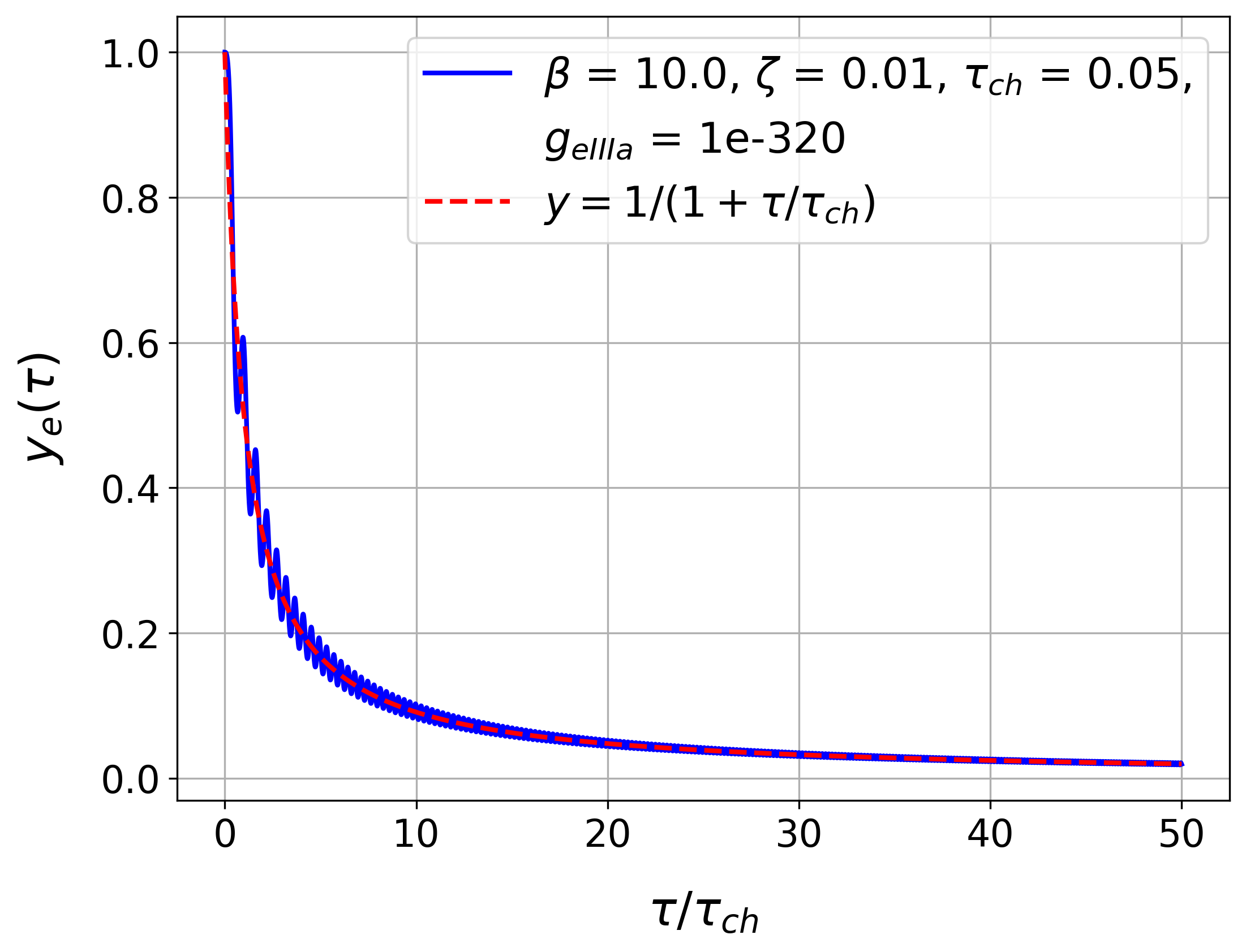} 
				\caption{} 
				\label{fig:subfig3_alpha2}
			\end{subfigure}

			\vspace{1em} 
			
			\begin{subfigure}[b]{0.32\textwidth} 
				\includegraphics[width=\textwidth]{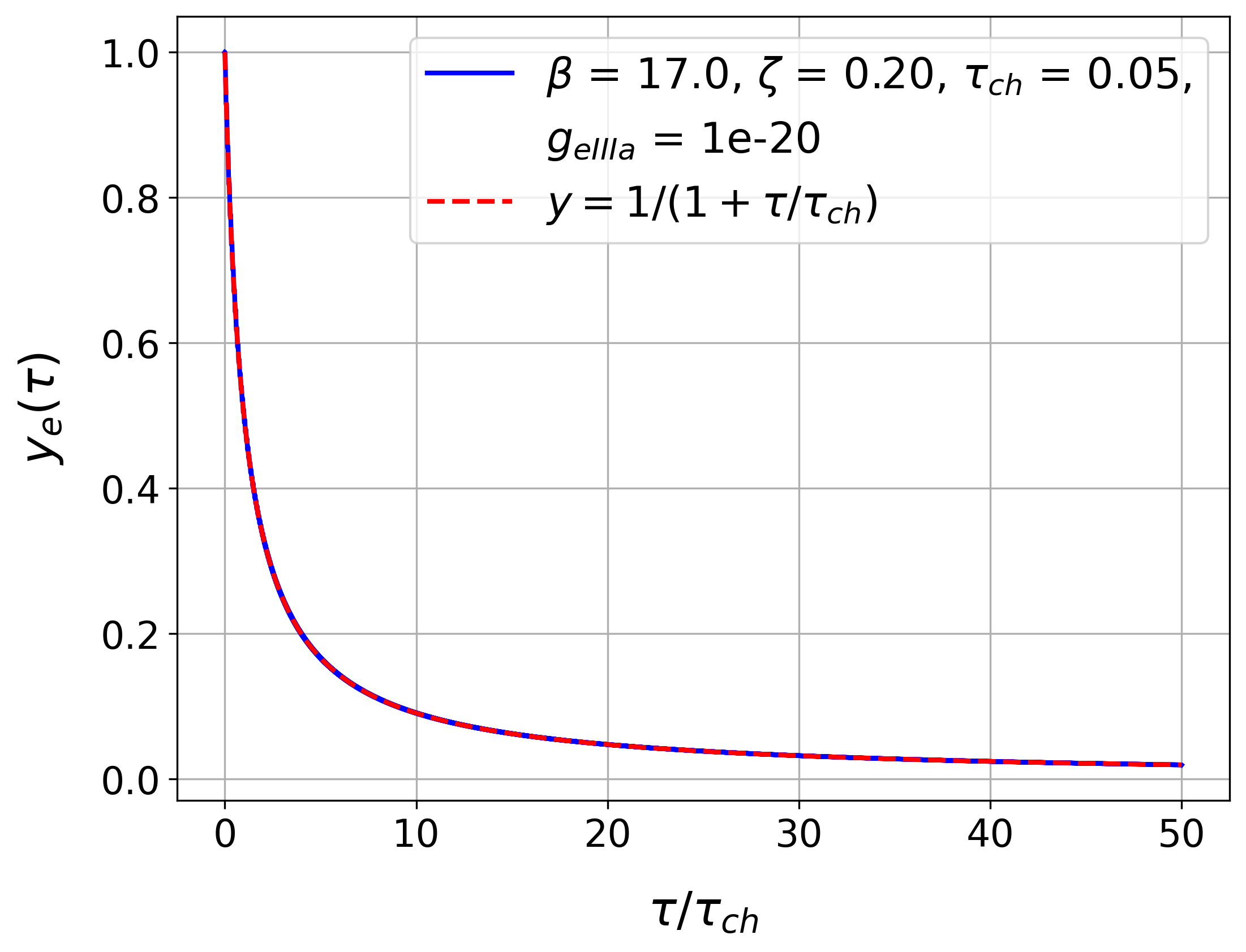} 
				\caption{} 
				\label{fig:subfig4_alpha2}
			\end{subfigure}
			\hfill 
			\begin{subfigure}[b]{0.32\textwidth} 
				\includegraphics[width=\textwidth]{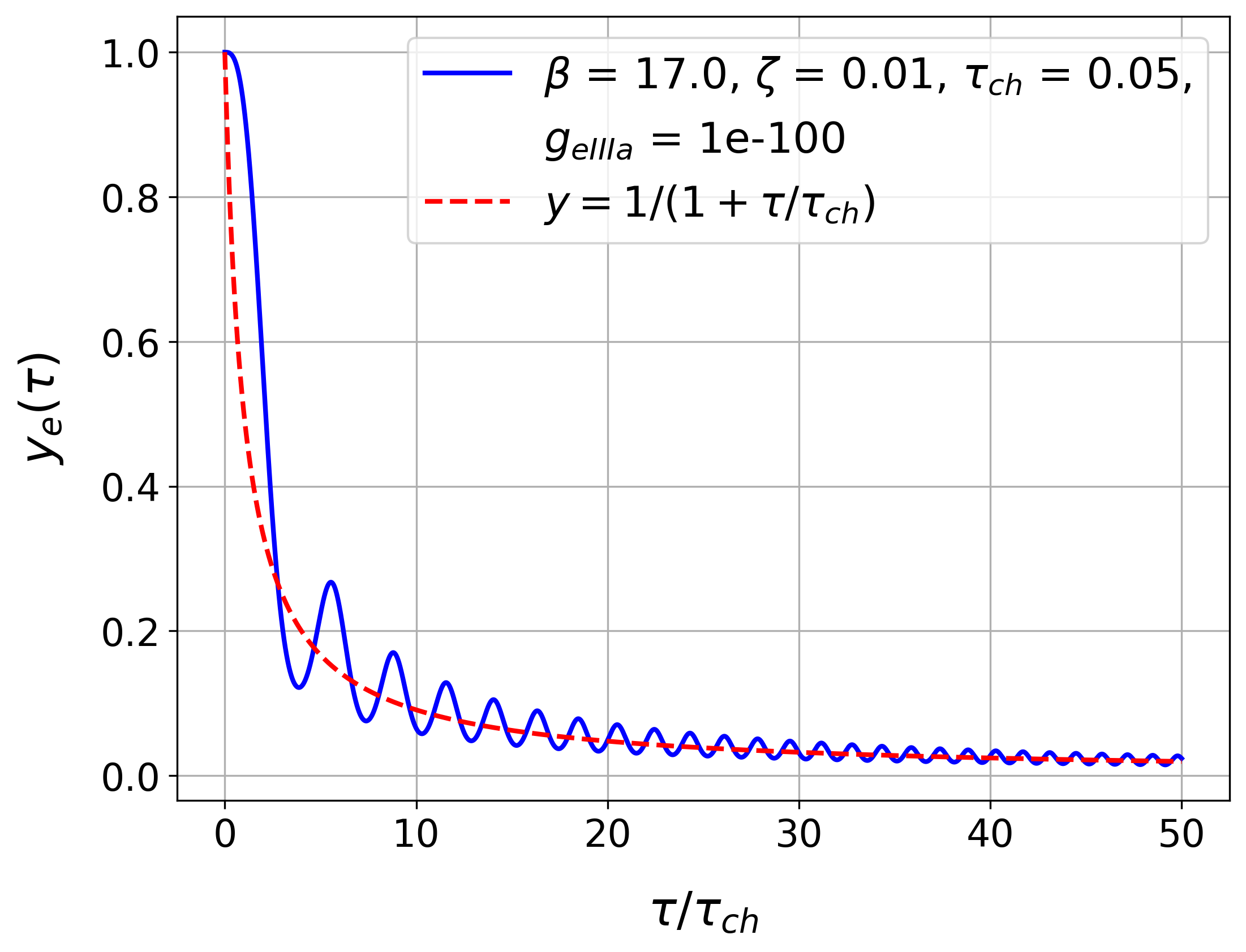} 
				\caption{} 
				\label{fig:subfig5_alpha2}
			\end{subfigure}
			\hfill 
			\begin{subfigure}[b]{0.32\textwidth} 
				\includegraphics[width=\textwidth]{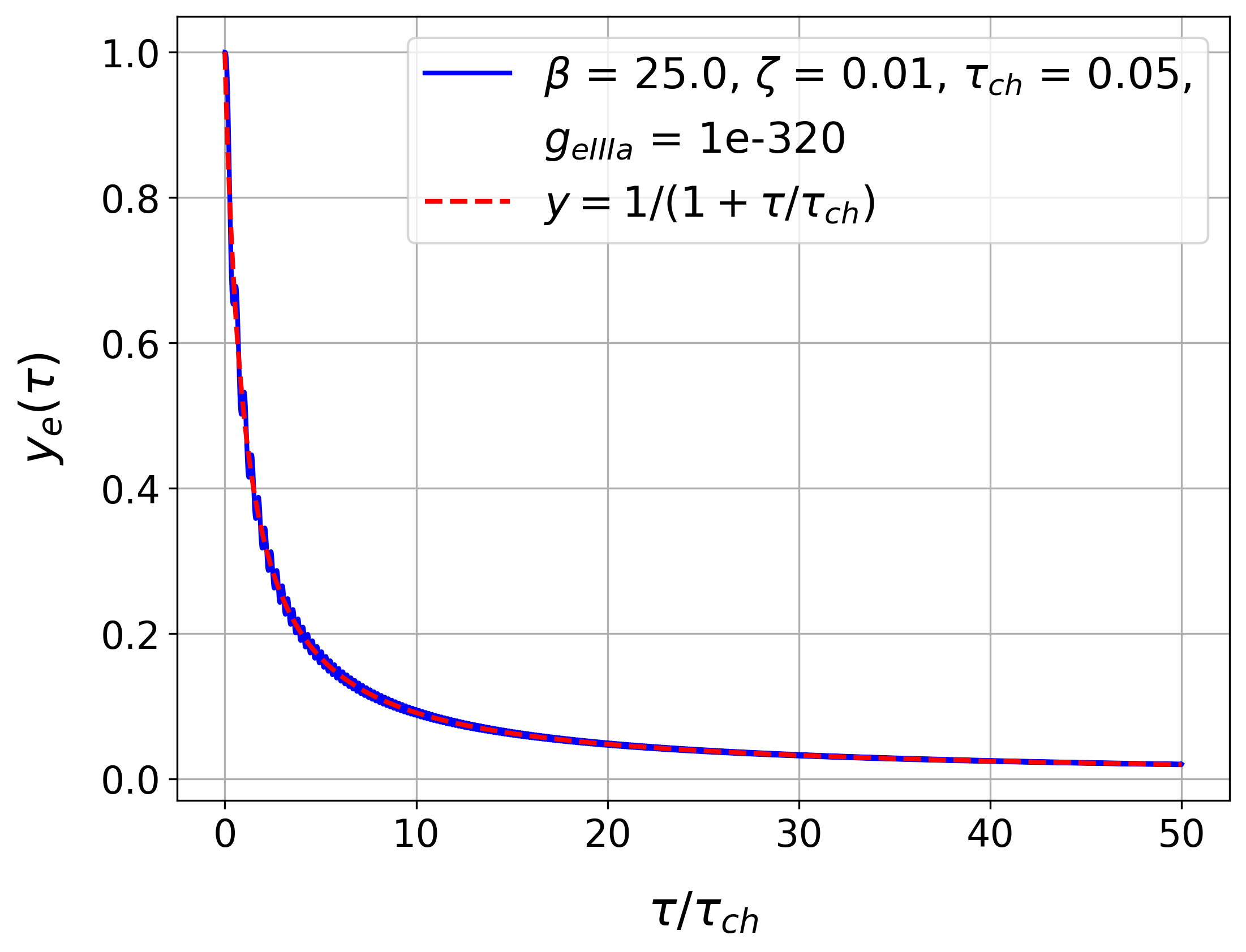} 
				\caption{} 
				\label{fig:subfig6_alpha2}
			\end{subfigure}
			\caption{Oscillations of $y_e(\tau)$ in the exponential-law dark energy model with $\alpha$-attractor correction.} 
			\label{fig:fe_alpha}
		\end{figure}
		
		Similar to the power-law dark energy case with $\alpha$-attractor corrections, the exponential-law dark energy model exhibits modifications to curvature oscillations through the $g_{eIIIa}$ term when $\zeta\neq 1$. Figure~\ref{fig:fe_alpha} displays the numerical solutions for $y_e(\tau)$ obtained from Eq.~\eqref{eq:6.9} across various parameter combinations of $\beta$, $\zeta$, $\tau_{ch}$, $\lambda$, and $g_{eIIIa}$. The left panels (a) and (d) illustrate the dependence on $\zeta$, while panels (b) and (e) demonstrate the effects of varying $g_{eIIIa}$ within the range $0<g_{eIIIa}<1$. The right panels (c) and (f) present results for different $\beta$ values under the extreme condition $g_{eIIIa}=10^{-320}$, which corresponds to $g_{eIIIb}$ values between $10^{2}$ and $10^{4}$. Notably, when $g_{eIIIa}$ approaches such extremely small values, the solutions begin to develop curvature singularities.
		
		The numerical results reveal several important features of the model. First, the amplitude of curvature oscillations in the exponential-law dark energy model decreases systematically with increasing $\zeta$ within the range $0<\zeta<1$. Remarkably, the solution $y_e(\tau)$ asymptotically approaches the equilibrium form $1/(1+\tau/\tau_{ch})$ even for very small $\zeta$ values ($\zeta=10^{-2}$). The solution profile further improves with larger values of the density parameter $\beta$ and smaller values of $g_{eIIIa}$, the latter leading to $g_{eIIIb}>10^2$. This behavior indicates that the $R^{1+\zeta}$ correction originating from the Starobinsky model - which emerges as a consequence of the $\alpha$-attractor effect - simultaneously enhances both the oscillation profile and the curvature singularity behavior in the exponential-law dark energy model. This improvement persists even for $\zeta\ll 1$, provided the condition $g_{eIIIa}\ll 1$ maintains $g_{eIIIb}>10^2$.
		
		Further numerical investigation shows that solution profiles qualitatively similar to panel (d) can be obtained for parameter combinations such as $\beta=30$ with $\zeta=0.4$ and $g_{eIIIa}\simeq 1.0$. However, when $\beta$ is reduced to values like $\beta=17$ while maintaining $\zeta=0.4$ and $g_{eIIIa}\simeq 1.0$, significant curvature fluctuations emerge, producing profiles resembling panel (c). This demonstrates that when $g_{eIIIa}\simeq 1.0$, proper selection of $\zeta$ within $0<\zeta<1$ can effectively suppress both singularities and curvature fluctuations in the exponential-law dark energy model.

\section{Conclusion}
		
Dark energy models formulated within $f(R)$ gravity frameworks inherently face the challenge of curvature singularities developing in high-density regimes. This pathological behavior emerges when examining the trace equation derived from $f(R)$ field equations, which governs the dynamics of curvature fluctuations. The singularity becomes apparent as the scalar curvature $R$ diverges ($R \to \infty$), mathematically equivalent to the condition where the rescaled variable $y \equiv \beta(R_{\text{ch}}/R)$ approaches zero. Numerical solutions of this system reveal that $y$ exhibits characteristic oscillations around the equilibrium trajectory $y = 1/(1 + \tau/\tau_{\text{ch}})$, where $\tau$ represents the rescaled time coordinate and $\tau_{\text{ch}}$ the rescaled characteristic timescale. These singularities manifest at finite times in both power-law and exponential-law $f(R)$ models when the local density parameter satisfies $\beta \gg 1$, with the physical singularity timescale $t_{\text{sing}} = \gamma_p\tau_{\text{sing}}$ falling within cosmologically relevant ranges of $10^{15}-10^{16}$ seconds.
		
To address this fundamental limitation, we implement curvature correction terms inspired by successful inflationary $f(R)$ models, specifically the Starobinsky $R^2$ term, its generalized extension $R^{\frac{m+2}{m+1}}$, and the $\alpha$-attractor polynomial extension. These corrections serve the dual purpose of resolving curvature singularities while preserving compatibility with inflationary dynamics, thereby offering a unified framework for cosmic acceleration across different epochs. The stabilization mechanism operates through distinct dimensionless parameters: the $R^2$ correction is governed by $g_{pI}$ for power-law models and $g_{eI}$ for exponential models, while the generalized extension introduces $g_{pII}$ and $g_{eII}$ parameters, and the $\alpha$-attractor modification involves $g_{pIIIa,b}$ for power-law models and $g_{eIIIa,b}$ for exponential models.
		
Comprehensive numerical simulations demonstrate that curvature singularities can be completely avoided when specific threshold conditions are met: $g_{pI,eI} \geq 1$ for the Starobinsky case, $g_{pII,eII} \geq 1$ for the generalized extension, and $g_{pIIIb} \approx 1 $ and $ g_{eIIIa} \approx 1$ for the $\alpha$-attractor model. These stability conditions necessitate inflationary parameters that substantially differ from conventional inflation-constrained values. Notably, the effective mass scale $m_R$ required for successful $R^2$ corrections must be approximately 33 orders of magnitude lower than the standard Starobinsky scale. The exponent $m$ in generalized extensions shows optimal performance when approaching the Starobinsky limit ($m\to 0$), while the $\alpha$-attractor parameter $\zeta = 1-\frac{1}{\sqrt{\alpha}}$ demonstrates maximum efficacy within the range $0<\zeta<1$.
		
The analysis yields three principal insights regarding model construction. First, the Starobinsky $R^2$ term emerges as the most effective singularity resolution mechanism, particularly when other extensions asymptotically approach its functional form. Second, successful singularity avoidance requires the introduction of new mass scales distinct from traditional inflationary parameters. Third, polynomial extensions of the form $\propto R^{\zeta+1} + R^2$ demonstrate promising capacity to simultaneously describe early-universe inflation and late-time dark energy while maintaining mathematical regularity. These findings collectively establish that incorporating inflation-motivated corrections provides a viable pathway to singularity-free $f(R)$ dark energy models, with the necessary parameter adjustments preserving the models' ability to unify the description of cosmic acceleration across different cosmological epochs.

\bibliographystyle{ws-ijmpa}
\bibliography{references}

\end{document}